\documentclass[letterpaper,english,aps, nofootinbib, prd, twocolumn, lengthcheck, superscriptaddress]{revtex4-2}
\usepackage[T1]{fontenc}

\UseRawInputEncoding 

\usepackage[colorlinks=true,linktocpage=true,linkcolor=blue,citecolor=blue]{hyperref}
\newcommand{\Slash}[1]{{\ooalign{\hfil/\hfil\crcr$#1$}}}
\usepackage{verbatim}
\usepackage{bm}
\usepackage{amsmath}
\usepackage{amssymb}
\usepackage{graphicx}
\usepackage{slashed}
\usepackage{wrapfig}
\usepackage{bbm}
\usepackage[caption=false]{subfig}
\usepackage{verbatim}

\newcommand \beq{\begin{eqnarray}}
\newcommand \eeq{\end{eqnarray}}

\newcommand{\la}{\langle}
\newcommand{\ra}{\rangle}

\newcommand{\Tr}{{\rm Tr}}

\newcommand{\Nc}{N_{\rm c}}
\newcommand{\Nf}{N_{\rm f}}

\newcommand{\lqcd}{\Lambda_{\rm QCD}}

\newcommand{\calL}{\mathcal{L}}

\newcommand{\calP}{\mathcal{P}}

\newcommand{\vq}{\vec{q}}

\newcommand{\rmd}{\mathrm{d}}
\newcommand{\rmi}{\mathrm{i}}
\newcommand{\rme}{\mathrm{e}}

\usepackage[colorlinks=true,linktocpage=true,linkcolor=blue,citecolor=blue]{hyperref}

\begin{document}
\title{Phenomenological QCD equations of state for neutron star dynamics:  \\
Nuclear-2SC continuity and evolving effective couplings 
}

\author{Toru Kojo}
\author{Defu Hou}
\author{Jude Okafor}
\affiliation{Key Laboratory of Quark and Lepton Physics (MOE) and Institute of Particle Physics, Central China Normal University, Wuhan 430079, China}
\author{Hajime Togashi}
\affiliation{Department of Physics, Kyushu University, Fukuoka, 819-0395, Japan}
\date{\today}

\begin{abstract}
We delineate the quark-hadron continuity by constructing QCD equations of state for neutron star dynamics, covering the wide range of charge chemical potential ($\mu_Q$) and temperatures ($T$). 
Based on the nuclear-2SC continuity scenario, we match equations of state for nuclear and two-flavor color-superconducting (2SC) quark matter, where the matching baryon density is $n_B\simeq 1.5n_0$ ($n_0\simeq 0.16\, {\rm fm^{-3}}$: nuclear saturation density). The effective vector and diquark couplings in a quark matter model evolve as functions of 
$n_B$ or ($n_B, \mu_Q, T$), 
whose low density values are constrained by the nuclear matter properties and neutron star radii, with the high density behavior by the two-solar mass ($2M_\odot$) constraint. 
With couplings dependent on $n_B$,
we examined how smooth the nuclear-2SC continuity can be, and found problems in matching nuclear and 2SC entropies at low temperatures; 
they differ unless the baryon Fermi velocity significantly increases to match with the quark's, or the 2SC matter allows low energy collective modes whose velocities are as low as the baryon's.
This implies that the realization of the nuclear-2SC continuity, if possible, demands additional ingredients to the conventional nuclear and 2SC descriptions.
To proceed with the continuity scenario, we enforce smooth matching by making the couplings ($n_B, \mu_Q, T$)-dependent.
In effect, this adds phenomenological contributions which we call ``X'' to emphasize our ignorance on the practical description.
After the phenomenological matching, we take the rest as our predictions. The 2SC and color-flavor-locked (CFL) phases computed with these evolving couplings are called 2SCX and CFLX.
The CFLX appears around $n_B\simeq 2$-$4n_0$ and, in contrast to the conventional CFL, has non-negligible dependence on $(\mu_Q,T)$. 
To examine the astrophysical consequences of our modeling, we add charged leptons and neutrinos, and study the composition of matter for lepton fractions relevant for protoneutron stars and neutron star mergers.
The abundance of neutrinos and thermal effects reduce the strangeness fraction and stiffen equations of state. For a neutrino trapped neutron star at $T\simeq 30 $ MeV with a lepton fraction $Y_L\simeq 0.05$, the mass is larger than its cold static counterpart by  $\sim 0.1M_\odot$.
\end{abstract}

\maketitle

\section{Introduction}

\begin{figure*}[t]
\begin{center}
\vspace{-1.6cm}
\includegraphics[scale=0.4]{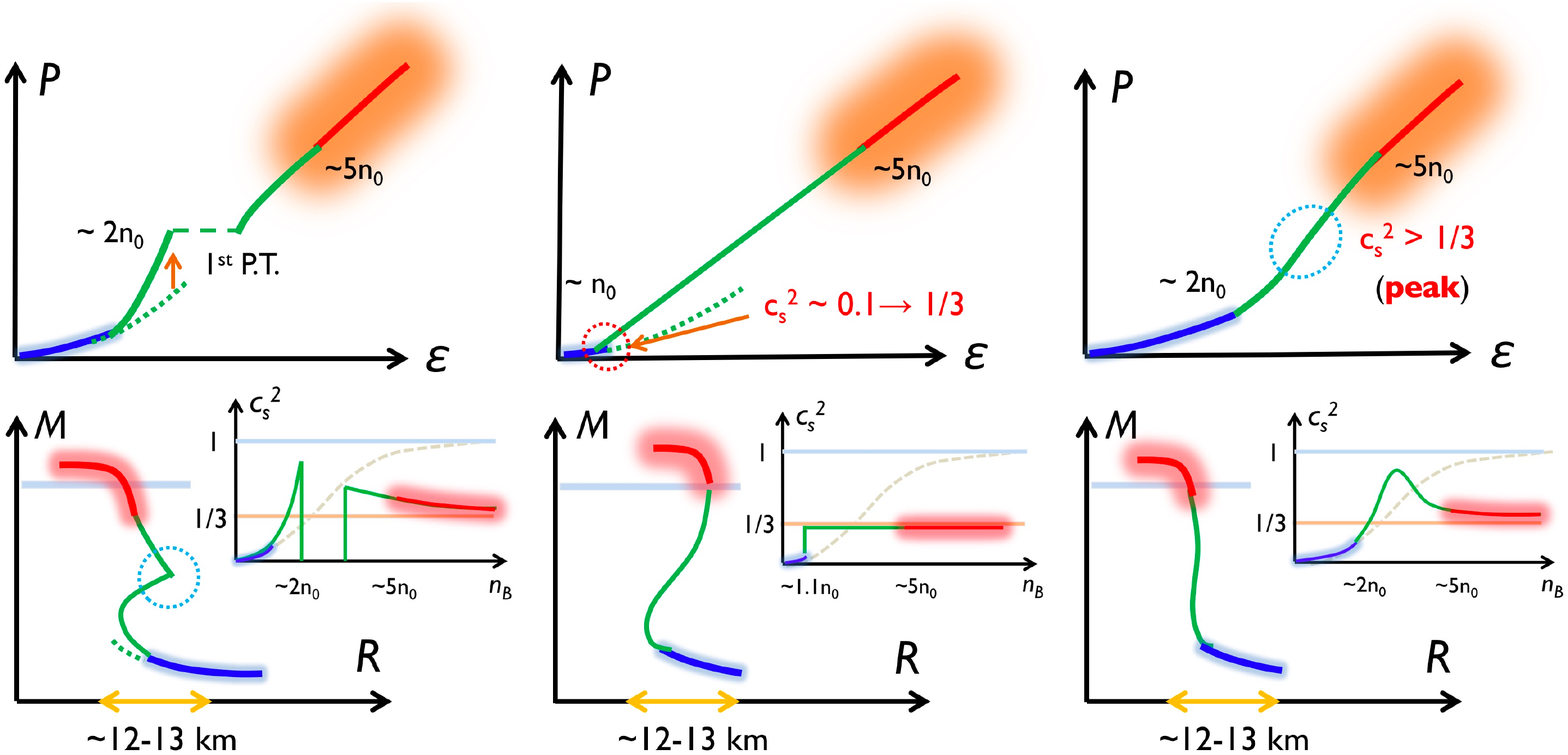}
\end{center}

\vspace{-1.6cm}
\caption{Schematic descriptions for the relation between $M$-$R$, $P$-$\varepsilon$, and $c_s^2$-$n_B$ for three types of equations of state compatible with the current observations. See the main text.
}
\label{fig:schematic_cs2}
\end{figure*}

Recent observations in neutron stars provide us with hints to delineate the properties of dense matter in Quantum Chromodynamics (QCD) \cite{Fukushima:2010bq,Fukushima:2013rx}. 
The mass-radius ($M$-$R$) relations of neutron stars are determined by the relation between the pressure $P$ and the energy density $\varepsilon$ of matters, and have one-to-one correspondence with the QCD equation of state. 
Finding two solar mass ($2M_\odot$) neutron stars \cite{Arzoumanian:2017puf,Antoniadis:2013pzd,Cromartie:2019kug} demands equations of state at high density to be stiff ($P$ is large at given $\varepsilon$), 
while the X-ray observations \cite{Watts:2016uzu,Miller:2019cac,Riley:2019yda}
and the tidal deformability \cite{Abbott:2018wiz,Annala:2017llu,De:2018uhw}
suggest the low density part to be relatively soft, leading to $R_{1.4} \simeq $12-13 km ($R_{1.4}$: the radius of $1.4M_\odot$ neutron stars). This soft-to-stiff evolution of equations of state leads to a steep growth in the speed of sound, $c_s = ( \partial P/\partial \varepsilon)^{1/2} $, but the growth must be moderate enough not to violate the causality constraint, $c_s \le 1$ (natural unit) \cite{Bedaque:2014sqa,Tews:2018kmu}. 
For the interplay between the low and high density constraints, see Ref.\cite{Drischler:2020fvz} for recent comprehensive studies.

Typical calculations suggest that a neutron star with its mass larger than $2M_\odot$ has the core density $\gtrsim 5n_0$ ($n_0\simeq 0.16{\rm fm}^{-3}$: nuclear saturation density) which is presumably too high for purely hadronic descriptions. 
Quark matter \cite{Itoh:1970uw,Collins:1974ky} is a natural alternative for the high density part.   
In light of the soft-to-stiff evolution of equations of state, there are at least three possible descriptions for the soft-to-stiff evolution, as shown in  Fig.\ref{fig:schematic_cs2} (for more details, see Refs.\cite{Kojo:2020krb,Alford:2013aca}): 
(i) Matter quickly gets stiffened beyond $n_0$, and the equations of state around $\sim 1.5$-$2n_0$ are so stiff that they can remain stiff even after the first order phase transition \cite{Benic:2014jia,Alvarez-Castillo:2016oln}.
(ii) Just above $\sim n_0$ matter quickly evolves into a matter stiffer than nuclear matter, with a radical change in $c_s^2$ from $\sim 0.1$ to $\simeq 1/3$ at low density, 
$\simeq 1.1$-$1.5n_0$ \cite{Freedman:1977gz,Witten:1984rs,Annala:2019puf}.
(iii) The nuclear matter picture is valid to $\sim 2n_0$, and then, the nuclear matter begins to transform to quark matter continuously, leading to a peak in $c_s^2$  \cite{Masuda:2012kf,Kojo:2014rca,McLerran:2018hbz}.

Based on case (iii), seminal works used the concept of the quark-hadron continuity \cite{Baym:1976yu,Schafer:1998ef,Hatsuda:2006ps,Yamamoto:2007ah}
as baselines to construct unified equations of state \cite{Masuda:2012kf,Masuda:2012ed,Masuda:2015kha,Kojo:2014rca,Baym:2017whm,Baym:2019iky}. 
Given that relevant degrees of freedom are uncertain for $\sim 2$-$5n_0$, the previous works have constructed equations of state for static neutron stars by interpolating the nuclear equations of state at $n_B\simeq 2n_0$ to those of quark matter at $n_B=4$-$7n_0$ \cite{Kojo:2014rca}. The quark matter is in the color-flavor-locked (CFL) phase with $u, d, s$-quarks forming condensed diquark pairs for a color-superconductor (CSC) \cite{Alford:2007xm}. 
For the nuclear part, we used the Akmal-Pandharipande-Ravenhall (APR) \cite{Akmal:1998cf} for QHC18 (Quark-Hadron-Crossover) equation of state \cite{Baym:2017whm}, and the Togashi \cite{Togashi:2017mjp} for QHC19 \cite{Baym:2019iky}, and found that these equations of state are consistent with available neutron star constraints. (The tables and manuals can be found, e.g., at [https\://compose.obspm.fr/home].)

Meanwhile, the previous interpolation method did not answer the composition of matter in the intermediate region, and did not describe how the degrees of freedom change from hadronic to quark matter. 
In this respect it is useful to extend equations of state for static neutron stars to those with general charge chemical potential $\mu_Q$ and temperature $T$ (see, e.g., a comprehensive review \cite{Oertel:2016bki}); 
in fact, more constraints can be found through this extension. 
Like transport quantities \cite{Schmitt:2017efp}, 
perturbing equations of state by $(\mu_Q,T)$ gives us insights on the degrees of freedom near the Fermi surface. The matter composition affects observables 
from neutron star dynamics \cite{Alford:2017rxf,Most:2018eaw,Bauswein:2020aag}
and neutron star cooling \cite{Yakovlev:2004iq,Page:2004fy}.
(For hadronic equations of state, see, e.g., Refs.\cite{Lattimer:1991nc,Shen:1998gq,Steiner:2012rk,Typel:2009sy}.)

To compute equations of state, in practice we are forced to put some assumptions on the degrees of freedom. 
Although current methodologies cannot pin down which assumption is correct, it should be useful to make a catalog of various scenarios and summarize the characteristic features.
Within the quark-hadron continuity scenario, one can consider, e.g., nuclear-2SC-CFL, or nuclear-hypernuclear-CFL type continuity, or else.
As our first step for the specific realization of the quark-hadron continuity, in this work
we focus on the nuclear-2SC continuity \cite{Fukushima:2015bda,Fujimoto:2019sxg}, where only $u$ and $d$ quarks are involved at the matching region. In the 2SC quark matter, two colored quarks (say, $R$ and $G$) participate in the diquark pairing and acquire the gaps, while $uB$ and $dB$ quarks remain gapless \cite{Alford:2007xm}. 
Using the Togashi equation of state \cite{Togashi:2017mjp} for the nuclear part, we consider the matching at relatively low density, $n_B\simeq 1.5n_0$. Such a matching was attempted previously along the $\beta$-equilibrium line at $T=0$ \cite{Fukushima:2015bda}, while this work extends the matching to more general cases.

To describe the nuclear-2SC continuity we let the effective couplings of quark models, the vector coupling $g_V$ and diquark coupling $H$ \cite{Kojo:2014rca}, evolve as functions of $(n_B, \mu_Q, T)$. Its high density ($n_B \gtrsim 5n_0$) behavior is constrained by the $2M_\odot$ constraint.

In the first analysis we allow only the $n_B$-dependence of the couplings and examine how reasonable the nuclear-2SC continuity can be. The most significant problem is found in the entropy matching. Interestingly, nuclear and 2SC matters have the same number of gapless fermionic modes which, at first glance, look supportive of the continuity scenario. However, actually smooth matching requires also the matching of the Fermi velocities of baryons and quarks, and this is not satisfied in conventional modeling. Therefore the first conclusion of this paper is that the nuclear-2SC continuity is not realized by direct matching of {\it conventional} nuclear and 2SC calculations.

At this stage, there are several alternative scenarios to take. 
The first is to give up the continuity scenario and allow the first order phase transition between conventional nuclear and 2SC phases. 
If we take the nuclear-2SC transition as real, we also take the 2SC-CFL phase boundary as of the first order, and then, we would have the first order phase transitions twice (at the nuclear-2SC and 2SC-CFL boundaries). 
The resulting equations of state would have trouble with the $2M_\odot$ constraints, unless extreme stiffening takes place between the phase boundaries. 
Another scenario (which we adopt in this paper) is that, around the matching region, 
both nuclear and 2SC matter are modified from the conventional ones in such a way that nuclear matter contains the 2SC pair correlation and 2SC matter includes baryonic three particle correlations. 
This sort of picture is what we had in mind in seminal works \cite{Kojo:2014rca,Baym:2017whm,Baym:2019iky} for the quark-hadron continuity.
Unfortunately we are still unable to demonstrate such computations. We are forced to use some phenomenological schemes; in this paper we let the couplings depend on ($n_B, \mu_Q, T$) so that thermodynamic quantities in the matching region are smooth in all directions of $(\mu_B, \mu_Q, T)$.

The CSC phases computed in our evolving couplings differ from the ordinary ones as they generate extra contributions. We use conventional CSC phases as baselines, but there are phenomenological corrections from the derivatives of evolving couplings. For this reason we call the CSC phases computed in this work ``CSCX'', where X emphasizes our ignorance.
The X needed here should be able to react to moderate changes in $(\mu_Q,T)$, and hence is likely to be gapless. 

After imposing nuclear-2SC matching, the rest is taken as our model predictions.
An important issue is how the strangeness appears. 
Within our modeling, the strangeness begins to appear within the 2SC quark matter and it drives the first order transition from 2SC to CFL around $n_B\simeq 2$-$4n_0$. 
So, it turns out that in this study we are working for case (i) in Fig.\ref{fig:schematic_cs2}. 
For the CFL equations of state (after the first transition) to be consistent with the $2M_\odot$ constraint, 
the 2SC equations of state must be stiff. In particular the speed of sound exceeds the conformal value, $(1/3)^{1/2}$, already around $n_B \simeq 2n_0$. 
Since  the location and strength of the 2SC-CFL transition have large impacts on neutron star structures, we perturb equations of state by adding flavor-dependent density-density repulsions. 
These short-range effects have been indicated by the lattice QCD simulations for baryon-baryon interactions  \cite{Ishii:2006ec,Iritani:2018sra} 
and in line with the  predictions of constituent quark models \cite{Oka:1980ax,Oka:1982qa}.

To test our descriptions, it is important to look for possible astrophysical consequences. In this context we study equations of state for cold and dynamic neutron stars.
For the neutron star dynamics, we consider equations of state in the neutrino trapping regime. The abundance of neutrinos (anti-neutrinos) reduces (enhances) the strangeness fraction and stiffens (softens) equations of state at $n_B\sim 2$-$4n_0$. For a neutrino trapped neutron star at $T\simeq 30 $ MeV with a lepton fraction $Y_L\simeq 0.05$, the mass is larger than its cold static counterpart by  $\sim 0.1M_\odot$.
In this paper we present the equations of state for particular sets of temperatures and neutrino densities. More extensive results and numerical tables are being prepared.

This paper is structured as follows. 
In Sec.\ref{sec:model} we explain the physics in our quark model and how to implement the scheme of evolving effective couplings. 
In Sec.\ref{sec:matching} we discuss the range of $(n_B,\mu_Q,T)$ of our interest, and discuss the low and high density constraints.
In Sec.\ref{sec:nuclear-2SC} we discuss QCD equations of state with evolving couplings whose low density values are fixed at $\mu_Q=T=0$ and do not depend on $(\mu_Q,T)$.
The purpose here is to examine the quality of matching and identify what physics is relevant.
In Sec.\ref{sec:nuclear-2SCX}, we let the evolving couplings depend on $(\mu_Q,T)$, and construct the resulting equations of state, the ``CSCX''.
In Sec.\ref{sec:EoS_for_NS} we add leptons to construct equations of state for neutrino trapped, hot neutron stars.
Section \ref{sec:summary} is devoted to the summary.

\section{Quark model}\label{sec:model}

\subsection{General remarks}

The composition of quark matter is sensitive to the effective interactions at high density. We use the Nambu-Jona-Lasinio (NJL) model \cite{Nambu:1961tp} to express various interactions at intermediate energy scale, $0.2$-$1$GeV, relevant for chiral symmetry breaking and semishort-range gluon exchange effects \cite{Manohar:1983md}. The obvious drawback in this model is the lack of confining effects relevant at $\lesssim 0.2$ GeV, so we restrict the use of the model to high density. In the previous studies we use the NJL model only for $n_B \gtrsim 5n_0$. In this work we take a more aggressive standpoint. We note that the meson (or quark) exchange interactions are important already at $n_B \gtrsim 1$-$2n_0$, so there is a chance that both the nuclear and quark descriptions have the validity and play complementary roles. With this viewpoint the nuclear and quark equations of state are matched around $n_B\simeq 1.5$-$2n_0$ by introducing a scheme of evolving couplings in the NJL model, as done in Ref.\cite{Fukushima:2015bda}.

Motivated by the quark-hadron continuity picture \cite{Baym:1976yu,Schafer:1998ef,Hatsuda:2006ps,Yamamoto:2007ah}, our description for the interactions is inspired by quark descriptions for the hadron spectroscopy \cite{DeRujula:1975qlm} as well as baryon-baryon interactions at short distance \cite{Oka:1980ax,Oka:1982qa}. In the context of quark matter in neutron stars, the roles of the color-magnetic interactions deserve special attentions; it has been known to play important roles for the level splitting between hadrons, such as the $N$-$\Delta$ splitting of $\simeq 200$ MeV. Moreover the lattice Monte Carlo simulations show that the color-magnetic interactions, together with the quark Pauli blocking, are essential to explain the channel dependence of short-range correlations between baryons \cite{Park:2019bsz}. In the nucleon-nucleon interactions such correlations result in the hard core repulsion \cite{Ishii:2006ec}; meanwhile for some channels, e.g., $N$-$\Omega$, involving the strangeness, the short-range correlations turn into attraction \cite{Iritani:2018sra}. Therefore the short-range repulsion is not universal. Also the lattice simulations showed that the short-range correlations are stronger for smaller current quark masses; this feature is consistent with the magnetic interactions which become more important in the relativistic regime.

We arrange our effective interactions in such a way that the above-mentioned short-range effects can be mimicked within the mean-field treatments. 
A simple and flexible model for this purpose is a three-flavor model with flavor-dependent vector repulsions and diquark attractions. 
Compared to the preceding works \cite{Baym:2019iky}, we update the vector repulsion from the flavor universal one to the flavor-dependent version; 
this update is to examine the channel dependence of hard core repulsions among baryons, especially nucleon-hyperon interactions. 
This modification has large impacts on the onset of the strangeness, as we examine later.

\subsection{Lagrangian}

\begin{center}
\begin{table}[t]
\caption{\footnotesize{Three common parameter sets for the three-flavor NJL model: the average up and down bare quark mass $m_{u,d}$, strange bare quark mass $m_s$, coupling constants $G$ and $K$, and three-momentum cutoff $\Lambda$~\cite{Hatsuda:1994pi}.}}
\begin{tabular}{|c|c|c|c|c|c|}
\hline 
   & $\Lambda$ (MeV) & \ $m_{u,d}$ (MeV) & \ $m_s$ (MeV) & \ $G \Lambda^2$ \ & \ $K \Lambda^5$ \ \\ \hline \hline
HK  & 631.4 & 5.5 & 135.7 & 1.835 & 9.29 \\ \hline
\end{tabular}
\label{tab:couplings}
\end{table}
\end{center}

We work with a Lagrangian\footnote{
Notations: we use the Gell-Mann matrices $\lambda_{A=1,\cdots, 8}$ and $\tau_{F=1,\cdots,8}$ for colors and for flavors, respectively. We also use the identity elements $\lambda_0= \tau_0 = {\bf 1}_{3 \times 3} \sqrt{2/3 \,}$. The charge matrix is ${\rm diag.}(2/3, -1/3,-1/3)$ for $u,d,s$-flavors. For quark fields $q_f^a$, indices $a$ and $f$ refer to colors $(R,G,B)$ and flavors $(u,d,s)$, respectively. $\Tr[\cdots]$ refers to the sum over momenta and all other indices (spinors, colors, flavors).
}
\beq
\calL = \calL_{\rm NJL} + \calL_d + \calL_V \,.
\label{eq:L}
\eeq
The first term in the RHS is the standard NJL Lagrangian for the hadron physics \cite{Hatsuda:1994pi}. For the parameters we use the Hatsuda-Kunihiro (HK) parameter set summarized in Table.\ref{tab:couplings}. 
Including finite chemical potentials, the Lagrangian is
\beq
\calL_{\rm NJL}
 =&&\, \bar{q} (\rmi \Slash{\partial} - \hat{m} + \hat{\mu} \gamma_0)q
 \nonumber \\ 
 &&
 +\, G_s \sum^8_{F=0} \left[ (\overline{q} \tau_F q)^2 + (\bar{q} \rmi \gamma_5 \tau_F q)^2 \right] 
 \nonumber \\
&&-\, 8 K ( \det\,\!\!_{f} \bar{q}_R q_L + \mbox{h.c.})  \,,
\eeq
where $\hat{m}={\rm diag.}(m_u, m_d, m_s)$, and $\hat{\mu}$ is the chemical potential matrix which is diagonal in color and flavor quantum numbers,
\beq
\hat{\mu} = \mu_B/3 + \mu_Q Q + \mu_{c3} \lambda_3 + \mu_{c8} \lambda_8 \,.
\eeq
The color chemical potentials are tuned to satisfy the color neutrality conditions \cite{Iida:2000ha}. 
The second term in (\ref{eq:L}) is responsible for diquark correlations,
\beq
\calL_d
 = H \sum_{A,F=2,5,7} &&\big[\, 
\left(\bar{q} \rmi \gamma_5 \lambda_{A} \tau_F  q_C \right)
\left(\bar{q}_C \rmi \gamma_5 \lambda_{A} \tau_F q \right) 
\nonumber \\
&&+\left(\bar{q} \lambda_{A} \tau_F q_C \right)
\left(\bar{q}_C \lambda_{A} \tau_F q \right) 
\, \big] \,.
\eeq
The last term in (\ref{eq:L}) is responsible for repulsive density-density interactions\footnote{For numerical computations, it is convenient to further add the color density replusion,
\beq
\calL^{\rm color}_V = \sum_A g_{A} \left(\overline{q} \gamma_\mu \frac{\, \lambda_A \,}{2} q \right)^2\,,
\eeq
and treat it in the same way as $V_q$ and $V_{3,8}$. These terms do not affect the mean field results when the color-neutrality conditions are satisfied. But these terms accelerate numerical searches for solutions satisfying the neutrality conditions, especially when the effective potential becomes somewhat flat.
},
\beq
\calL_V
= - g_V \left(\overline{q} \gamma_\mu  q \right)^2
- 2 \Nf\sum^8_{F=1} g_F \left(\overline{q} \gamma_\mu \frac{\, \tau_F \,}{2} q \right)^2  \,.
 \label{eq:L^color_V}
\eeq
The first term is the repulsion universal for all flavors which has been used in our previous works, while the second is the newly added flavor-dependent repulsion with the factor $2\Nf$ introduced for convenience. 
Below we write
\beq
g_F = c_F g_V \,,
\eeq
and vary $c_F$ for the range $[0.0, 1.0]$. The $U(\Nf)$ symmetric limit corresponds to the case $c_F = 1$.

Special remarks should be given for the flavor-asymmetric repulsion. 
In general $g_F$'s depend on the flavor channels which reflect the flavor asymmetry associated with the mass splitting and electric charges. 
Toward high density, the flavor charges are reduced and the model is effectively reduced to the model used for QHC equations of state \cite{Baym:2019iky}. 
Meanwhile, the flavor-asymmetric terms become important at lower densities and larger  charge chemical potentials.

\subsection{Mean fields}

We apply the mean-field approximations by introducing condensation fields $\eta = (\sigma_{f=1,2,3}, d_{f=1,2,3}, V_q, V_3, V_8)$ by dropping higher orders of fluctuation terms. 
Here, the condensation fields are defined through
\beq
 \left( \bar{q} \Gamma_\eta q \right)^2 = \left( \bar{q} \Gamma_\eta q - \eta + \eta \right)^2 
~ \rightarrow ~
2 \eta \big( \bar{q} \Gamma_\eta q \big) + \eta^2 \,,
\eeq
where $\Gamma_{\sigma_f} = ( 1_u, 1_d, 1_s)$ for scalar fields, $\Gamma_{d_f} = \rmi \gamma_5 (R_1, R_2, R_3)$ for diquark fields with $(R_1, R_2, R_3) = ( \tau_7 \lambda_{7}, \tau_5 \lambda_{5} , \tau_2 \lambda_{2} )$ for the $(ds, su, ud)$-diquark pairings, and $\Gamma_{V_q, V_3, V_8} = \gamma_0 ( {\bf 1}_{3\times 3}, \tau_3/2, \tau_8/2)$ for the vector fields. 
We apply the same approximation for the determinant term and drop off the cubic fluctuation terms. 
We note that condensation fields $\eta$ do not necessarily coincide with the mean-field contributions $\la \bar{q} \Gamma_\eta q \ra_{\rm MF} \equiv - \Tr[ S_{\rm MF} \Gamma_\eta] $ 
when density-dependent couplings are included, and to emphasize this point we attach the index ``MF'' to the expectation value.

With the condensation fields in the background the quarks acquire the effective mass and gap parameters, 
\begin{eqnarray}
M_f &=& m_f - 4G \sigma_f - 2 K \frac{\, \partial \,}{\, \partial \sigma_f \,}  \big( \sigma_u \sigma_d \sigma_s \big) \,,
 \\
\Delta_f &=& -2 H d_f \,,
\label{eq:delta-d}
\end{eqnarray}
and the effective chemical potential matrix,
\beq
\hat{\mu}_{\rm eff} 
= \hat{\mu}  - 2 g_V V_q - 12 g_3 V_3 \frac{\, \tau_3 \,}{2} - 12 g_8 V_8 \frac{\, \tau_8 \,}{2}  \,.
\eeq
The values of $\eta$'s are optimized by solving the gap equations. Using the Nambu-Gor'kov spinors,
\beq
\Psi = \frac{1}{\, \sqrt{2} \,} \left( q, q_C \right)^T \,,
\eeq
the quark bilinear terms have the components
\beq
\calL 
= \bar{\Psi} \left[\begin{matrix}
~~ \rmi \Slash{\partial} - \hat{M} + \hat{\mu}_{\rm eff} \gamma_0 ~&~
 \gamma_5 \Delta_f  R_f ~\\
~ -  \gamma_5  \Delta_f  R_f ~&~
\rmi \Slash{\partial} - \hat{M} - \hat{\mu}_{\rm eff} \gamma_0 ~\,
\end{matrix}\right] \Psi \,.
\eeq
This expression is used to construct single particle propagators having the poles $q_0 = \epsilon_{i=1,\cdots, 72} (\vq)$ at a spatial momentum $\vq$. There are degeneracies in Nambu-Gor'kov bases and spins, so only 18 components are independent.

\subsection{Thermodynamic functionals}

We write a pressure functional 
\beq
\calP (\lambda; \eta, g ) 
= \calP_{\rm sp} + \calP_{\rm cond} \,, 
\eeq
where ({\rm sp}) refers to the single particle contribution (see below), and $\lambda =  (\mu_B, \mu_Q, T ) $. The coupling $g$ is assumed to depend on the densities through the following form
\beq
g [V_q; \mu_Q, T] \,,~~~~~~~g = (g_V, g_3, g_8, H) \,,
\eeq
where the ratio between $g_V$ and $g_3, g_8$ are fixed by constants $c_3$ and $c_8$, respectively. 
The reason why we treat the $\mu_B$- and $(\mu_Q, T)$-directions in an asymmetric way 
and use $V_q$ instead of $n_B$ or $\mu_B$ is due to competing demands from technical simplicity and physical clarity. 
It is more intuitive to use $n_B$ than $\mu_B$ as we can estimate the distance between particles, 
but such a choice makes the self-consistent calculations computationally more demanding as we have to refer to the neighboring tables in $(\mu_B,\mu_Q,T)$. 
To save the intuitive clarity we use $V_q$ which has the same qualitative trends as $3n_B$\footnote{In some beyond-mean field treatments $V_q$ can differ from $3n_B$ substantially, see e.g, Ref.\cite{Zhang:2017icm}. This situation was not found in this paper.}, while make the calculations much simpler. Then we will constrain the form of $g[V_q]$ along the $\mu_B$-direction for a given $(\mu_Q, T)$, and later combine those tables to calculate the derivatives of the pressure functionals in the $(\mu_Q, T)$-directions.

The thermodynamic pressure (before the vacuum subtraction) is obtained by substituting the solutions of gap equations, $\eta_* (\lambda) $, 
\beq
P_{\rm bare} = \calP \big(  \lambda, \eta_*, g_* \big) \,,~~~~~~~~ g_* \equiv g [ V_q^*; \mu_Q, T ] \,.
\eeq
The number densities and entropy are derived from the derivatives of $P(\lambda)$ with respect to $\lambda$'s. 
Later, we identify extra terms which are absent in the mean-field expressions without running couplings.

The expressions of the pressure functional remain the same as those with fixed couplings; the difference, which we discuss later, emerges only when we consider the derivatives.
The single particle contribution is
\beq
\calP_{\rm sp}
= 
\sum_{i=1}^{18} \int_0^\Lambda \!\! \frac{\, \rmd q \,}{\, 2\pi^2 \,} 
\,\vq^{\,2}
\left[\, |\epsilon_i| + 2T \ln \left(1+\rme^{- |\epsilon_i|/T } \right)
\right] .
\eeq
where the integral is cutoff by $\Lambda$. The condensation energy is
\beq
\calP_{\rm cond} 
&= - 2 G_s \sum_{f=1}^3 \sigma_f^2 + 4K \sigma_u \sigma_d \sigma_s
	- H \sum_{f=1}^3  \left| d_f \right|^2 \
 \nonumber \\
&+ g_V V_q^2 + 6 g_3 V_3^2 + 6 g_8 V_8^2 \,.
\label{eq:P_cond}
\eeq
In the last step we have to normalize the pressure by subtracting the vacuum contributions,
\beq
P (\lambda) = P_{\rm bare} (\lambda) - P_{\rm bare} (\lambda=0) \,.
\eeq
%

\subsection{Derivatives of the pressure functional}

\subsubsection{The gap equations}

The gap equations are derived by differentiating the pressure functional with fixed $\lambda$'s,
\beq 
\frac{\, \partial \calP \,}{\, \partial \eta \,} \bigg|_{ \lambda}
= 
\frac{\, \partial \calP \,}{\, \partial \eta \,} \bigg|_{ \lambda, g}
+ \delta_{\eta, V_q}  \frac{\, \partial g \,}{\, \partial \eta \,} \bigg|_{ \lambda}  \frac{\, \partial \calP \,}{\, \partial g \,} \bigg|_{ \lambda, \eta} 
= 0 \,.
\eeq
where sums over $g$ are implicit. For fields $(\sigma_f, d_f, V_3, V_8)$, the gap equations take the usual form,
\beq
\la\, \bar{q}_f (\Gamma_{\sigma_f, d_f, V_3,V_8} ) q_f \, \ra_{\rm MF} = (\sigma_f, d_f, V_3, V_8) \,.
\eeq
For the field $V_q$, an extra term appears through the density-dependent couplings, 
\beq
V_q = n_{\rm sp}
- \frac{1}{\, 2g_V \,} 
 \frac{\, \partial g \,}{\, \partial V_q \,} \bigg|_{ \lambda} 
	  \frac{\, \partial \calP \,}{\, \partial g \,} \bigg|_{ \lambda, \eta} \,,
\eeq
with $n_{\rm sp} =  \la \bar{q} \gamma_0 q \ra_{\rm MF}$, and
\beq
 \frac{\partial \calP}{\partial g_V}  \bigg|_{ \lambda, \eta}
 &=& - V_q \big( 2 n_{\rm sp}  - V_q \big) \,,
 \\
  \frac{\, \partial \calP \,}{\, \partial g_{3,8} \,}  \bigg|_{ \lambda, \eta}
 & = &- 6 V_{3,8}^2 \,,
 \\
 \frac{\partial \calP}{\partial H}  \bigg|_{ \lambda, \eta}
 &=&  \sum_{f=1}^3 |d_f |^2  \,.
\eeq
(We have used the gap equation for diquark terms.) 
The dependence of $g$ on $V_q$ is set up in Section \ref{sec:matching}.

\subsubsection{The number densities and entropy}

%
The number density is computed as
\beq
n_{ \lambda } 
= \frac{\, \partial \calP \,}{\, \partial  \lambda \,} \bigg|_{ g, \eta}
+ \frac{\, \rmd \eta_* \,}{\, \rmd  \lambda \,} \frac{\, \partial \calP \,}{\, \partial \eta \,} \bigg|_{ \lambda, g} 
+ \frac{\, \rmd g_* \,}{\rmd  \lambda } \frac{\partial \calP}{\partial g} \bigg|_{ \lambda, \eta} 
\,.
\label{eq:number}
\eeq
With fixed couplings, we can drop off the terms with derivatives of condensates with respect to $\lambda$ by using the gap equations. This is practically useful as the self-consistent calculations to determine condensates can be closed in a local form, i.e., we do not have to refer to the data in the neighborhood in $\lambda$. This nice property does not readily follow for running couplings as $\partial \calP/\partial V_q\neq 0$, but after some extra cares we can make self-consistent calculations into the local form. Using the gap equations, the sum over $\eta$ in the second term is nonzero only for $V_q$,
\beq
\frac{\, \rmd \eta_* \,}{\, \rmd  \lambda \,} \frac{\, \partial \calP \,}{\, \partial \eta \,} \bigg|_{ \lambda, g} 
= - \frac{\, \rmd V_q^* \,}{\, \rmd  \lambda \,}
 \left( \frac{\, \partial g \,}{\, \partial V_q \,} \bigg|_{ \lambda}  \frac{\, \partial \calP \,}{\, \partial g \,} \bigg|_{ \lambda, \eta} \right)^{\eta\rightarrow \eta_*} \,.
\eeq
where  $\eta\rightarrow \eta_*$ emphasizes that we substitute $\eta_*$ only after we take the derivative. Meanwhile the the third term in Eq.(\ref{eq:number}) is
\beq
\frac{\, \rmd g_* \,}{\rmd  \lambda } \frac{\partial \calP}{\partial g} \bigg|_{ \lambda, \eta} 
=
\bigg( \frac{\, \rmd V^*_q \,}{\, \rmd \lambda \,}  \frac{\, \partial g_* \,}{\, \partial V^*_q \,} \bigg|_{\lambda}
+  \frac{\, \partial g_* \,}{\, \partial  \lambda \,} \bigg|_{V_q}
 \bigg)
 \frac{\partial \calP}{\partial g} 
 \bigg|_{ \lambda, \eta} \,.
 \eeq
Now we note that 
\beq
 \frac{\, \partial g_* \,}{\, \partial V^*_q \,} =  \left( \frac{\, \partial g \,}{\, \partial V_q \,} \bigg|_{ \lambda}  \right)^{\eta\rightarrow \eta_*} \,,
\eeq 
with which we can eliminate $ \rmd V^*_q / \rmd \mu_B $  from the expressions for thermodynamic quantities as in usual fixed coupling calculations. (Still $ \rmd V^*_q / \rmd \mu_Q $ and $ \rmd V^*_q / \rmd T $ still appear from $\partial g_*/\partial \lambda$, but they do not show up in the self-consistent calculations.) Now the quark, charge, and entropy densities can be expressed as
\beq
n_B &=& n_B^{\rm sp} \,, 
\\
n_Q &=& n_Q^{\rm sp} 
+ \frac{\, \partial g_* \,}{\, \partial  \mu_Q \,} \bigg|_{V_q} \frac{\partial \calP}{\partial g}  \bigg|_{ \lambda, \eta} \,,
\\
s &=& s^{\rm sp} + \frac{\, \partial g_* \,}{\, \partial T \,} \bigg|_{V_q} \frac{\partial \calP}{\partial g}  \bigg|_{ \lambda, \eta} \,.
\label{eq:n-nQ-s}
\eeq
The quark number density is saturated by the single particle contribution, while the charge and entropy densities are not. 
As we mentioned, we constrain the form of $g[V_q]$ for given $(\mu_Q,T)$ along the $\mu_B$-axis and from which we prepare data for $ \partial g_* / \partial  \mu_Q $ and $ \partial g_* /\partial  T $.

\subsubsection{The susceptibilities}

Finally we briefly mention the susceptibilities. The computations of the susceptibilities cannot be closed in a local form. The baryon number susceptibility is
\beq
\chi_{B} = \frac{\, \rmd n_B \,}{\, \rmd \mu_B \,} 
=  \chi^{\rm sp}_{B}
	+ \frac{\, \rmd \eta_* \,}{\, \rmd  \mu_B \,} \frac{\, \partial n_B^{\rm sp}  \,}{\, \partial \eta \,} \bigg|_{ \lambda, g} 
	+ \frac{\, \rmd g_* \,}{\rmd  \mu_B } \frac{\partial n_B^{\rm sp} }{\partial g} \bigg|_{ \lambda, \eta} \,,
\eeq
where $\chi^{\rm sp}_{B} = \partial n_B^{\rm sp} /\partial \mu_B \,|_{ \lambda, \eta} $, and this single particle contribution does not saturate the total susceptibility as $ \partial n_B^{\rm sp} / \partial \eta |_{ \lambda, g} $ is generally nonzero. If the couplings run, there is an additional term proportional to $\rmd g_*/\rmd \mu_B$. For this nonlocal property it is difficult to construct a simple scheme to achieve  very precise matching of the susceptibilities between hadronic and quark models. Nevertheless, as we see that the susceptibilities can be matched in reasonable accuracy.

\subsection{The form of running couplings}

We consider a model in which low density couplings are tuned to reproduce the hadronic pressure and number density, and they evolve into the high density values required from the constraints on the maximal mass of neutron stars.

As the running of couplings generate extra pressure and density, the forms involving very radical changes often cause problems to maintain the thermodynamic stabilities; the second derivatives of the physical pressure must be non-negative in any directions of $(\mu_B, \mu_Q, T)$. In addition there is the causality condition, $c_s^2 = \partial P/\partial \varepsilon |_{s/n={\rm const} } \le 1$. For constant couplings, we have checked that all these constraints are satisfied. Thus we start with modest departure from the constant couplings. 

With these remarks we consider a model in which the low density coupling $g_{\rm low}$ relaxes to the high density values $g_{\rm high}$ monotonically. One particular realization is to let $g=(g_V, g_3, g_8, H) =(g_V, c_3 g_V, c_8 g_V, H)$ depend on $V_q$ as
\beq
g (V_q) &= g_{\rm low} \rme^{-V_q/V^g_{\rm trans} }  + g_{\rm high}  \big( 1-\rme^{-V_q/V^g_{\rm trans} } \big) \,, 
\eeq
where $V^g_{\rm trans}$ characterizes the transition density from the low to high density couplings, 
and we vary $V^{g_V}_{\rm trans} \sim V^{H}_{\rm trans} \sim 2$-$5n_0$. In the low density limit the couplings linearly depend on $V_q$,
\beq
g (V_q) \rightarrow  g_{\rm low} + \big( g_{\rm high} - g_{\rm low} \big) V_q/V^g_{\rm trans} \,,
\eeq
while in the high density limit $g$ approaches $g_{\rm high}$ exponentially fast. 

In principle $g_{\rm high}$ can be further arranged to reproduce the perturbative QCD results which are supposed to be valid for $n_B \gtrsim 40 n_0$ \cite{Annala:2019puf}, but we restrict our attention to domains of $n_B \lesssim 10n_0$ and have not attempted such matching. 
This is partly because the cutoff effects inherent to the NJL type models introduce more artifacts at higher density (in fact we cannot go beyond $\simeq 20n_0$), and partly because our parametrization will be more complicated.

For a given $(\mu_Q, T)$, the model contains eight parameters: $(c_3, c_8)$ and $( g_V^{\rm low}, H^{\rm low})$  largely responsible for matching to hadronic equations of state; 
$( g_V^{\rm high}, H^{\rm high})$ largely correlated with the maximal mass of neutron stars; and $( V^{g_V}_{\rm trans}, V^H_{\rm trans} )$ to characterize the transition density. 
The transient regime from hadronic to quark matter is constrained by the causality and thermodynamic stability conditions. 

Among our eight parameters, two of them are fine-tuned to reproduce the pressure $P$ and number density $n_B$ in hadronic models, 
while for the rest of parameters we pick up some samples to extract generic trends. 
In this work we fine-tune the values of $(g_V^{\rm low}, H^{\rm low})$ 
as they are strongly correlated with the physics at low density, and choose samples for $( c_3, c_8, V^{g_V}_{\rm trans}, V^H_{\rm trans}, g_V^{\rm high}, H^{\rm high})$.

This procedure is repeated for various $(\mu_Q, T)$. In our choice of the tuning parameters, $(g_V^{\rm low}, H^{\rm low})$ are responsible for the $(\mu_Q, T)$ dependence of $(g_V, g_3, g_8, H)$, while $( c_3, c_8, V^{g_V}_{\rm trans}, V^H_{\rm trans}, g_V^{\rm high}, H^{\rm high})$ are kept fixed with respect to changes in $(\mu_Q, T)$. As a result,
\beq
 \frac{\, \partial g_* \,}{\, \partial \mu_Q \,} \bigg|_{V_q} &=&  \frac{\, \partial g^{\rm low}_* \,}{\, \partial \mu_Q \,} \, \rme^{-V_q/V^g_{\rm trans} }  \,,~~~~~
 \nonumber \\
 \frac{\, \partial g_* \,}{\, \partial T \,} \bigg|_{V_q} &=&  \frac{\, \partial g^{\rm low}_* \,}{\, \partial T \,} \, \rme^{-V_q/V^g_{\rm trans} } \,.
 \label{eq:dg_dlam}
\eeq
These extra contributions which come from the phenomenological matching disappear at high density.
These derivatives are used in determination of the charge and entropy densities.

\section{Low and high density constraints on quark models} \label{sec:matching}

\subsection{Domains of interest} \label{sec:domains_of_interest}


We first mention the range relevant for dynamic neutron star phenomena in terms of $(n_B,\mu_Q,T)$. 
In nuclear equations of state, the charge chemical potential is the difference between the neutron and proton chemical potentials, $\mu_Q=\mu_p -\mu_n $. 
For the baryon density, we limit our discussions to $ n_B \lesssim 10n_0$, 
which is sufficient unless we describe the collapse of neutron stars to black holes (see Refs.\cite{Nakazato:2010ue,Nakazato:2010qy,Fischer:2010wp,Fischer:2017lag} for such studies). 
In this section, we often quote the estimates in literatures, which are frequently given in terms of charged and neutral lepton fractions, 
$Y_e = n_e/n_B$, $Y_\nu=n_\nu/n_B$, the sum $Y_L=Y_e +Y_\nu$, and entropy per baryon $s/n_B$.

For static neutron stars, the core region typically has $Y_e \sim 0.1$, $Y_\nu \simeq 0$, and $s/n_B \simeq 0$. 
The lepton fraction is dominated by charged leptons as neutrinos have already diffused out. 
The estimate $n_{e} \sim 0.1$-$0.2n_0$ requires us to cover the range of $\mu_Q$ from $0$ to $-140$ MeV. 
If we consider only nucleonic degrees of freedom for hadrons for $n_B \gtrsim 2n_0$, $\mu_Q$ reaches even lower values, to $\lesssim  -200$ MeV. 
Meanwhile, the positive $\mu_Q$ does not show up.

In dynamical processes of neutron stars, the neutrinos are produced and trapped during the time scale shorter than the diffusion time which is sensitive to the matter properties. As neutrinos are trapped, the lepton number changes adiabatically, and the charged lepton and neutrino chemical potentials are established as $\mu_{e} = - \mu_Q + \mu_L$ and $\mu_\nu= \mu_L$. The $\mu_e$ is tuned to satisfy the charge neutrality condition. Considering the initial conditions, the net lepton number should be overall positive (more leptons than anti-leptons); $\mu_L > 0$ appears preferentially and approaches zero as neutrinos leak out. Then, $\mu_Q$ tends to be larger than in static neutron stars, and can be even positive in some domains. 

As the dynamical processes of neutron stars are fairly complex, the most reliable way to estimate the relevant range of $(Y_L, s/n_B)$ is to refer to available simulation data, see, for instance, Ref.\cite{PhysRevLett.86.5223,Camelio:2017nka} for protoneutron stars after supernovae and Ref.\cite{Vincent:2019kor} for neutron star mergers. 
In the former, the simulations typically lead to $\simeq 1.4 M_\odot$ protoneutron stars with $n_B \sim 2$-$3n_0$, $Y_e \sim 0.3$-$0.4$, $Y_\nu \sim 0.05$-$0.1$, and $s/n_B \sim 1$-$2$. 
The conservative choice of the range is $-200 $ MeV $ \lesssim \mu_Q \lesssim +40$ MeV and $T =0$-$100$ MeV which are covered in 
the Togashi equation of state. 
Meanwhile, in the denser regime, the range of $(Y_L, s/n_B)$ seems closer to that in static neutron stars. 

Meanwhile, neutron star mergers accommodate matter at higher density and lower temperature; 
the cores of merging neutron stars remain cool as heats are mainly produced in the outer core region and is not quickly delivered to the core. 
The core heats up more for a binary with the larger asymmetric mass ratio where the collision becomes more head on, and the two neutron stars have more direct contact. 
Recent simulations \cite{Vincent:2019kor} for a $1.2M_{\odot}$-$1.44 M_\odot$ merger suggest that the temperature is raised to $\sim 30$ MeV and $Y_e \sim 0.1$ for nucleonic equations of state.

We note that the above estimates are based on nucleonic equations of state. 
Below, we consider the cases with non-nucleonic degrees of freedom for $n_B \gtrsim 2n_0$ whose details affect the relevant domain substantially. 
One of important effects beyond pure nucleonic descriptions is the appearance of strangeness \cite{PhysRevLett.86.5223,PhysRevD.97.094023}. 

For {\it gapless} hadronic or quark matters, the strangeness makes the relevant range of $(\mu_Q, T)$ narrower. 
The particles with strangeness are negatively charged and reduce the abundance of charged leptons, so we need only $\mu_Q $ closer to $\mu_Q=0$ than prepared for a pure nucleonic regime. 
The temperature range also needs no extension, because for a given entropy the temperature reduces as more active degrees of freedom become available. 
For a normal three-flavor quark matter, the temperature is about a half of hadronic matter for a given entropy density (see, for instance, Ref.\cite{Masuda:2015wva}).

The situation considerably differs for matter with the pairing gaps, especially in the CFL phase. 
In the CFL phase the ($u,d,s$)-quarks all participate in the pairing and make the matter charge neutral. 
Due to the pairing gaps, quarks in this phase hardly react to changes in $(\mu_Q, T)$ until they become large enough to break the pairs apart. 
When some lepton number or entropy densities are given in the CFL phase, they must be saturated by leptonic contributions. 
As the QCD matter is charge neutral by itself, there should be no charged leptons, 
so the charged lepton chemical potentials should be $\simeq 0$, or $\mu_Q \simeq \mu_L$, which leads to a neutrino density (at low temperatures) of
$n_\nu \simeq N_\nu \mu_Q^3 /6\pi^2$, 
with $N_\nu$ being the number of the trapped neutrino species. 
We set $N_\nu =2$ at high density to include $\nu_e$ and $\nu_\mu$, 
while the chemical potential for $\nu_\tau$'s is set to zero as they appear only through the pair production processes and hence $n_{\nu_\tau} = n_{\bar{\nu}_\tau}$. 
As the initial condition has more leptons over anti-leptons, $\mu_Q$ is expected to be positive. 
For example, for $n_\nu = 0.05n_0$ (or $Y_\nu =0.01$ for $n_B=5n_0$), we need $\mu_Q \simeq 120$ MeV for $T\sim 0$, and smaller $\mu_Q$ at finite temperature. 
As for the range of temperature, less degrees of freedom than in nuclear matter contribute, and the temperature can be about twice as large for a given entropy density. 
The core temperature may become larger than $\sim 30$ MeV for neutron star mergers with asymmetric masses.

Taking these considerations into account, in this paper, we mainly explore the range $-180 $ MeV $\le \mu_Q \le 100$ MeV and $0 \le T \lesssim 50$ MeV within the mean-field approximation. 
No mesonic excitations are included in this paper, although we are already aware that their contributions can be important in the CFL phase; 
they will narrow the relevant range of $(\mu_Q, T)$. 
More comments are given in Sec.\ref{sec:summary}.

\subsection{Low density constraints}

For hadronic equations of state for $n_B \lesssim 2n_0$ we use the Togashi nuclear equations of state. 
By construction this equation of state is consistent with laboratory experiments at $n_B \simeq n_0$.
The physics of $n_B \lesssim 2n_0$ is largely correlated with neutron star radii with which one can infer equations of state around $2n_0$. The Togashi predicts $R_{1.4} \simeq 11.5$ km.  
There are two trends in the estimates of the radii based on astrophysical observations (for the methodology and general overview, see, e.g., Ref.\cite{Watts:2016uzu}). The discovery of the neutron merger event GW170817 and calculations lead to the upper bound $R_{1.4} \lesssim 13$ km, and typical estimates of the radii are $R_{1.4} \simeq 12 \pm 1$ km \cite{Annala:2017llu,De:2018uhw}. 
Meanwhile the x ray timing observations by the NICER lead to larger radii, $R_{1.4} \simeq 13 \pm 1$ km, $R = 13.02^{+1.24}_{-1.02}$ km for $M = 1.44^{+0.15}_{-0.14} \, M_{\odot}$ (68\%) \cite{Miller:2019cac}
and $12.71^{+1.14}_{-1.19}$ km for $1.34^{+0.15}_{-0.16} M_\odot$ \cite{Riley:2019yda}.
Thus, the Togashi equation of state belong to a soft class of low density equations of state.

The physics at low density in $\beta$-equilibrated matter is constrained in the above-mentioned way, 
but the domain with a more general lepton fraction and finite temperature has not been well-constrained from observations. 
So our discussions are based on theoretical predictions. 
The Togashi covers sufficiently wide domains for $n_B$ and $T$, while $Y_Q$ is covered up to $0.65$,
 which corresponds to $\mu_Q \sim 40$ MeV for $n_B \sim 2n_0$. 
 As we use $\mu_Q$ to $100$ MeV for the quark model,  
 our coupling interpolations need the nuclear tables to $\mu_Q =100$ MeV, and we have to extrapolate the nuclear data.

Our extrapolation at a given $T$ is guided by the isospin symmetry, taking into account its breaking only up to the neutron-proton mass difference $m_n-m_p$ in the energy density from the rest mass. The details are given in the Appendix.\ref{app:proton-rich}, and here, we just quote approximate relations
\beq
\varepsilon(1-Y_p) ~&\simeq&~ \varepsilon(Y_p) + ( m_p -  m_n ) (1-2Y_p) n_B \,,
\nonumber \\
s(1-Y_p) ~&\simeq&~ s(Y_p) \,,
\nonumber \\
P (1-Y_{p} ) ~&\simeq&~  P (Y_{p} ) \,,
\nonumber \\
\mu_Q ( 1-Y_{p} ) ~&\simeq&~ - \mu_Q (Y_{p} ) - 2 \big( m_n - m_p \big) \,,
\nonumber \\
\mu_B (1-Y_{p} ) ~&\simeq&~  \mu_B (Y_{p} ) + \mu_Q (Y_{p} ) + m_n -  m_p 
\,.
\eeq
%
We have checked that the relations hold in good accuracy for $0.4 \lesssim Y_p \lesssim 0.6$ using the nuclear tables, and we assume its validity for $Y_p \gtrsim 0.6$. 
We use the original data if they are available, and if not use the data created from the approximate relations. 
Here is an example: to construct tables for $n_B = 2n_0$ and $Y_Q = 0.9$, in the Togashi's 
we use the data $Y_Q = 0.1$ which corresponds to $\mu_B (2n_0, 0.1) \simeq 1020$ MeV and $\mu_Q (2n_0, 0.1) \simeq -120$ MeV, 
and produce the results $\mu_B (2n_0, 0.9) \simeq 900$ MeV and $\mu_Q (2n_0, 0.9) \simeq 120$ MeV.


After these preparations, now the tables are used to constrain quark models. We choose the matching point to $n_B \simeq 1.5n_0$.\footnote{More precisely, to avoid additional interpolation between data points, for all ($\mu_Q, T)$ we use tables given at $n_B$ closest to $1.5n_0$; for Togashi's it is $n_B \simeq 1.498n_0$. } 
We have also tried to match at higher densities such as $2n_0$. 
But for the negative $\mu_Q \lesssim -100$ MeV domain the strangeness can appear below $n_B = 2n_0$ for some parameters in our quark model, and it often accompanies the first order transition. 
This introduces additional technical complications and we simply avoid them by choosing lower values of $n_B$ for the matching procedure. 
Then, the nuclear equations of state are matched with the 2SC phase in quark models as in Ref.\cite{Fukushima:2015bda}. 
The other possible scheme is to choose the matching point at a larger $n_B$ and use hadronic equations of state with hyperons, see, for instance Refs.\cite{Togashi:2016fky,Ishizuka:2008gr,Fortin:2017dsj,Marques:2017zju,Burgio:2011wt}; in this case, we may interpolate the CFL equations of state to the hyperonic ones. 
We leave the analyses of such schemes for future works.

\subsection{High density constraints}

\begin{figure}[t]
\begin{center}
\vspace{0.cm}
\includegraphics[scale=0.2]{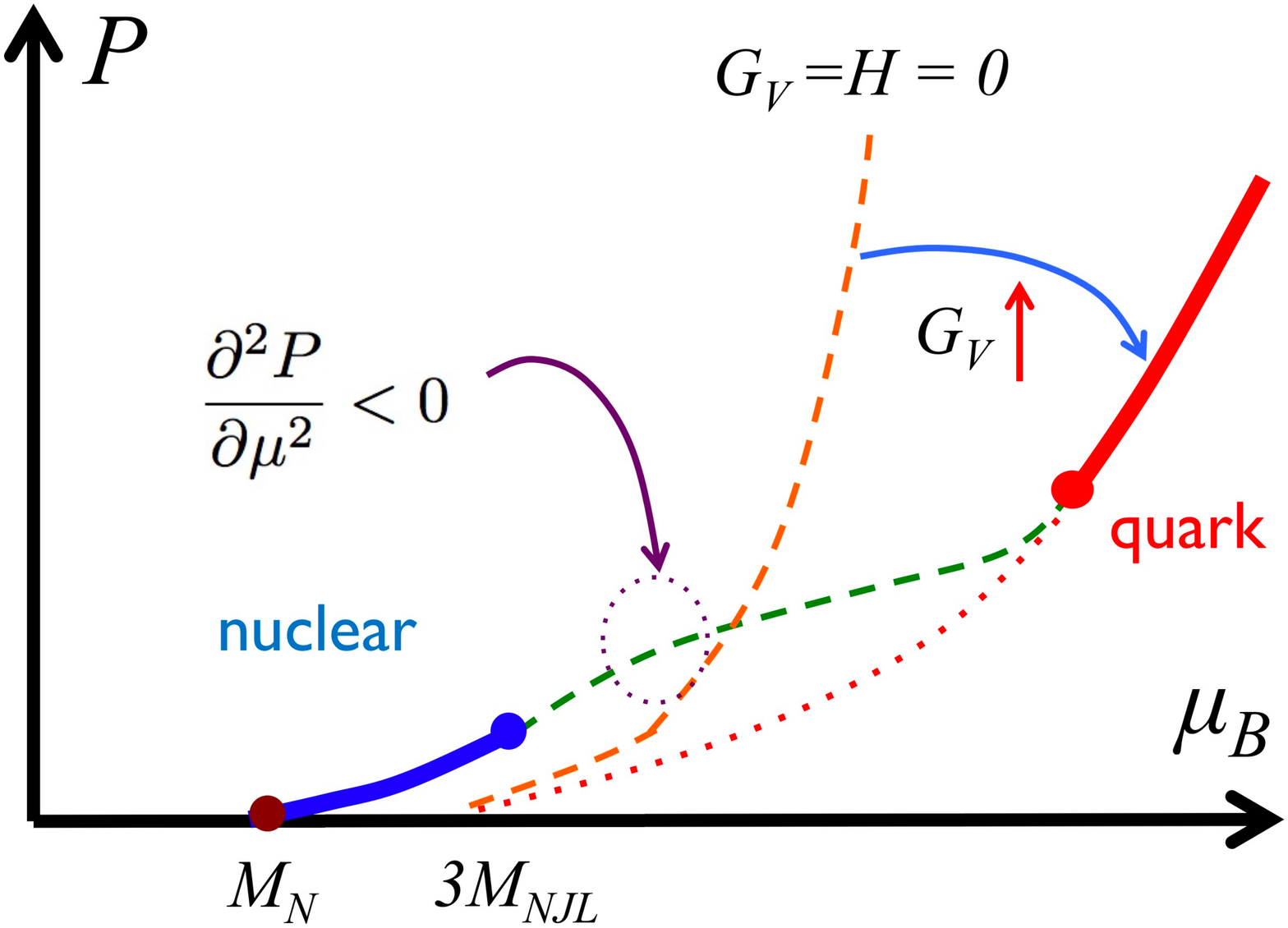}
\includegraphics[scale=0.2]{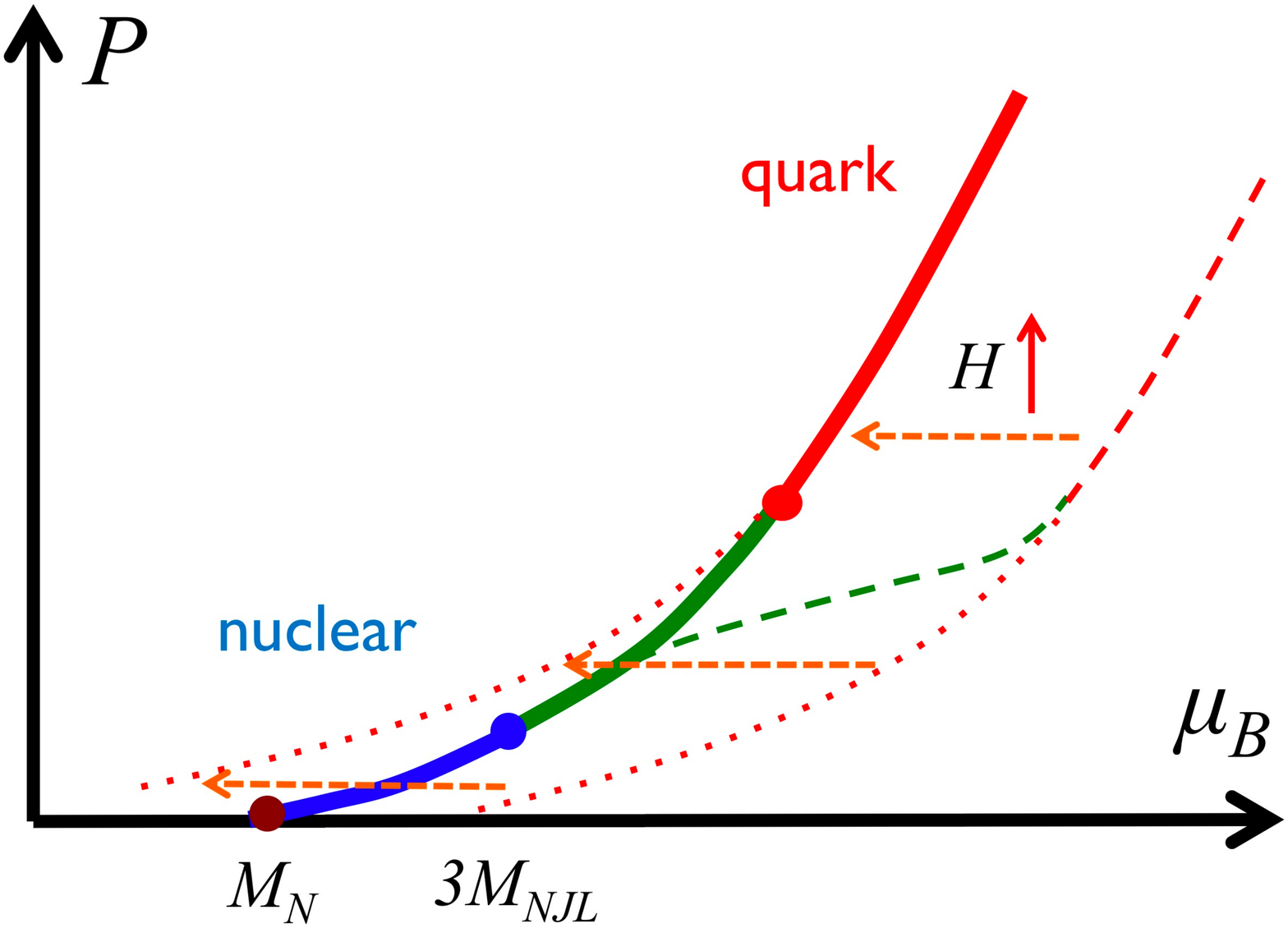}
\end{center}

\vspace{.0cm}
\caption{Schematic figure to explain the impacts of $(g_V, H)$. (Upper) A NJL quark matter equation of state with $g_V=H=0$ is stiffened by increasing $g_V$. But it enhances the danger to introduce the unstable region ($\partial^2 P/\partial \mu_B^2 <0$) between quark and nuclear equations of state. (Lower) Increasing $H$ overall shifts the pressure curve to a lower $\mu_B$ domain, allowing us to connect quark and nuclear equations of state without introducing the unstable region.
}
\label{fig:schematic_gV-H}
\end{figure}

The high density part of equations of state at $n_B \gtrsim 5n_0$ must be sufficiently stiff to pass the so-called two solar mass ($2M_\odot$) constraint; the accurately measured masses are
$M=1.908^{+0.016}_{-0.016} M_\odot$ \cite{Arzoumanian:2017puf}, $M=2.01^{+0.04}_{-0.04} M_\odot$ \cite{Antoniadis:2013pzd}, and $2.14^{+0.10}_{-0.09} M_\odot $ \cite{Cromartie:2019kug}.

Although the maximum mass is strongly correlated with the high density equations of state, 
low density equations of state serve important constraints as the low and high density domains must be connected in a causal and thermodynamically stable way. 
In general, the connection becomes more problematic for softer hadronic equations of state because the growth of the stiffness, characterized by $c_s^2 = \partial P/\partial \varepsilon $, is bound by the light velocity. 
If we include the first order phase transitions in modeling, the difficulty is further enhanced as the region other than the first order domain must have an even larger speed of sound.

The previous series of QHC equations of state for static neutron stars smoothly interpolate a nucleonic pressure at $2n_0$ and a (CFL) quark matter one at $5n_0$ by polynomials of $\mu_B$. 
Within this interpolation scheme the range of model parameters $(g_V, H)$, used for $n_B \ge 5n_0$, is constrained for given nucleonic equations of state. The impacts of $g_V$ and $H$ are schematically illustrated in Fig.\ref{fig:schematic_gV-H}; a larger $g_V$ is necessary to make equations of state stiff enough to pass the $2M_\odot$ constraint, but too large $g_V$ makes thermodynamic and causal interpolation to nuclear equations of state impossible. This problem is relaxed by increasing $H$. 
Clearly, the allowed values of $g_V$ and $H$ are strongly correlated. 
The difficulty of the interpolation depends on nuclear equations of state and more difficulties for softer nuclear equations of state. 
Most comprehensive studies were done for the Togashi which is relatively soft \cite{Baym:2019iky}.  
The absolute maximum allowed mass in the QHC19 \cite{Baym:2019iky} is $\simeq 2.35 M_\odot$ at $(g_V, H)/G_s \simeq (1.30, 1.65)$ with the core baryon density $\simeq 6n_0$. 
Overall, the analyses suggest $g_V \gtrsim 0.6 G_s$ and $H \gtrsim 1.4 G_s$. 
This estimate is consistent with the analyses \cite{Song:2019qoh} based on nonperturbative gluon propagators \cite{PhysRevD.84.045018,Suenaga:2019jjv}.
These analyses motivate us to pick up samples from $g_V^{\rm high} \gtrsim 0.6G_s$ and $H^{\rm high} \gtrsim 1.4 G_s$.

Actually, the constraints on $(g_V, H)$ should be stronger than the previously found one, because the previous analyses treated only $\beta$-equilibrated matter, 
and do not guarantee that the interpolation can be done for general $(\mu_Q, T)$.
Indeed, we found that the constraints for $(g_V, H)$ become significantly tighter if we demand the interpolation for wide range of $(\mu_Q,T)$ \cite{preparation1}. 
In short, this is due to the disparity between the CFL quark matter and hadronic matter in their response to changes of $(\mu_Q, T)$. 
Hadronic equations of state change considerably while the CFL one does not, so the acceptable domains of $(g_V, H)$ change considerably as we vary $(\mu_Q, T)$. 
We do not show the systematic analyses here, but just pick up a particular set of $(g_V, H)^{\rm high} $ which passes the above mentioned constraint. 

Finally we mention how the evolution of effective couplings affect the stiffness, or more precisely how $\rmd g/\rmd V_q \sim \rmd g/\rmd n_B$ impacts the relation between $P$ and $\varepsilon$. 
We consider a simple parametrization of energy density for a given number density as
\beq
\varepsilon (n_B) = a n_B^{4/3} + b n_B^\alpha \,,~~~~~(a,b: {\rm constant})
\eeq
where the first term comes from a relativistic kinetic energy and the second from interactions. The chemical potential is
\beq
\mu (n_B) = \frac{\, 4 \,}{3} a n_B^{1/3} + b \alpha n_B^{\alpha -1 } \,.
\eeq
Using the thermodynamic relation $P= \mu_B n_B - \varepsilon$, and eliminating $a$, we get \cite{Kojo:2014rca}
\beq
P = \frac{\, \varepsilon \,}{3} + b \bigg( \alpha - \frac{4}{3} \bigg) n_B^\alpha \,.
\eeq
The interaction modifies the $P$ vs $\varepsilon$ relation from the conformal limit. Whether equations of state are stiffened or softened depend on not only the sign of interactions but also the powers in $n_B$. To stiffen equations of state, repulsive interactions $(b>0)$ must have $\alpha > 4/3$; for attractive interactions $(b<0)$, $\alpha < 4/3$ is necessary.

For constant $(g_V, H)$, they add the energy density terms, roughly $\sim g_V n_B^2$ and $\sim - \Delta^2 (n_B) /H$ 
[see Eqs.(\ref{eq:delta-d}) and (\ref{eq:P_cond}), here $\Delta^2 (n_B)$ grows more slowly than powers of $4/3$], that stiffen equations of state compared to the conformal limit. 
This trend changes for evolving couplings that yield additional powers in $n_B$. 
For example, the vector repulsion softens equations of state for $g_V(n_B) \sim n_B^{-\beta}$ with $\beta > 2/3$. 
Such power is expected if we deduce $g_V$ from a one-gluon exchange which is expected to scale as $\sim \alpha (p_F) /p_F^2$ at large density. 
For $H$, details depend on how $\Delta(n_B)$ depends on $n_B$. If we assume that the $n_B$ dependence is weak, the reduction of $H$ for larger $n_B$ softens the equation of state.



\begin{figure}[t]
\begin{center}
\includegraphics[scale=0.6]{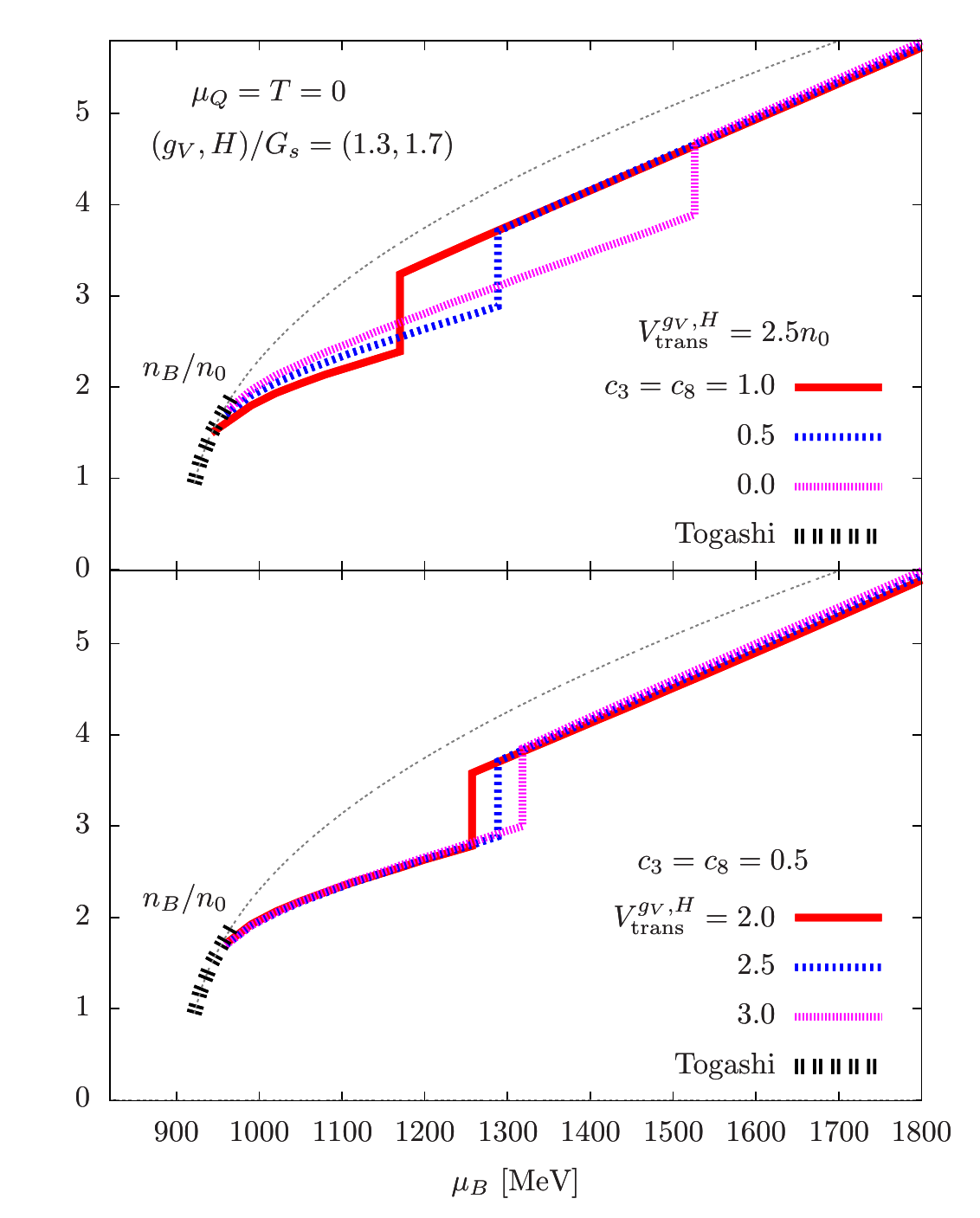}
\end{center}
\vspace{-0.5cm}
\caption{The number density $n_B$ (normalized by $n_0$) at $\mu_Q=T=0$ for (top) $c_3=c_8=0.0,0.5,1.0$ with $V_{\rm trans}^{g_V,H}=2.5n_0$; and (bottom) $V_{\rm trans}^{g_V,H}/n_0=2.0, 2.5$, $3.0$ with $c_3=c_8=0.5$. 
}
\label{fig:nB-mu_g38_v0conf_compare_Gv130H170}
\end{figure}

\begin{figure}[t]
\begin{center}
\includegraphics[scale=0.6]{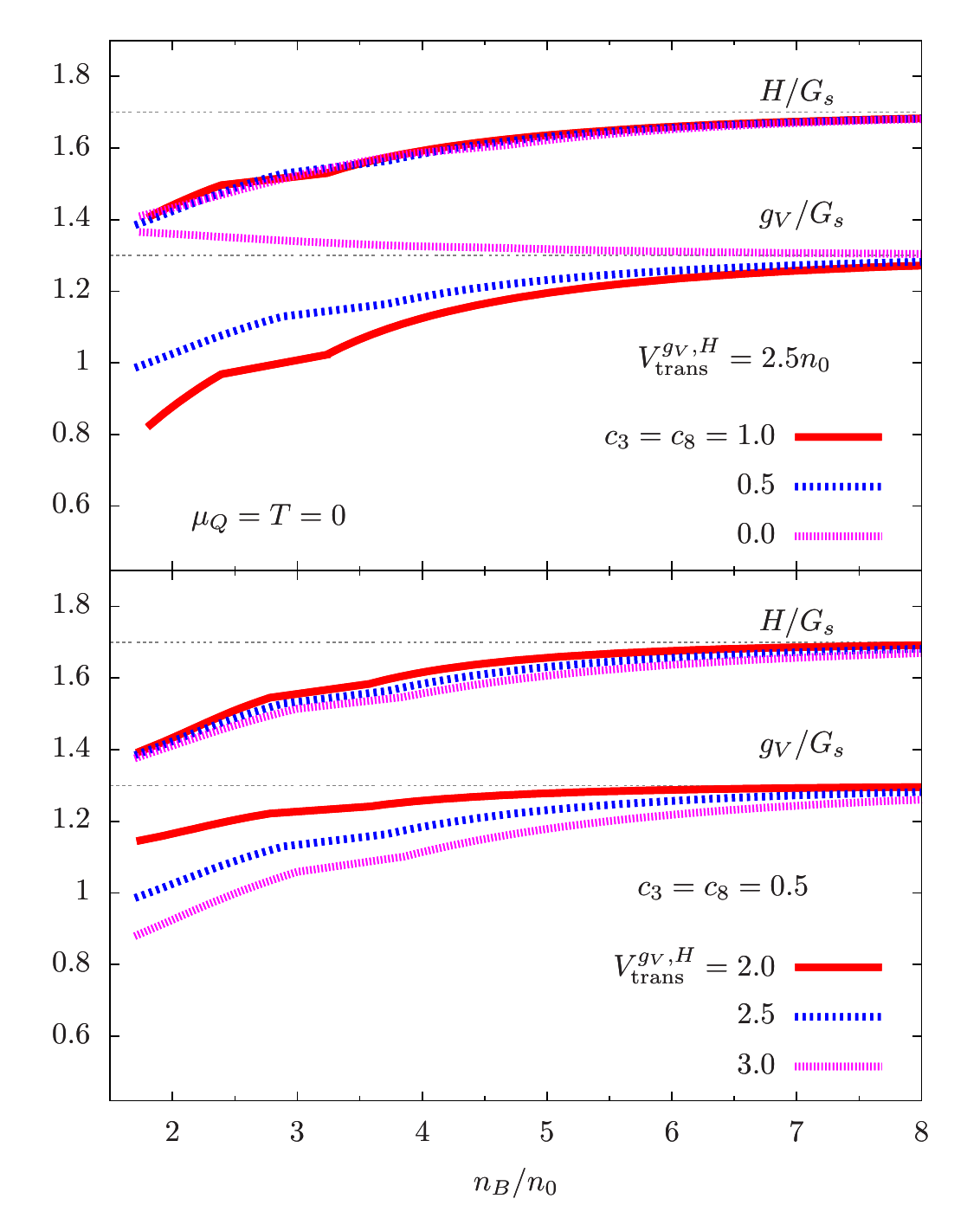}
\end{center}
\vspace{-0.5cm}
\caption{The evolving effective couplings $g_V$ and $H$ as functions of $n_B$. The parameter set is the same as in Fig.\ref{fig:nB-mu_g38_v0conf_compare_Gv130H170}.
}
\label{fig:g-mu_Gv130H170}
\end{figure}

\section{Matching nuclear to 2SC}\label{sec:nuclear-2SC}

In this section we analyze the continuity between the nuclear and 2SC phases, assuming that descriptions based on nuclear and quark pictures are reasonably valid around $n_B\simeq 1.5$-$2n_0$.
Specifically we choose $n_B=1.5n_0$ at $\mu_Q=T=0$ as a matching point and then fix the quark model couplings to reproduce the pressure and number density of the nuclear equations of state. 
In this section, the evolving coupling constants are functions of $V_q$ only; they do not depend on $(\mu_Q,T)$.  
The purpose in this section is to see to what extent the matching works for general $(\mu_Q,T)$ within this simplest setup. 
The $(\mu_Q, T)$ dependence is introduced in the next section.

Unless otherwise stated, we choose the following set of parameters:
\beq
\big( g_V^{\rm high}, H^{\rm high} \big)/G_s &=& (1.3, 1.7)\,,~~~~
\nonumber \\
 V^{g_V}_{\rm trans} =V^H_{\rm trans} &=& 2.5 n_0 \,,
\nonumber \\
c_3 =c_8 &=& 0.5\,.
\label{eq:set_parameters}
\eeq
This choice of $( g_V^{\rm high}, H^{\rm high} )$ is based on the guideline presented in the previous section.
 The specific values of $V_{\rm trans}^g$ are chosen after some trials and errors; with too small values, we typically found instabilities in searching the solutions of gap equations. 
 Meanwhile, with too large values, $g$ is not dominated by $g_{\rm high}$, obscuring the meaning of our framework. 
 Finally, the values of $(c_3,c_8)$ are chosen as an intermediate between the universal $U(1)$ repulsion $(c_3=c_8=0)$ and $U(\Nf)$-symmetric repulsion $(c_3=c_8=1)$. 
 The difference strongly correlates with the strength of phase transitions associated with production of the strangeness, as is shown below.

\subsection{Zero temperature}

\subsubsection{Onset of strangeness}

First we analyze the zero temperature results. We begin with studies of the flavor-asymmetric repulsions for $c_3=c_8=0.0, 0.5, 1.0$ to check their impacts. 

Shown in Fig.\ref{fig:nB-mu_g38_v0conf_compare_Gv130H170} are the number density $n_B$ for various $c_{3,8}$ parameters. 
We have checked that the impact of $c_3$ is negligible for the range of $\mu_Q$ we have explored. 
Meanwhile, the value of $c_8$ has a dramatic impact. 
As we increase $c_8$ from 0 to 1, the onset density of strangeness is reduced from $\simeq 4n_0$ to $\simeq 2.5n_0$, or in $\mu_B$, from $\mu_B\simeq 1550$ MeV to $\simeq 1200$ MeV.
To understand this tendency, it is useful to note the structure of the repulsive terms in the thermodynamic potential; in the mean field,
\beq
\Omega_V^{\mu_Q\simeq 0} &\simeq& g_V (n_u+n_d+n_s)^2 + \frac{\, c_8 \,}{2} g_V\big( n_u + n_d - 2n_s \big)^2
\nonumber \\
&= &  g_V \bigg(1+\frac{c_8}{2} \bigg) (n_u+n_d)^2 + g_V(1+2c_8) n_s^2 
\nonumber \\
&& ~~~ + 2g_V(1-c_8) (n_u+n_d) n_s \,.
\eeq
For $c_8=1$, the mean-field repulsion between $u,d$-quarks and $s$-quarks vanishes; as a consequence the onset of the strangeness is not disturbed by the effective repulsions from the $u,d$-quark densities.

Another aspect of $c_8$ is that it suppresses the susceptibility at low density; the repulsive terms in the absence of $s$-quark becomes
\beq
\Omega_V^{\mu_Q\simeq 0} 
~\xrightarrow{n_s \rightarrow 0}~  g_V \bigg(1+\frac{c_8}{2} \bigg) (n_u+n_d)^2\,,
 \eeq
so larger $c_8$ tempers the growth of $n_B$ more strongly. 
Figure \ref{fig:nB-mu_g38_v0conf_compare_Gv130H170} shows that the choice $c_8=1$ suppresses the growth of $n_B$ a bit too much; 
as a result, the matching to the Togashi is good only at a single point, $n_B/n_0\simeq 1.5$. 
For smaller $c_8$ the matching is better over the range $n_B\simeq 1.0$-$1.5n_0$. 
Below, we use the intermediate values $c_3=c_8=0.5$. 

We also check the impact of variation of $V^{g}_{\rm trans} $. 
Its impact on equations of state on $V^{g}_{\rm trans} $ is not as large as $c_{3,8}$ for the range we are interested in, as can be seen from Fig.\ref{fig:nB-mu_g38_v0conf_compare_Gv130H170}.

\subsubsection{Evolving couplings $g(V_q)$}

\begin{figure}[t]
\begin{center}
\vspace{-0.5cm}
\includegraphics[scale=0.7]{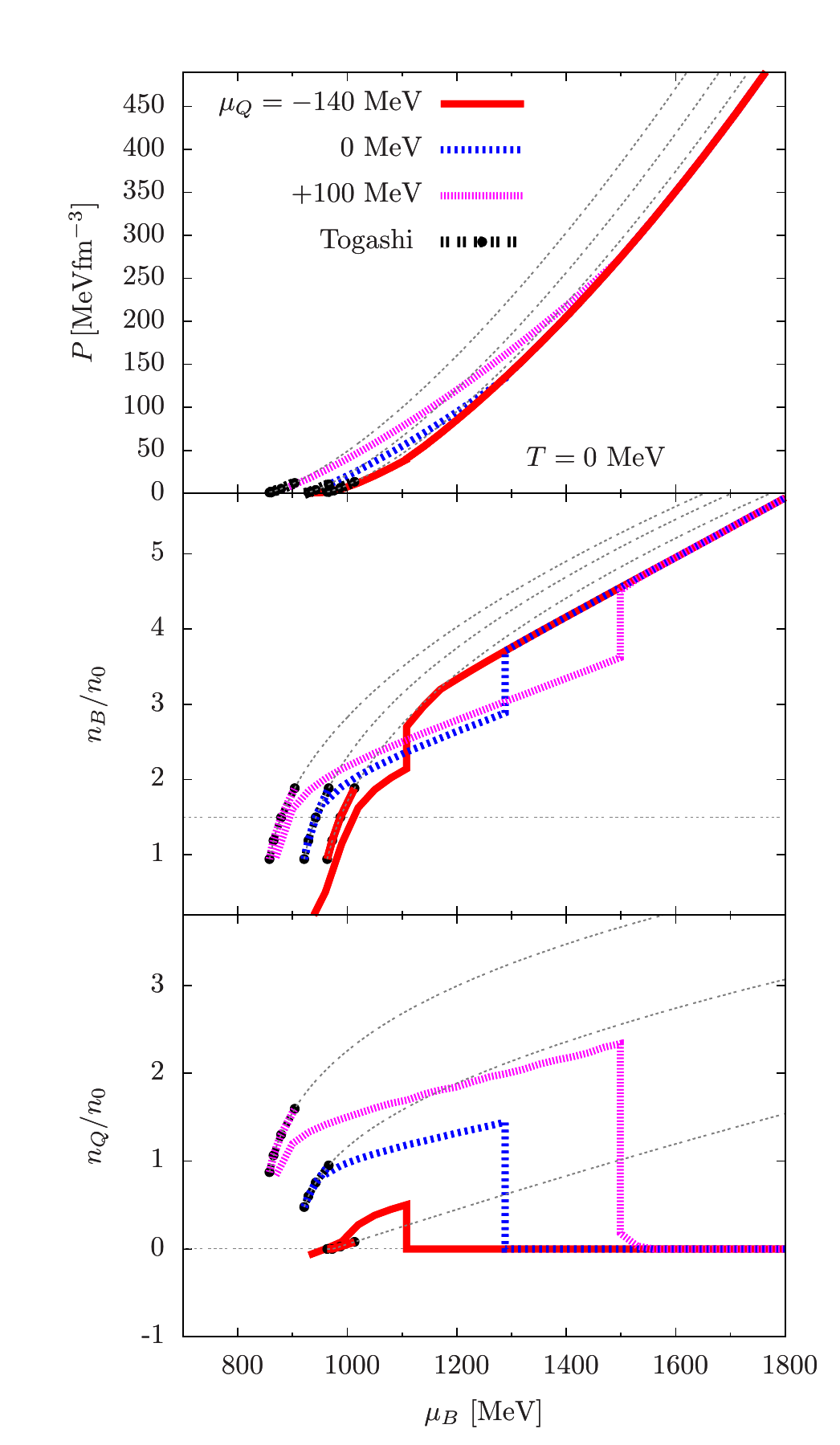}
\end{center}
\vspace{-0.5cm}
\caption{Zero temperature equations of state, $P$, $n_B/n_0$, $n_Q/n_0$ as functions of $\mu_B$ for $\mu_Q=-140, 0, 100$ MeV.  The extrapolation of the Togashi is also shown with thin lines. The quark model equations of state approach those in the CFL phase at high density.
}
\label{fig:eos_T001_Gv130H170G3G8_v0conf_gv2.5_H2.5}
\end{figure}

\begin{figure}[t]
\begin{center}
\includegraphics[scale=0.7]{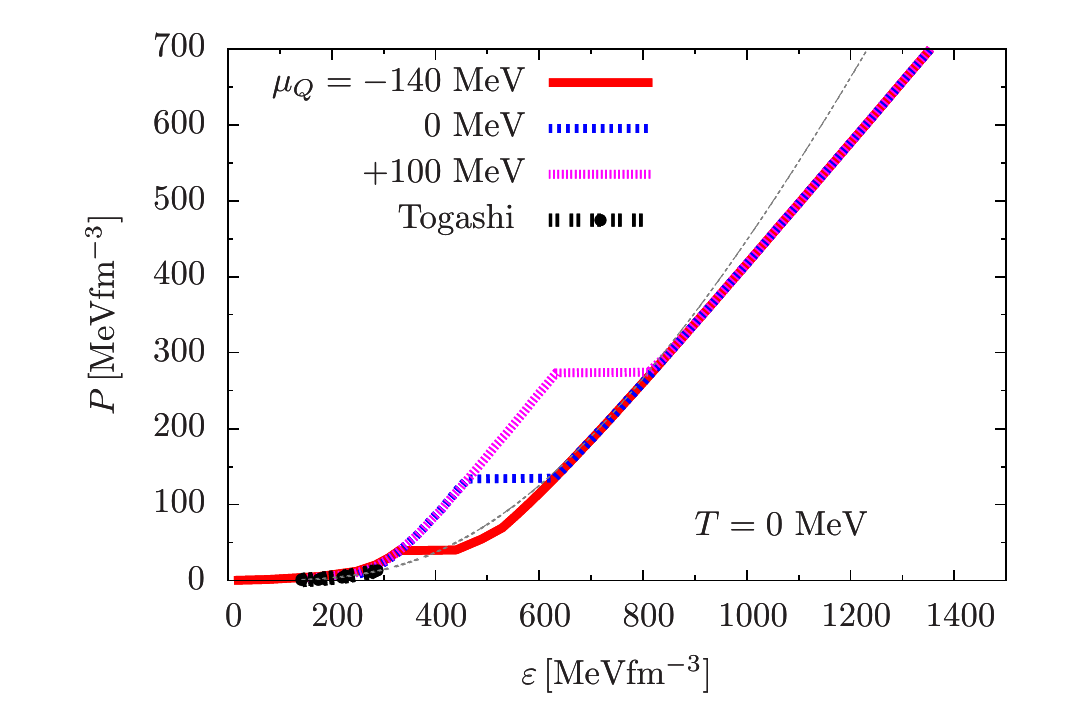}
\end{center}
\vspace{-0.5cm}
\caption{The zero temperature pressure vs energy density for $\mu_Q=-140, 0, 100$ MeV. (For the Togashi equations of state, the $\mu_Q$-dependence is not visible.)
}
\label{fig:P-e_T001_Gv130H170G305G805_v0conf_gv25H25}
\end{figure}

Having learned the impact of parameters $c_{3,8}$ and $V^g_{\rm trans}$, now we examine how the effective couplings evolve. We pick up $\mu_Q=-140, 0, 100$ MeV as samples. The domain around $\mu_Q \simeq -140$ MeV is important for the neutron-rich matter as in static cold neutron stars, while $\mu_Q\simeq 100$ MeV may be realized for matter with large neutrino density.

Shown in Fig.\ref{fig:g-mu_Gv130H170} are the evolving couplings $g = (g_V, H )$ plotted as functions of $n_B/n_0$. 
The coupling $g_V$ at low density is sensitive to the choice of $c_{3,8}$ and $V^{g}_{\rm trans} $. 
For our baseline $c_{3,8}=0.5$ and $V^{g}_{\rm trans} =2.5n_0$, we found $g_V$ to be an increasing function of $n_B$. 
Meanwhile, the value of $H$ is remarkably insensitive to the choice of $c_{3,8}$ and $V^{g}_{\rm trans} $.

The interpretation of this qualitative tendency is not straightforward and hence deserves special remarks. 
If we had kept $H$ constant everywhere and just extrapolated a $H=H^{\rm high}$ from the high density domain, the quark equations of state did not match with the nuclear one, as we have illustrated in the lower panel of Fig.\ref{fig:schematic_gV-H}. At a given $\mu_B$, there would be too much pressure and number densities compared to the nucleonic case  \cite{Kojo:2014rca,Kojo:2015fua}. Without confining effects in quark models,  quarks are overpopulated by attractive pairings, and we regard it as an artifact of using our quark models in dilute regime. 
From this perspective, one way to suppress overpopulated quarks is to take a very large value for $g_V$ at low density, as done in Ref.\cite{Fukushima:2015bda} where $g_V$ depends on $\mu_B$. 
This descriptions, however, are found to be problematic if we let $g_V$ depend on $V_q \sim n_B$. 
In this case $g_V$ needs to behave singular at low density; 
otherwise the terms $\sim g_V V_q^2$ simply decouple from the analyses as $V_q\rightarrow 0$, and cannot eliminate the artifacts. 
Then we found that such a singular function makes numerical solutions for self-consistent equations typically unstable. 
It turns out that letting $H$ density dependent allows us more efficient matchings. 
In fact, reducing $H$ at low density eliminates the overpopulated quarks; in physical terms colored diquarks are not allowed to be stable in the dilute regime, as their isolated color charges should cost the energy. 
For our current model, the confining effects are not explicitly included, so we reduce $H$ to allow less number of diquarks.
Meanwhile, in a denser regime diquarks can find another quark to get neutralized, 
and we expect that $H$ may have the magnitude which is roughly those expected inside of a baryon, $H\gtrsim 1.4 G_s$ \cite{Song:2019qoh}.

\subsubsection{Equations of state; nuclear-2SC-CFL}

We have set up the parameters for evolving couplings at $\mu_Q=T=0$. Next we use them to construct equations of state.
In Fig.\ref{fig:eos_T001_Gv130H170G3G8_v0conf_gv2.5_H2.5} we show equations of state, $P$, $n_B/n_0$, and $n_Q/n_0$, as functions of $\mu_B$, and 
in Fig.\ref{fig:P-e_T001_Gv130H170G305G805_v0conf_gv25H25} 
we also show the $P$ vs $\varepsilon$. We chose the cases with $\mu_Q=-140, 0, 100$ MeV as samples. 
Some remarks are in order:

(i) First we note its overall structure of the pressure curves. 
At low density the nucleonic equations of state vary significantly as functions of $\mu_Q$.
For larger $\mu_Q$, the matter becomes more proton rich, the onset chemical potential lower, and the baryon density higher at given $\mu_B$. 
As density increases, these different sets of pressure curves approach a single curve; the matter becomes the CFL phase, which is insensitive to changes in $\mu_Q$. 
Thus, there is strong disparity in the $\mu_Q$-dependence of low and high density equations of state.

(ii) This strong disparity is washed out through the transition from the 2SC to the CFL phases.
 The 2SC phase is sensitive to changes in $\mu_Q$, and so is the location of the phase transition. 
For larger $\mu_Q$, the 2SC phase persists to higher density, and the CFL phase radically sets in with a large jump in the energy density. 
The strength of the transition becomes weaker for a negative $\mu_Q$, as such a $\mu_Q$ assists the population of $s$-quarks, and then, the disparity between $u,d$ and $s$-quarks becomes smaller. 

(iii) Around $n_B\simeq 1$-$2n_0$, the baryon number and charge densities at $\mu_Q=-140, 100$ MeV in nuclear and 2SC descriptions seem reasonably consistent even before tuning of the evolving couplings. The discrepancy is the order of $10\%$-$20\%$ of the total. 
Nevertheless, it is surprisingly difficult to fill this gap unless we introduce the physics beyond the quasiparticle descriptions of the 2SC. 
We come back to this point later. The quality of matching for the nuclear and 2SC equations of state is not as good as it may look.

\begin{figure}[t]
\begin{center}
\includegraphics[scale=0.6]{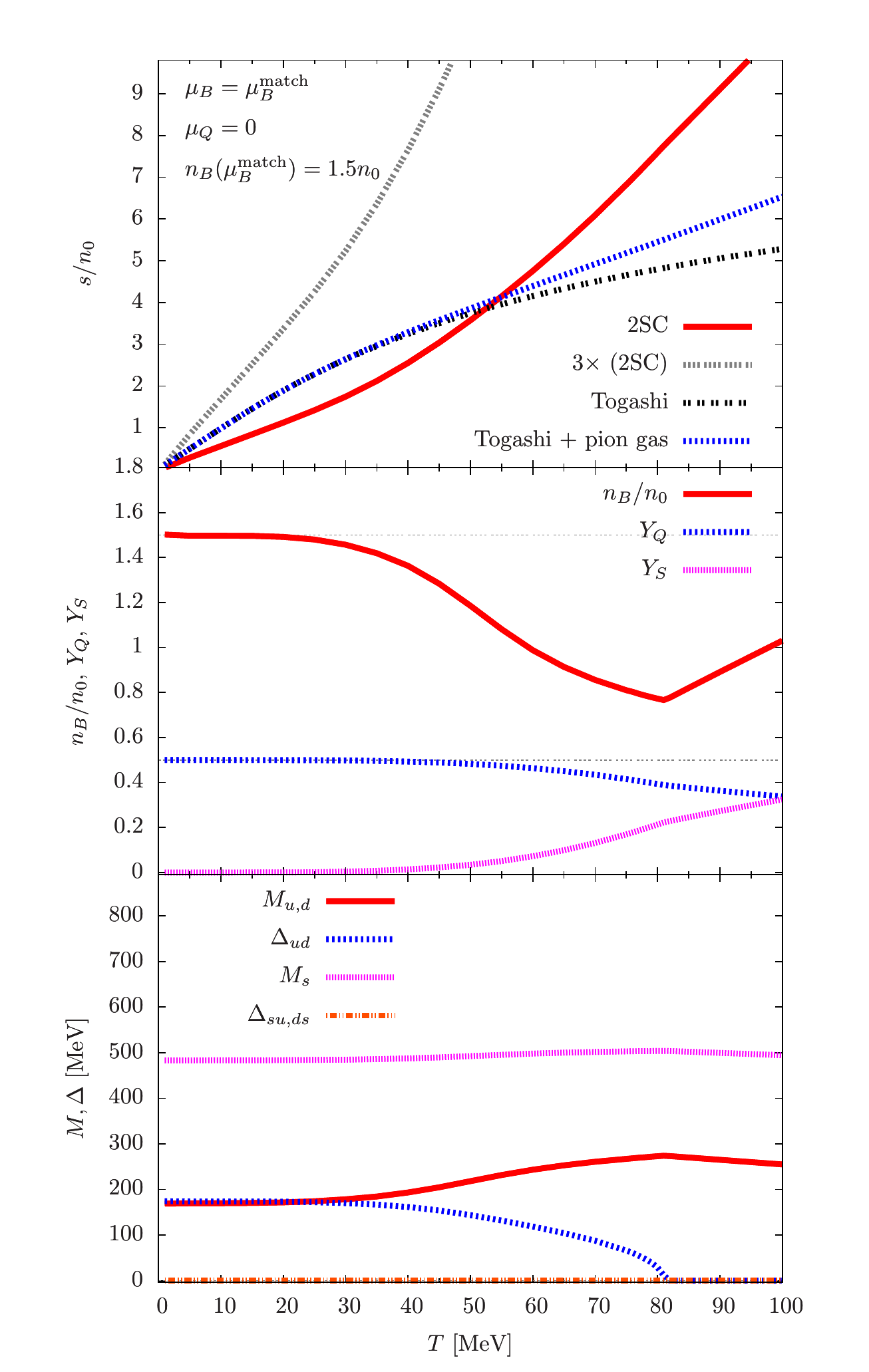}
\end{center}
\vspace{-0.5cm}
\caption{The comparison of thermal quark equations of state with the Togashi at  $\mu_B=\mu_B^{\rm match}$ and $\mu_Q=0$: 
(top) the entropy density $s/n_0$ of the quark model and the Togashi with and without a pion gas; 
(middle) the composition, $n_B/n_0$, $n_Q/n_0$, and $Y_S$. 
By the definition of $\mu_B=\mu_B^{\rm match}$, the Togashi has $n_B/n_0=1.5$, $Y_Q=0.5$, and $Y_S=0$; (bottom) the effective masses and pairing gaps. 
The $ud$-diquark gap ($\Delta_{ud}(T=0) \simeq 174$ MeV) disappears at $T\simeq 81$ MeV $\simeq 0.47 \Delta_{ud}(T=0)$.
}
\label{fig:s-n-con_T_Gv130H170G305G805_v0conf_gv25H25}
\end{figure}

\subsection{Finite temperature; nuclear-2SC-CFL}

\begin{figure*}[ht]
\begin{center}
\vspace{-1.0cm}
\includegraphics[scale=0.68]{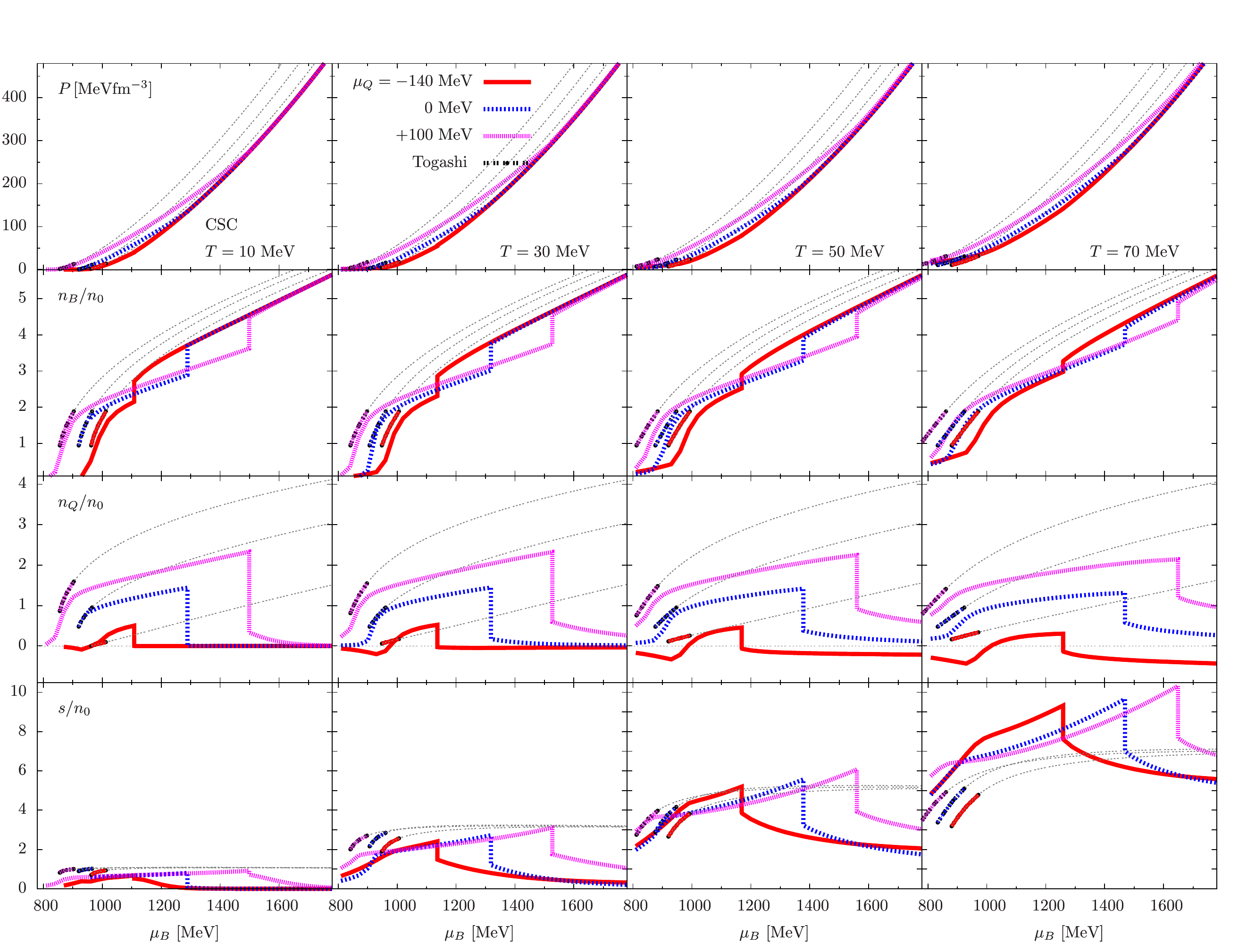}
\end{center}
\vspace{-0.5cm}
\caption{Thermal equations of state, $P$, $n_B/n_0$, $n_Q/n_0$, and $s/n_0$ as functions of $\mu_B$. The temperatures are $T=10, 30, 50$, and $70$ MeV, and the charge chemical potentials are $\mu_Q=-140, 0$, and $100$ MeV. The Togashi equation of state is used to tune the evolving couplings at $\mu_Q=T=0$ MeV and $n_B\simeq 1.5n_0$ for the setup given in Eq.(\ref{eq:set_parameters}).
}
\label{fig:eos_Tvary_g305g805}
\end{figure*}

We further examine the nuclear-2SC continuity including thermal corrections. Theoretically, there are amusing similarities in the nuclear and 2SC descriptions which are encouraging to push the continuity idea, but there are also notable differences associated with their kinematics. After all,  we need to consider supplemental correlation effects that would complete the continuity scenario.

First we mention the similarities. 
The first amusing fact is that the nuclear and 2SC phases have the same number of gapless fermions. 
A nuclear matter has four gapless modes, protons and neutrons with 1/2. 
Meanwhile, in the 2SC matter, $uR$, $dG$, $uG$, and $dR$ quarks participate in the diquark pairing and get gapped, 
while $uB$ and $dB$ are left gapless; taking into account their spins, there are four gapless modes in the 2SC phase as in the nuclear matter. 
Second, the number density for protons plus neutrons and for $uB$ plus $dB$ are equal; to see this, we note 
\beq
 n_B = n_p + n_n = \frac{1}{\, \Nc \,} \sum_{f=u,d} \left( n_{fR} + n_{fG} + n_{fB} \right) \,,
 \eeq
where $n_{ f(RGB) } $ are the quark density with flavor $f$ and colors $(RGB)$. We further note that the color neutrality condition sets 
\beq
\sum_{f=u,d} n_{fR} = \sum_{f=u,d}  n_{fG} = \sum_{f=u,d}  n_{fB} \,,
\eeq
with which one can write
\beq
 n_B = n_p + n_n = n_{uB} + n_{dB}  \,.
 \eeq
In particular, the isospin symmetric matter has $n_p = n_n = n_{uB} = n_{dB}$ with which the Fermi momenta for $p,n, uB, dB$ are all the same.

Now, we turn into the difference between the nuclear and 2SC descriptions. 
The difference comes from kinetic and dynamical reasons; the densities of states near the Fermi surface, or the Fermi velocities, turn out to be different for these two descriptions. 
This can be seen by looking at the entropies. Applying the Fermi liquid descriptions for gapless fermions, we may write entropy at low temperatures as
\beq
s \simeq N_{\rm dof}  \frac{\, p^2_B \,}{\, 3 v^{\rm R}_F \,} T \,,
\eeq
where $N_{\rm dof}$ is the number of the gapless fermion species. 
Here, we slightly generalize the Fermi velocity in nonrelativistic framework into $v_F^{\rm R}=p_F/E_F$ where $E_F=\sqrt{m_*+p_F^2}$; 
in the nonrelativistic limit, it reduces to $ v_F^{\rm R} \rightarrow v_F^{\rm NR} = p_F/m_*$ where $m_*$ is the effective mass, 
while in the relativistic limit $ v_F^{\rm R} \rightarrow 1$ recovering the result of a massless fermi gas.
For the nuclear and 2SC descriptions, their $N_{\rm dof}$'s are equal, but the $v^{\rm R}_F$'s are likely different, as the masses of gapless fermions are considerably different. 
Neglecting the effects of interactions, the mass in the nuclear case should be $m_* \sim m_N$, about three times larger than the 2SC case, 
and which in turn suggests $s_{\rm nuclear} \sim 3 s_{{\rm 2SC}}$. 
Thus, in this simplest consideration, the nuclear and 2SC entropies do not match.

For more detailed inspections, one must include the effects of interactions. 
Figure \ref{fig:s-n-con_T_Gv130H170G305G805_v0conf_gv25H25} shows the temperature dependence of the Togashi and 2SC equations of state, 
including entropies, compositions, and dynamical masses and gaps at $\mu_B=\mu_B^{\rm match}$ (with which $n_B\sim 1.5n_0$) where we match the nuclear and 2SC equations of state. 
At $n_B\simeq 1.5n_0$ the Fermi momentum is $p_F\simeq 300$ MeV. Several remarks are in order:

(i) The Fermi velocity in the nuclear phase around $n_B\simeq 1.5n_0$ is $v_F^R\simeq 0.50$, enhanced from the free gas limit $v_F^{\rm free} \simeq 0.32$. 
This means the interactions effectively make nucleons more relativistic, and this tends to close the gap between the nuclear and 2SC descriptions.

(ii) The Fermi velocity in the 2SC phase around $n_B = 1.5n_0$ is $v^R_F \simeq 0.87$, close to the velocity of light. 
Such a large velocity is first due to the effective quark mass less than the nucleon mass and second due to the chiral restoration effects. 
In our setup the effective quark mass around $n_B\simeq  1.5n_0$ is $M_{u,d}\simeq 170$ MeV (in vacuum $M_{u,d} \simeq 336$ MeV). 

(iii)  It is interesting to see what value of the effective quark mass can reproduce the nuclear result, $v_F^R\simeq 0.50$. 
Setting $0.5= p_F/E_F$, one obtains $M_{u,d} = \sqrt{3} p_F \simeq 510$ MeV, which is too heavy to be satisfied within the 2SC description. 
This suggests that, if the nuclear-2SC continuity takes place around $n_B\simeq 1.5n_0$, there must be substantial corrections in both nuclear and 2SC results.

(iv) If the 2SC pairings are absent, $N_{\rm dof}$ in normal quark matter is about three times greater than the 2SC case. 
In this case, the quark entropy is too large compared to the nuclear's.

(v) To see how matching works at higher temperatures, we must look at modifications of the condensation effects. Up to $T\simeq 30$ MeV, we do not see substantial changes. 
The pairing gap for  $ud$-quarks is about $\Delta_{ud}(T=0) \simeq 174$ MeV at $T=0$, 
and it decreases for larger $T$ and vanishes at $T\simeq 81$ MeV $\simeq 0.47 \Delta_{ud}(T=0)$. 
After $u,d$-quarks are released from the diquark pairing, they in turn join the chiral pairing to enhance the chiral effective mass. 
This reduces the Fermi velocity, and the resulting entropy becomes closer to the nuclear one.

(vi) Around $T\simeq 30$ MeV, thermally excited $s$-quarks, which are gapless, start to make significant contributions. 
While the density of $s$-quarks are not as large as $u,d$-quark's, all colors can contribute, making the roles of $s$-quarks substantial.

(vii) Around $T\simeq 50$ MeV, the relation between the nuclear and 2SC entropies is reversed; the 2SC entropy becomes larger than the nuclear entropy, 
and grows faster. One possibility to reduce the gap is to add contributions from thermally excited mesons, such as pions, to the nuclear entropy. 
But it turned out that such corrections are not large enough to catch up the growth of the quark entropy. 
A possible way to achieve the thermal nuclear-2SC continuity is to suppress thermally overpopulated quarks by introducing Polyakov loops in quark models \cite{Fukushima:2003fw,Roessner:2006xn}.

Having seen in detail the tendency at $\mu_Q=0$, we further extend our survey to a wider domain in $\mu_Q$. 
Shown in Fig.\ref{fig:eos_Tvary_g305g805} are $P$, $n_B/n_0$, $n_Q/n_0$, and $s/n_0$ for $\mu_Q=-140, 0, 100$ MeV and $T=10, 30, 50, 70$ MeV. 
The mismatch found in the $\mu_Q=0$ result persists for general $\mu_Q$. 
The trend of entropies is similar to the already discussed $\mu_Q=0$ case. 
Another trend is that $n_B$ and $n_Q$ in the 2SC react to changes in $\mu_Q$ more strongly than in nuclear descriptions. 
In contrast, the CFL domain hardly reacts to changes in $\mu_Q$. 
The main difference between the 2SC and CFL is the existence of gapless quarks which can react to small perturbations.

Looking over the results for $T\lesssim 50$ MeV, one might think the nuclear and 2SC results are reasonably consistent.
For the pressure curves, the mismatch is not so apparent, and the number densities in the 2SC descriptions deviate from the nuclear's by $\sim 30\%$ or so. 
These would give impressions that the mismatches can be readily eliminated.
 We found this is not the case. 
For this reason, we are forced to allow phenomenological corrections with which we depart from the conventional quasiparticle descriptions of CSCs. 
They are discussed in the next section.

\section{Phenomenological corrections: CSCX}\label{sec:nuclear-2SCX}

\begin{figure}[th]
\begin{center}
\vspace{-0.0cm}
\includegraphics[scale=0.7]{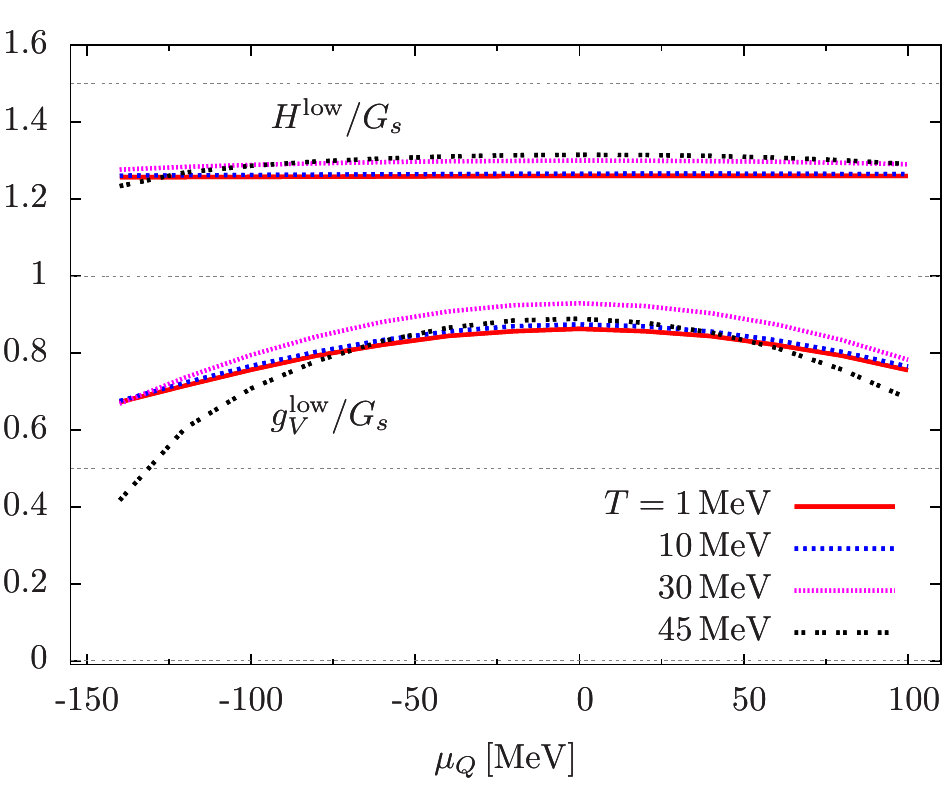}
\end{center}
\vspace{-0.5cm}
\caption{The low density limit of evolving effective couplings $g_V^{\rm low}$ and $H^{\rm low}$ as functions of $\mu_Q$ and $T$. (At $T\simeq 46$ MeV, $g^{\rm low}_V$ at $\mu_Q\simeq -140$ MeV changes the sign, preventing us from getting stable solutions for self-consistent equations.)
}
\label{fig:g_muQ-dep_T}
\end{figure}

\begin{figure*}[ht]
\begin{center}
\vspace{-1.0cm}
\includegraphics[scale=0.68]{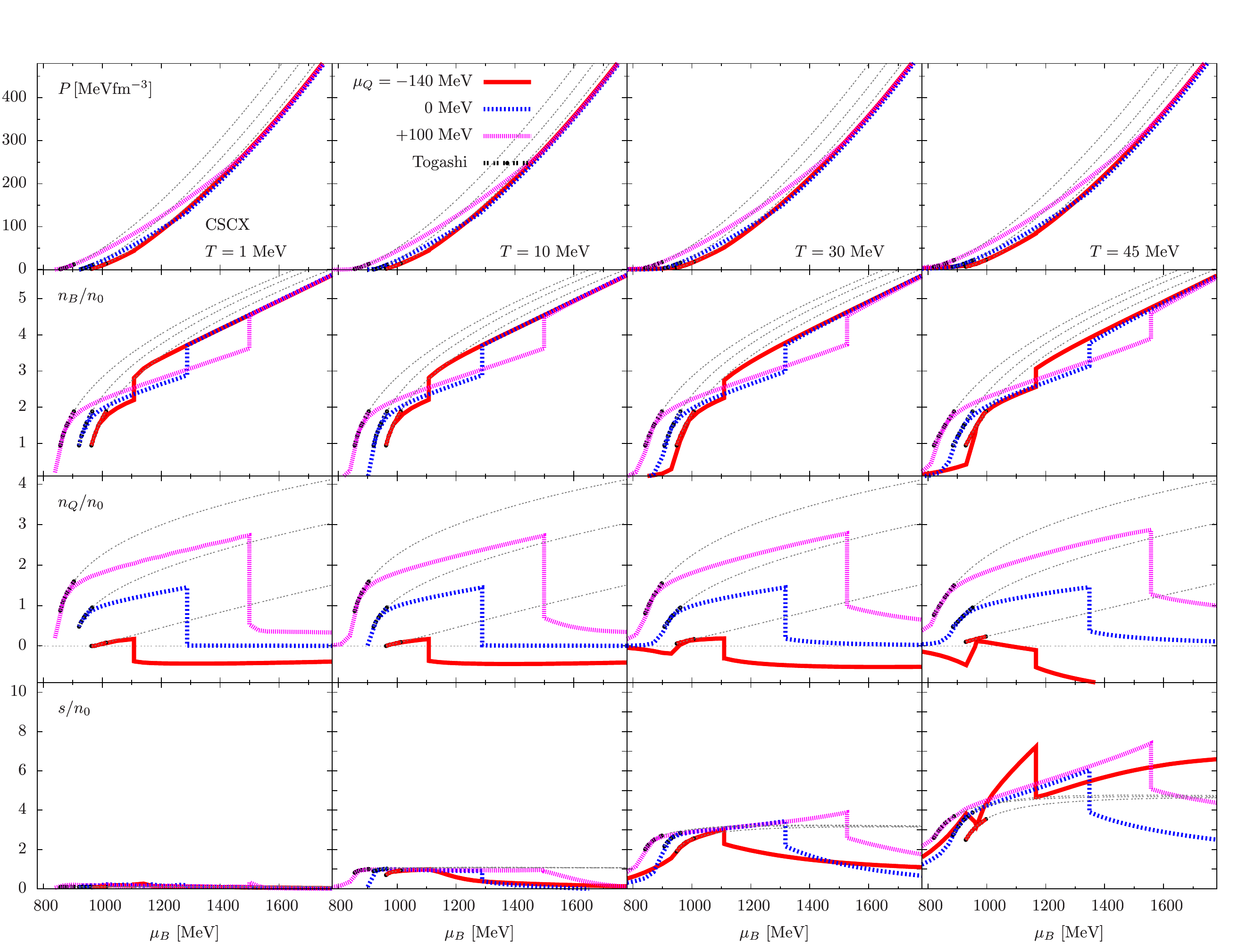}
\end{center}
\vspace{-0.5cm}
\caption{Thermal equations of state, $P$, $n_B/n_0$, $n_Q/n_0$, and $s/n_0$ as functions of $\mu_B$. The temperatures are $T=1, 10, 30, 45$ MeV, and the charge chemical potentials are $\mu_Q=-140, 0, 100$ MeV. The Togashi equation of state is used to tune the evolving couplings for all $(\mu_Q,T)$ and at $n_B\simeq 1.5n_0$ for the setup given in Eq.(\ref{eq:set_parameters}). At $T\gtrsim 45$ MeV, a matching between the nuclear and 2SC begins to be impossible within the current framework, as we can expect from the entropy at low density for $\mu_Q = -140$ MeV. For this difficulty the results beyond $T =45$ MeV are not displayed.
}
\label{fig:eos_Tvary_g305g805_CSCX}
\end{figure*}

In this section, we construct unified equations of state which cover from the nuclear to quark matter domains. 
As we have seen, the matching between the nuclear and 2SC equations of state needs some phenomenological corrections. 
A possible conclusion is that the nuclear-2SC continuity is simply impossible. 
Another possibility is that our descriptions of nuclear and 2SC are insufficient.
The nuclear model does not describe the 2SC correlation, and the 2SC calculation does not include three-particle correlations for baryons.
In this section we explore the latter possibility and examine how far we can go with this assumption. 
We introduce the $(\mu_Q,T)$-dependence in the evolving couplings and allow the 2SC to react to changes in $(\mu_Q,T)$ in the same way as the nuclear matter.
Although our quark matter uses the ordinary 2SC as a baseline, it now has the qualitative properties different from the ordinary 2SC. 
So we call the 2SC with evolving couplings ``CSCX'', with X emphasizing our ignorance of the practical descriptions.
The X contributes to the baryon number, charge, and entropy densities, but does not contribute to the color density, as our evolving couplings are taken independent of color chemical potentials. 

As before, we use evolving couplings for the 2SC pressure and baryon density to match the nuclear ones at $n_B=1.5n_0$, but now repeat the procedure for each $(\mu_Q,T)$. 
After $g$ is extended to $(\mu_Q,T)$-dependent parameters, they add extra contributions to the charge and entropy densities,
\beq
\Delta n_Q 
&=& \frac{\partial \calP}{\partial g} \bigg|_{ \lambda, \eta}  \frac{\, \partial g^{\rm low}_* \,}{\, \partial \mu_Q \,} \, \rme^{-V_q/V^g_{\rm trans} } 
 \,,~~~~~
 \nonumber \\
\Delta s 
&=&  \frac{\partial \calP}{\partial g} \bigg|_{ \lambda, \eta} \frac{\, \partial g^{\rm low}_* \,}{\, \partial T \,} \, \rme^{-V_q/V^g_{\rm trans} }
 \,.
\eeq
as we have discussed in Eqs.(\ref{eq:number}) and (\ref{eq:dg_dlam}). These contributions die out as $\sim \rme^{-V_q/V^g_{\rm trans} }$ as baryon density increases, but to some extent they survive in the CFL phase. In particular charge density is nonzero in the CFLX, in contrast to the usual CFL phase.

To begin with, we examine the ($\mu_Q$, $T$)-dependence of $g_*^{\rm low}$. In Fig.\ref{fig:g_muQ-dep_T} we show the behaviors of $(g_V^{\rm low}, H^{\rm low})$ as functions of $\mu_Q$ for $T=1, 10, 30,45$ MeV. The other parameters are the same as in the previous section, see Eq.(\ref{eq:set_parameters}).

We first remark on the $\mu_Q$-dependence. 
The behavior of $g$ is approximately symmetric with respect to $\mu_Q \leftrightarrow -\mu_Q$, as our quark model (nuclear models) has the (approximate) isospin symmetry. 
Next we notice that $H$ is insensitive to changes in $\mu_Q$, while $g_V$ considerably reduces as $\mu_Q$ deviates from $\mu_Q=0$. 
At this point we recall that the number density in the pure 2SC description was not sufficiently large at $\mu_Q=100$ and $-140$ MeV 
(see Figs.\ref{fig:eos_T001_Gv130H170G3G8_v0conf_gv2.5_H2.5} or \ref{fig:eos_Tvary_g305g805}). 
This over-reduced number density is enhanced back by reduction of $g_V$, or by weakening the repulsive density-density interactions.

Next we discuss the $T$-dependence. Both $g_V$ and $H$ depend on $T$ in a nonlinear way. To $T \sim 30$ MeV, both couplings increase as $T$ does. 
We recall that the pressure and number density in the pure 2SC descriptions are lower than the nuclear's. 
These mismatches are cured by increasing $H$ that enhances the pressure and number density at low density as well as the entropy. 
The increase of $g_V$ tends to counteract such tendency, but at low density the impact of $g_V$ is not as important as $H$ because the former appears as $\sim g_V n_q^2$. 
Beyond $T\sim 30$ MeV, the number density in the pure 2SC is still underestimated, but the entropy starts to increase faster than the nuclear's. 
This introduces the complex behaviors in $g_V$ and $H$. 
While $H$ increases a little, $g_V$ starts to decrease substantially. 
The impact is greater when $\mu_Q$ deviates more from zero.
 In particular, around $T\simeq 50$ MeV, $g_V^{\rm low}$ for $\mu_Q \simeq -140$ MeV turns into negative, preventing us from the self-consistent solutions. 
 At higher temperature, the same happens for the other $\mu_Q$ domain. 
 This is in part because the entropy in the 2SC description becomes too large, 
 as seen in Fig.\ref{fig:eos_T001_Gv130H170G3G8_v0conf_gv2.5_H2.5}, and these discrepancies can no longer be compensated by just arranging the strengths of $(g_V, H)$. 
 It seems that more fundamental modifications must be introduced. 
 For this reason we stop our illustration at $T=45$ MeV.

Having seen these $(\mu_Q,T)$-dependences of effective couplings, 
we found that their behaviors are not quite natural within the conventional 2SC picture; 
the effective couplings describe nonperturbative dynamics whose typical scale is $\sim \lqcd$, and should not be substantially affected by small changes in $\mu_Q$ and $T$ of a few tens of MeV. 
With this consideration, we reassure that the problems of matching are not mere fine-tuning issues, but are related to problems in physical descriptions.

With this caution in mind, we now turn to the equations of state of the CSCX. 
As in Fig.\ref{fig:eos_Tvary_g305g805} of the previous section,
 in Fig.\ref{fig:eos_Tvary_g305g805_CSCX} we show $P$, $n_B/n_0$, $n_Q/n_0$, and $s/n_0$ for $\mu_Q=-140, 0, 100$ MeV and $T=1, 10, 30, 45$ MeV. 
 By construction, the CSCX reproduces the nuclear equations of state around $n_B\simeq 1.5n_0$. 
 But it is worth mentioning that the matching is good over a finite range of $1$-$1.8n_0$, although we have demanded the matching only at a single point, $n_B=1.5n_0$.

As we have mentioned, the CSCX has extra contributions from the $(\mu_Q, T)$-dependence of effective couplings. 
Most notably, in the CFLX domain, the charge density is also positive (negative) for a positive (negative) $\mu_Q$.
The CFLX is charge neutral only at $\mu_Q \simeq 0$. 
This can be understood by recalling that  $\rmd g/\rmd \mu_Q = 0$ at $\mu_Q \simeq 0$ (Fig.\ref{fig:g_muQ-dep_T}) and the pure CFL is charge neutral. 
The particle content is then $n_u=n_d=n_s$ at $\mu_Q \simeq 0$. 
As the CFLX is perturbed by $\mu_Q$, $n_d+n_s < n_u$ for positive $\mu_Q$, while $n_d+n_s > n_u$ for negative $\mu_Q$. 
In next section we see the consequence of this relation by coupling leptons. 

Up to $T\simeq 45$ MeV, the entropy at $n_B \gtrsim 1.5n_0$ is smaller in the CSCX than in the nuclear case. 
The CFLX has less entropy than the 2SCX, as in the relation between the pure CFL and 2SC.

\section{CSCX for neutron stars}\label{sec:EoS_for_NS}

\begin{figure}[t]
\begin{center}
\includegraphics[scale=0.7]{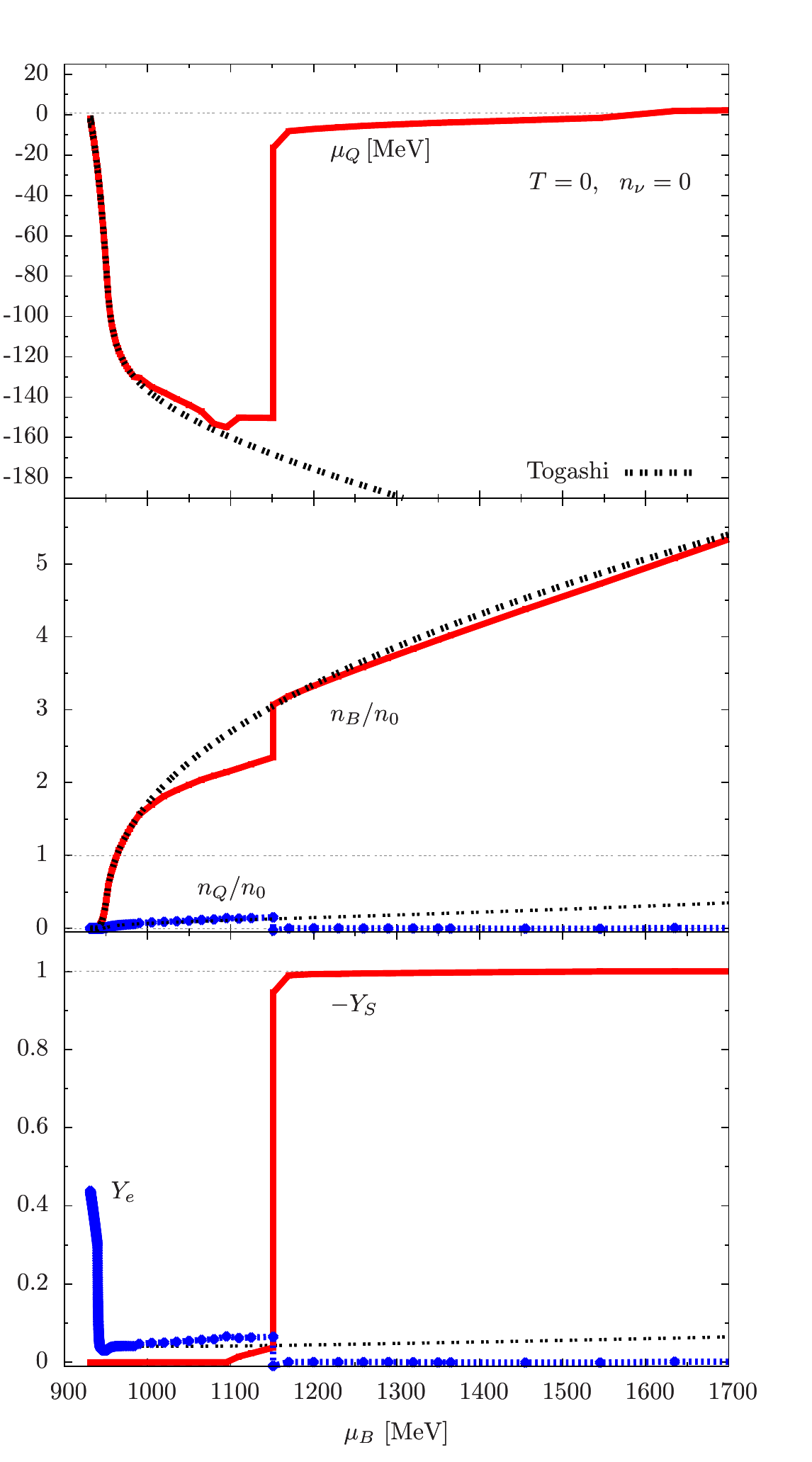}
\end{center}
\vspace{-0.5cm}
\caption{ Equations of state for static neutron stars, $\mu_Q$, $n_B/n_0$, $n_Q/n_0$, $-Y_S$, and $Y_Q$ for the CSCX+Togashi. The curves for the Togashi are also shown (except $Y_S$).
}
\label{fig:muQ-nB-Ys_muB_for_NS}
\end{figure}

\begin{figure}[th]
\begin{center}
\vspace{-0.0cm}
\includegraphics[scale=0.75]{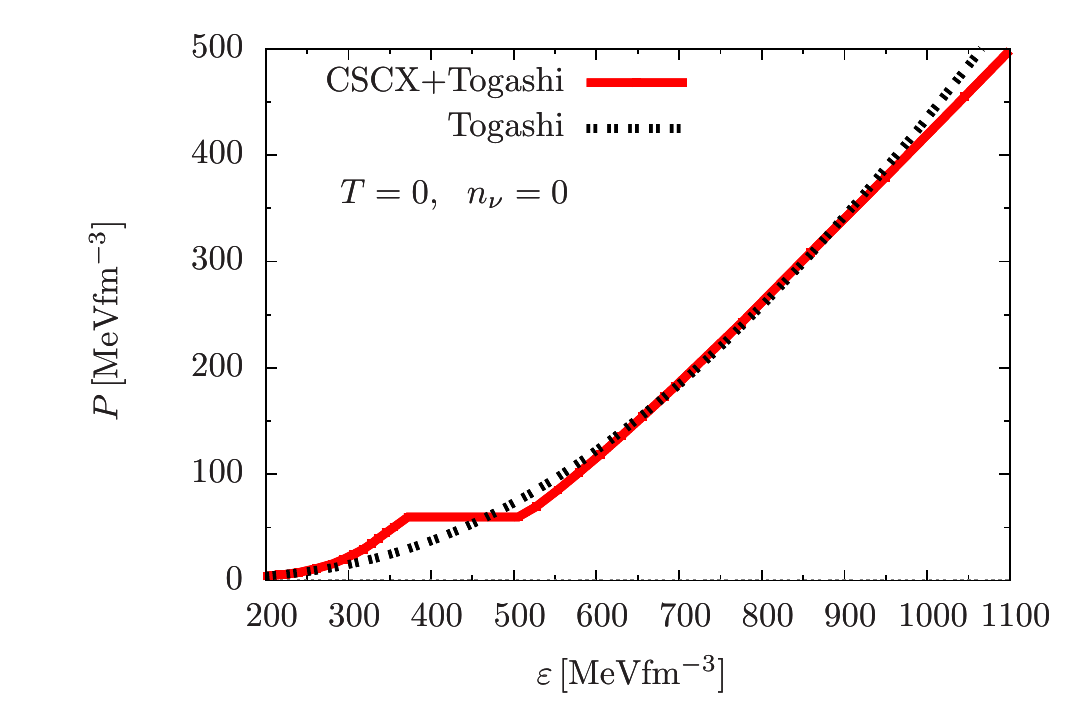}
\end{center}
\vspace{-0.5cm}
\caption{The $P$-$\varepsilon$ relations for the CSCX-Togashi and Togashi equations of state for static neutron stars. 
}
\label{fig:P-e_for_NS}
\end{figure}

\begin{figure}[th]
\begin{center}
\vspace{-0.0cm}
\includegraphics[scale=0.8]{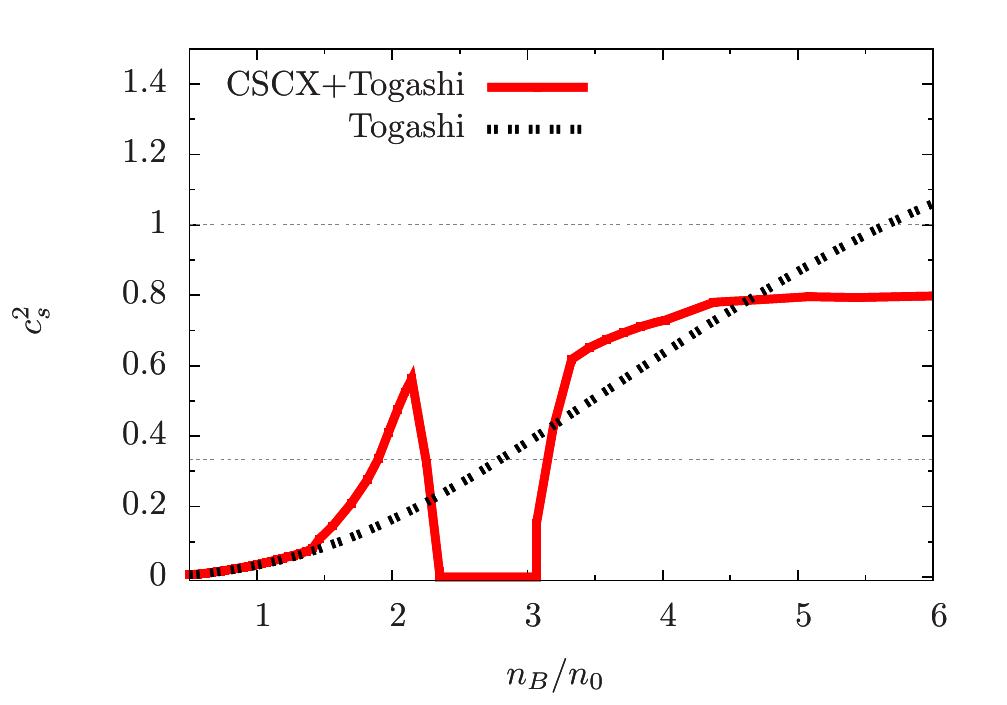}
\end{center}
\vspace{-0.5cm}
\caption{The speed of sound for the CSCX-Togashi and Togashi equations of state for static neutron stars. 
}
\label{fig:cs2-nB_for_NS}
\end{figure}

\begin{figure}[th]
\begin{center}
\vspace{-0.0cm}
\includegraphics[scale=0.7]{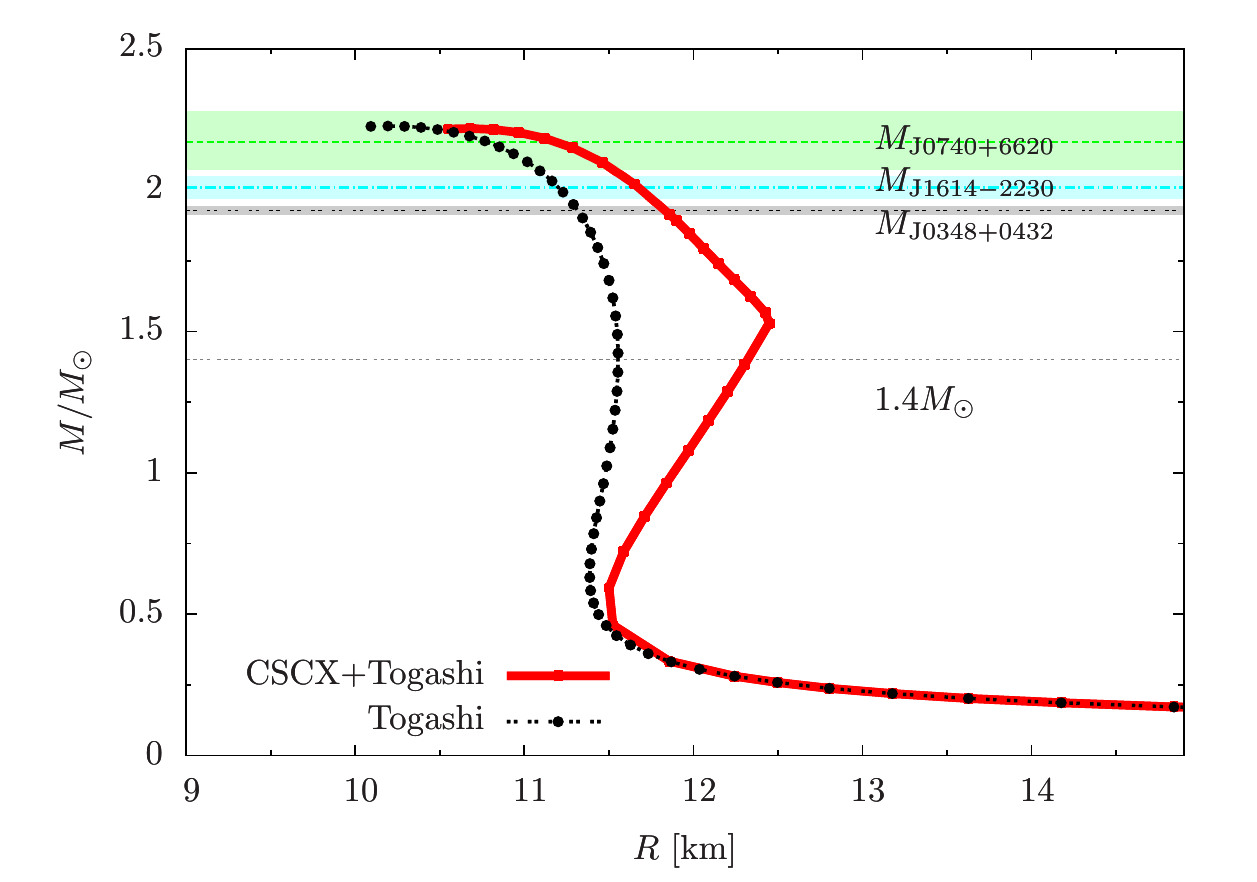}
\end{center}
\vspace{-0.5cm}
\caption{ The $M$-$R$ relations for the Togashi and CSCX-Togashi equations of state for $n_\nu=0$. 
The maximum masses are $2.23 M_\odot$ and $ 2.22 M_\odot$, and $R_{1.4}$ are $11.5$ and $12.3$ km, respectively. 
}
\label{fig:M-R}
\end{figure}

Finally, we examine the astrophysical implications of the equations of state of nuclear-``CSCX'', which were studied in the previous section.
Our main concern is the composition of matter at a given lepton and temperature which change during the dynamics of neutron stars. 
The pressure is given by ($T$ is hidden here)
\beq
P (\mu_B, \mu_Q ,\mu_L) = P_{\rm QCD} (\mu_B, \mu_Q) + P_e (\mu_Q ,\mu_L)  +P_{\nu} (\mu_L) \,,
\nonumber \\
\eeq
where the charged lepton has the chemical potential $\mu_e = \mu_L - \mu_Q$.

We first examine the equation of state ``CSCX+Togashi'' (with matching of the Togashi and CSCX) at $T=\mu_L=0$ for static neutron stars. 
The purpose is to check that the high density value for evolving couplings $g_{\rm high}$ are chosen to be consistent with the available neutron star observations. 
Next, we consider the neutrino trapping regime at finite temperature and lepton chemical potential.

\subsection{Static neutron stars}\label{sec:static_NS}

For static neutron stars we must tune $\mu_Q$ to satisfy the neutrality of electric charges\footnote{In contrast to the constant coupling cases, the determination of $\mu_Q^*$ is more cumbersome as the evolving couplings contain the $\mu_Q$-dependence. In particular we need to compute the $\mu_Q$-dependence of various quantities including condensates for which we do not have analytic expressions. For this reason we first prepare tables for various set of $(\mu_B, \mu_Q)$ and calculate the numerical derivatives.},
\beq
\frac{\partial P  }{ \partial \mu_Q} = n_Q (\mu_B, \mu_Q^* ) =0 \,,
\eeq
which determines the $\mu_Q^*$ as a function of $\mu_B$. We have set the lepton chemical potential to zero so that neutrinos are absent.

Shown in Fig.\ref{fig:muQ-nB-Ys_muB_for_NS} are $\mu_Q$, $n_B/n_0$, and $-Y_S$, for the CSCX+Togashi and the Togashi. 
All these quantities jump at a 2SCX-CFLX transition that takes place at $\mu_B\simeq 1150$ MeV; 
the baryon density jumps from $\simeq 2.3n_0$ to $\simeq 3.1 n_0$, $\mu_Q$  from $\simeq -150$ MeV to $\simeq -17$ MeV, and $-Y_S$ from $\simeq 0.03$ to $\simeq 0.94$. 

Shown in Figs.\ref{fig:P-e_for_NS} and \ref{fig:cs2-nB_for_NS} are $P$-$\varepsilon$ and $c_s^2$-$n_B/n_0$ relations for the CSCX+Togashi and the Togashi. 
The stiffness of the 2SCX phase grows faster than the Togashi, and the $c_s^2$ reaches beyond the conformal value $0.3$ already around $n_B\simeq 1.9n_0$. 
This stiffening effects are reflected in Fig.\ref{fig:M-R} for the $M$-$R$ relation. 
For low mass neutron stars the Togashi and CSCX-Togashi coincides to points around $(M, R)\simeq (0.5 M_\odot,\, 11.5\,{\rm km})$; 
for a heavier star the CSCX-Togashi leads to a larger radius; e.g., for $1.4M_\odot$ neutron stars $R_{1.4}\simeq 12.3$ km. 
A kink in the $M$-$R$ curve, with $M\simeq 1.53M_\odot$ and $R\simeq 12.4$ km, reflects the 2SCX-CFLX transition. 
After having the transition, the resultant CFLX matter must be sufficiently stiff to pass the $2M_\odot$ constraint. 
Our choice of the high density couplings, $(g_V, H)^{\rm high}/G_s = (1.3, 1.7)$, indeed satisfies the constraint. 
As the CSCX+Togashi and Togashi have similar $P$-$\varepsilon$ relations at high density, the maximum masses are similar. 
At the maximum mass, the CSCX+Togashi has the $M$-$R$ relations and the core density, $(M, R, n_B^{\rm core}) \simeq (2.22 M_\odot,\, 10.7\, {\rm km}, \, 6.5n_0)$,
and for the Togashi, $\simeq (2.23 M_\odot,\, 10.2\,{\rm km},\, 6.9n_0).$\footnote{Actually 
 $c_s^2$ in the Togashi begins to violate the causality at $n_B \simeq 5.6n_0$; if we stop calculations at this point then $M \simeq 2.18M_\odot$ and $R\simeq 10.7$ km.
}

\subsection{Neutrino trapping regime}\label{sec:nu_trapping}

\begin{figure*}[ht]
\begin{center}
\vspace{-1.0cm}
\includegraphics[scale=0.68]{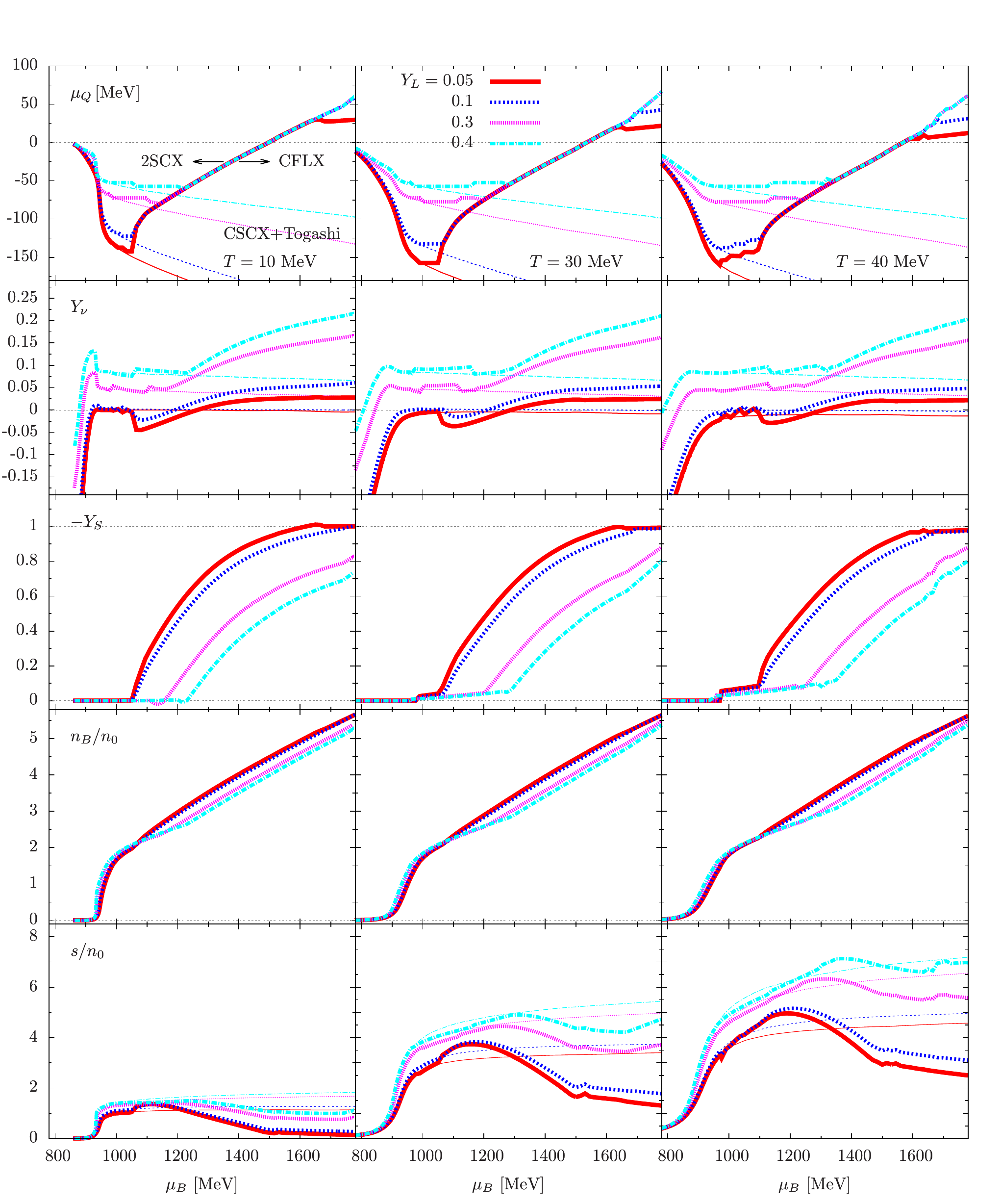}
\end{center}
\vspace{-0.5cm}
\caption{Equations of state (CSCX+Togashi) for charge neutral matter, $\mu_Q$, $Y_\nu$, $-Y_S$, $n_B/n_0$, and $s/n_0$ as functions of $\mu_B$. The temperatures are 
$T=10, 30, 40$ MeV, 
and the lepton fraction is $Y_L=0.05, 0.1, 0.3, 0.4$. For $\mu_Q$, $Y_\nu$, and $s/n_0$, we also plot the Togashi results with thin lines. Increasing lepton numbers broaden the 2SCX domain by shifting the 2SCX-CFLX boundary to the higher density.
}
\label{fig:eos_with_nu_YL}
\end{figure*}

\begin{figure*}[ht]
\begin{center}
\vspace{.0cm}
\includegraphics[scale=0.67]{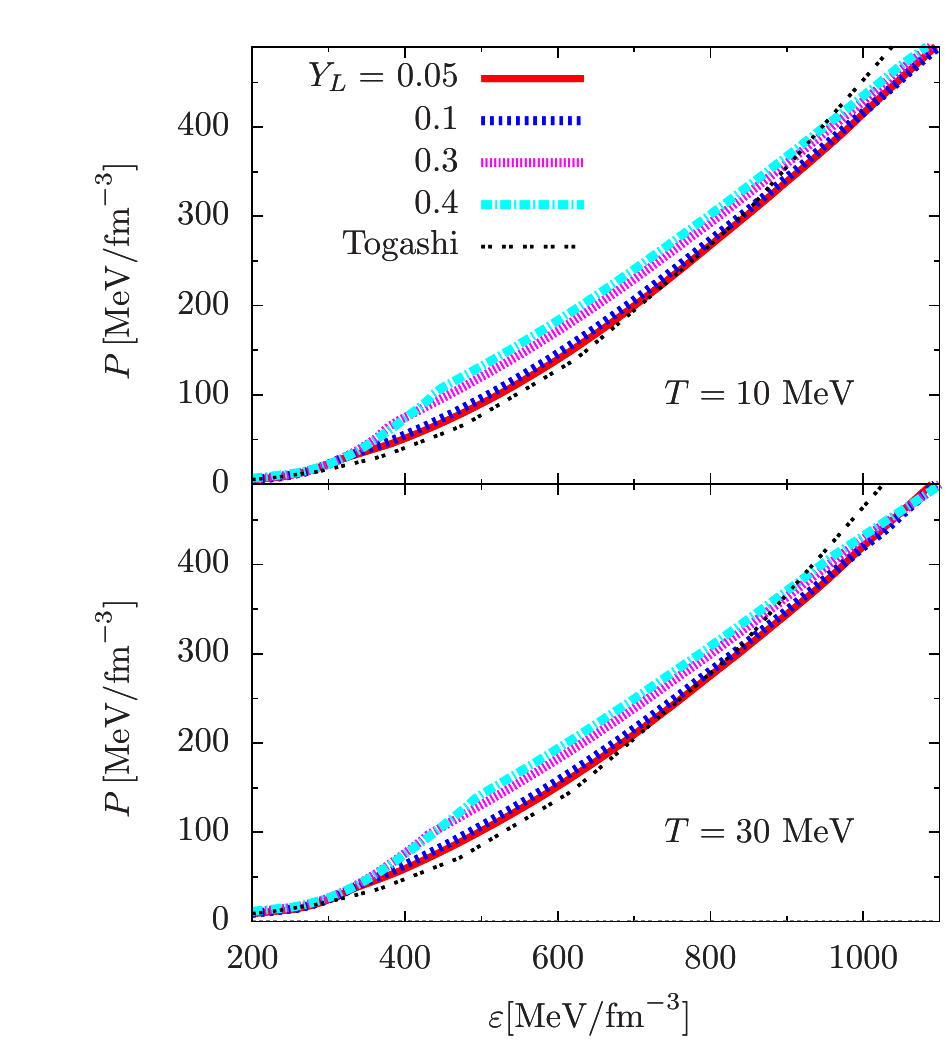}
\includegraphics[scale=0.67]{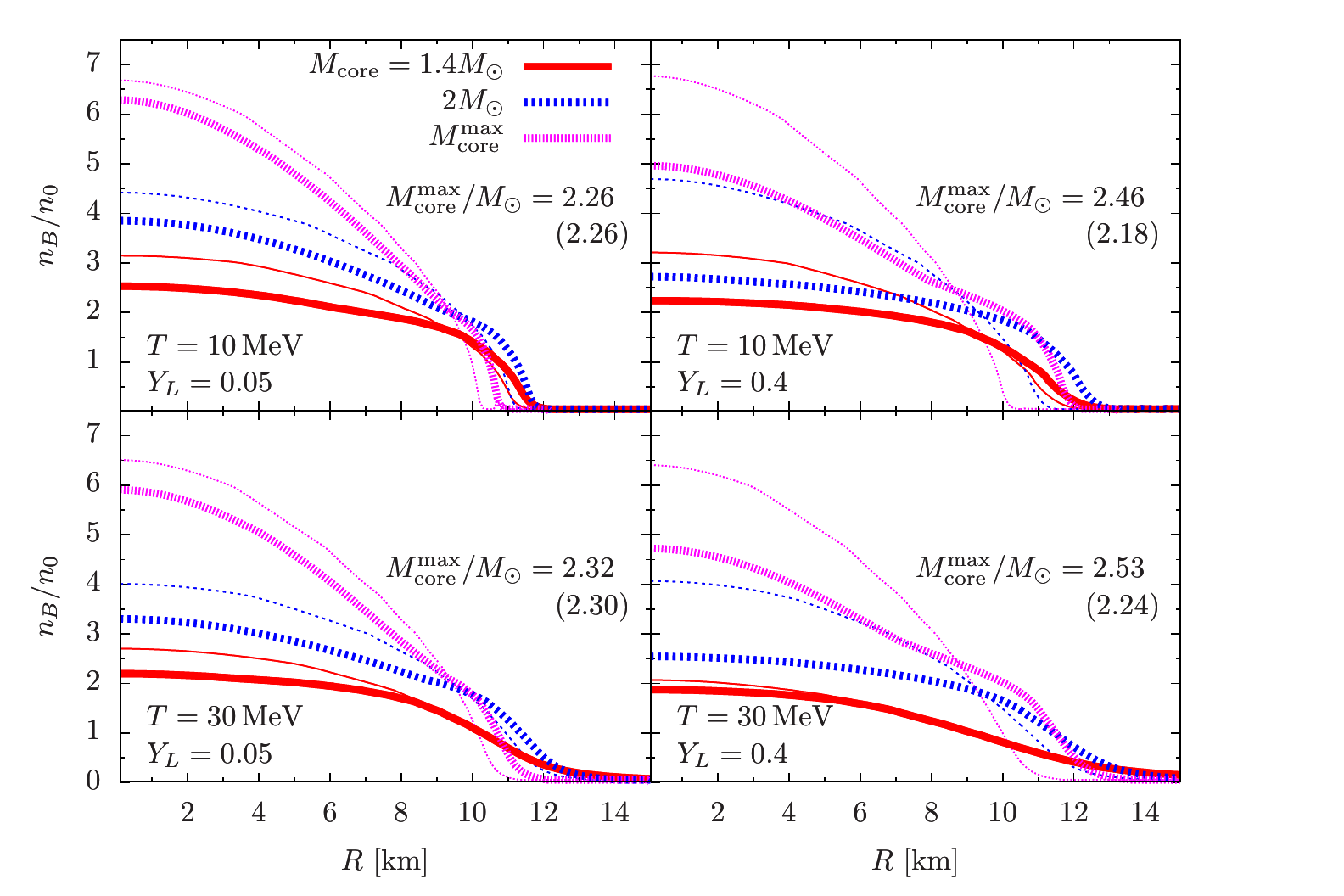}
\end{center}
\vspace{-0.5cm}
\caption{(Left) The CSCX+Togashi pressure as a function of energy density for $T=10, 30$ MeV and $Y_L=0.05, 0.1, 0.3, 0.4$. 
The Togashi case at $Y_L=0.05$ is also shown as a reference.
(Right) The baryon density distributions in neutron stars for given core masses, $M_{\rm core} =1.4, 2.0 M_\odot$ and $M_{\rm core}^{\rm max}$. 
The conditions are $T=10,30$ MeV and $Y_L=0.05, 0.4$. 
For the ``core mass'' $M_{\rm core}$, we integrated only matter at $n_B \gtrsim 0.05n_0$ and omitted loosely bound matter which has a large volume at finite $T$. 
The thin lines correspond to the results for the Togashi, and its maximal core mass is shown in the parenthesis.
}
\label{fig:nB_dist}
\end{figure*}

We next consider the neutrino trapping regime. This regime is possible only when there are substantial amounts of thermally excited states which interact with neutrinos. 
In the protoneutron star context the neutrino trapped matter is expected to have $s/n_B \simeq 1$-$2$. 
As we have seen in Fig.\ref{fig:eos_Tvary_g305g805}, this condition is met at $T\sim 10$ MeV for the Togashi-2SCX region, and at $T\sim 30$ MeV for the CFLX region. 
Below we consider $T \gtrsim 10$ MeV and assume the neutrino trapping regime for equations of state.

With neutrinos we have a lepton chemical potential. 
We determine the lepton chemical potential at a given $(\mu_B,\mu_Q, T)$ through the charge neutrality constraint,
\beq
\frac{\partial P  }{ \partial \mu_Q} = n_Q (\mu_B, \mu_Q ,\mu_L^*,T) =0 \,.
\eeq
%
Thus $\mu_L^*$ is a function of $(\mu_B,\mu_Q,T)$. 
Assuming massless neutrinos with a single helicity, a set of $(\mu_L,T)$ is readily converted into the neutrino equations of state
\beq
P_{\nu} = N_\nu \bigg( \frac{\, \mu_L^4 \,}{\, 24\pi^2 \,} + \frac{\, \mu_L^2 T^2 \,}{\, 12 \,} \bigg) + N_\nu' \frac{\, 7\pi^2 T^4 \,}{\, 360 \,} \,.
\eeq
where $N_\nu$-species of neutrinos have the chemical potential $\mu_L$ and $N_\nu'$ species of neutrinos contribute thermal pressure. We assume $\nu_e$ and $\nu_\mu$ have nonzero chemical potentials but $\nu_\tau$ does not, and set $N_\nu=2$ and $N_\nu'=3$. 

In general there can be the first order transitions from the 2SCX to CFLX. The location is determined by the condition
\beq
P_{\rm 2SCX} (\mu_B, \mu_Q, \mu_L^{\rm 2SCX} ) =  P_{\rm CFLX} (\mu_B, \mu_Q, \mu_L^{\rm CFLX} ) \,,
\eeq
where $\mu_L^{\rm 2SCX} $ and $ \mu_L^{\rm CFLX}$ are fixed by the condition
\beq
n_Q^{\rm 2SCX} = n_Q^{\rm CFLX} = 0 \,. 
\eeq
At the first order transitions, extensive quantities generally jump. 
The values of $n_\nu$ which we are investigating can be often found within the first order transitions. 
To find the corresponding equations of state, we consider a state at the first order transitions ($0 \le x \le 1$),
\beq
|\Psi_{\rm 1st} \ra = \sqrt{x\,}\, |\Psi_{\rm 2SCX} \ra + \sqrt{1-x \,}\, |\Psi_{\rm CFLX} \ra \,,
\eeq
where $|\Psi_{\rm 2SCX} \ra$ and $|\Psi_{\rm CFLX} \ra$ lead to the same pressure.
Local operators $\hat{O}$ are evaluated as
\beq
&& \la \Psi_{\rm 1st}  | \hat{O} |\Psi_{\rm 1st} \ra \nonumber \\
&& \simeq  x \la \Psi_{\rm 2SCX}  | \hat{O} |\Psi_{\rm 2SCX} \ra + (1-x) \la \Psi_{\rm CFLX}  | \hat{O} |\Psi_{\rm CFLX} \ra \,,
\nonumber \\
\label{eq:along_1st_PT}
\eeq
where we neglected the off-diagonal components, $ \la \Psi_{\rm 2SCX}  | \hat{O} |\Psi_{\rm CFLX} \ra$, which should vanish in the infinite volume limit \cite{weinberg1995quantum}. 
Using this relation, we first fix $x$ to reproduce a given lepton fraction, and then use its value to compute the other quantities.

Shown in Fig.\ref{fig:eos_with_nu_YL} are equations of state, $\mu_Q$, $-Y_S$, $n_B/n_0$, and $s/n_0$ as functions of $\mu_B$. For $\mu_Q$ and $s/n_0$, we also plot the Togashi results with thin lines as they are substantially different from the CSCX results.  
The temperatures are
$T=10, 30, 40$ MeV 
which cover $s/n_B \sim 1$-$3$, and the lepton fraction is $Y_L=0.05, 0.1, 0.3, 0.4$. 
The $Y_L=0.05$ and $0.1$ are suitable for neutron star mergers as static neutron stars before merging do not have many leptons, 
while larger values of $Y_L \gtrsim 0.3$ are typical for protoneutron stars as they are formed through contractions of stars with many nuclei. 

Below, we examine the quark composition and neutrino fractions, stiffness, and the structure of hot neutron stars.

\subsubsection{Quark composition}

First, we note changes in the phase structure with increasing lepton fractions or $\mu_L$. 
With a larger $\mu_L$, the system is electrically neutralized without invoking a large negative value of $\mu_Q$ in the charged lepton chemical potential $\mu_e = \mu_L-\mu_Q$; 
the value of $\mu_Q$ approaches a more positive value. 
Accordingly, the chemical potential for strange quarks is reduced, and the strangeness fraction is suppressed. 
This broadens the 2SCX domain with overall shifts of the 2SCX-CFLX boundary to the higher density. 
For a fixed $Y_L$ line, extensive quantities change smoothly everywhere. 
After the line meets the 2SCX-CFLX phase boundary, 
the line changes along the phase boundary line for a while and then departs when the CFL phase alone can satisfy the constraint of $Y_L$. 
On the phase boundary, the extensive quantities follow Eq.(\ref{eq:along_1st_PT}).

When the temperature is turned on, the basic features of the phase structure remain the same, 
except broadening of the 2SCX domain to higher density. 
For the temperature range in this work, the major impact of the temperature is on the strangeness fraction.

\subsubsection{Neutrino fraction}

Next, we consider the neutrino fraction. We divide the domain into five and examine the results shown in Fig.\ref{fig:eos_with_nu_YL}:

(i)  In the dilute regime, $Y_e\sim 0.4$, so $Y_\nu = Y_L - Y_e \lesssim 0$ for our choices of $Y_L$, leading to more antineutrinos than neutrinos.

(ii) Near the nuclear matter domain around $n_B \sim n_0$, our experience on static neutron stars with $Y_\nu =0$ indicates that $Y_e \sim 0.05$ (see Fig.\label{fig:muQ-nB-Ys_muB_for_NS}). 
Then, we can infer that, for the $Y_L=0.05$ and $0.1$ cases, these conditions are satisfied with $Y_\nu \sim 0$ and $\mu_L\sim 0$, 
as can be confirmed from Fig.\ref{fig:eos_with_nu_YL}. 
Then, for a larger $Y_L$, the $\mu_L$ should increase, so does $Y_\nu$.

(iii) Along the first order line at $\mu_Q < 0$, $Y_\nu$ can take both positive and negative values. To understand this, we examine the charge density in the QCD sector for the the CFLX phase. For $\mu_Q < 0 $, the charge density is negative, $n^{\rm QCD}_Q < 0$, so we need charged anti-leptons, $n_e < 0$. This means $\mu_e = \mu_L - \mu_Q < 0$, leading to $\mu_L < \mu_Q < 0$, and hence $Y_\nu <0$. Along the first order line at $\mu_Q <0$, the neutrino vs anti-neutrino fractions depend on the ratio between the nuclear-2SC and CFL phases, and the condition $Y_L$.

(iv) Along the first order line at $\mu_Q > 0$, the $Y_\nu$ turns out to be positive. In the CFLX phase at $\mu_Q >0$, the charge density is positive, $n^{\rm QCD}_Q > 0$, so we need $n_e > 0$. This means $\mu_e = \mu_L - \mu_Q > 0$, leading to $\mu_L > \mu_Q > 0$, and hence $Y_\nu >0$. Since the 2SCX phase also leads to $Y_\nu >0$, we have $Y_\nu >0$ at $\mu_Q >0$ from the 2SCX to CFLX domain. 

(v) At very large density, the QCD sector neutralizes by itself for a wide domain in $\mu_Q$. 
Thus, $\mu_e \sim 0$, and the lepton number is chiefly carried by neutrinos. As a result the neutrino abundance is greater than the Togashi by several factors.

\subsubsection{Stiffness and the core structure}

Finally we examine the structure of hot, neutrino-rich neutron stars within the isothermal picture of the core. 
The finite temperature effects significantly change the crust part as it is loosely bound to the core; 
this dilute domain can be widely spread to $\sim 100$ km or even more. 
Clearly, this crust part is dominated by the physics different from the core part. 
For this reason, we take into account only the $n_B \gtrsim 0.05n_0$ part of equations of state to integrate the Tolman-Oppenheimer-Volkoff equation. 
We call the resulting mass ``core mass'' $M_{\rm core}$ in this paper.

Shown in the left panel of Fig.\ref{fig:nB_dist} are $P$ vs $\varepsilon$ for $T=10, 30$ MeV and $Y_L=0.05, 0.1, 0.3, 0.4$. 
The Togashi case at $Y_L=0.05$ is also shown as a reference at a given temperature. 
As we can infer from the previous sections, the lepton fraction $Y_L$ controls the stiffness through the strangeness fraction. 
A large $Y_L$ leads to the stiffer equation of state. 
Meanwhile, the temperature effects are overall small in size, and its major impact seems to be in the shift of the phase boundaries.

Shown in the right panel of Fig.\ref{fig:nB_dist} are the baryon density distributions in neutron stars for given core masses, 
$M_{\rm core} =1.4M_\odot, 2.0 M_\odot$, and $M_{\rm core}^{\rm max}$. 
The conditions are $T=10,30$ MeV and $Y_L=0.05, 0.4$. Several remarks are in order:

(i) The increase in $T$ from 10 to 30 MeV enhances the maximal core mass, $M_{\rm core}^{\rm max}$, by $\sim 0.05 M_\odot$. 
The baryon number distribution at $R \lesssim 10$ km is not affected much by thermal effects. 
The impacts of thermal effects are more significant for a larger $R$ and a lighter star, due to its diluter structure which can be easily deformed by the gravity. 
The same is also applied to the Togashi.

(ii) The increase in $Y_L$ substantially affects $M_{\rm core}^{\rm max}$ and the density distribution. 
The change in $Y_L$ from $0.05$ to $0.4$ results in the enhancement of $M_{\rm core}^{\rm max}$ by $\simeq 0.2M_\odot$ in both the $T=10$ and 30 MeV cases. For $Y_L=0.05$ MeV,
\beq
M_{\rm core}^{\rm max}/M_\odot \simeq 2.26, 2.32 \,,~~~ ({\rm for}~T=10, 30\,{\rm MeV})
\eeq
and for $Y_L=0.4$,
\beq
M_{\rm core}^{\rm max}/M_\odot \simeq 2.32, 2.53 \,.~~~ ({\rm for}~T=10, 30\,{\rm MeV})
\eeq
This enhancement is due to stiffening at $n_B \simeq 1.5$-$4n_0$ which tempers the growth in baryon density. 
Accordingly the core density for the $M_{\rm core}^{\rm max}$ star is lower for the $Y_L=0.4$ case, $n_B^{\rm core} \simeq 5n_0$, 
than the $Y_L=0.05$ case, $n_B^{\rm core} \simeq 6n_0$. 
In contrast, in the Togashi case, changes in $Y_L$ do not lead to substantial increase in $M_{\rm core}^{\rm max}$, 
but a slight reduction by $\sim 0.07M_\odot$; this is due to the fact that a larger $Y_L$ makes pure nuclear matter more symmetric in isospin and reduces its stiffness.

\section{Summary}\label{sec:summary}


We have performed comprehensive analyses for equations of state based on the nuclear-2SC continuity picture. We elaborated a scheme of evolving couplings which are tuned to reproduce nuclear pressure and number density at $n_B=1.5n_0$ and $\mu_Q=T=0$, and they approach the high density values to reproduce the $2M_\odot$ constraint. 

Our analyses indicate that the nuclear and 2SC equations of state do not match well over phenomenologically relevant domains of $\mu_Q$ and $T$. This is most clearly seen in entropies whose low temperature behaviors are characterized by the number of gapless fermions and the Fermi velocities. The nuclear and 2SC phases have the same number of gapless fermions, but the Fermi velocities are different, as the effective masses for nucleons and gapless quarks differ by a factor $\sim \Nc$ if we neglect interaction effects in nuclear calculations, and a factor $\sim 2$ if interactions are taken into account. If we used the unpaired quark matter for the matching, the number of gapless fermions is different and the discrepancy in entropies becomes even larger.

These observations suggest that, to achieve the nuclear-2SC continuity in entropies, it is necessary to consider corrections to both the nuclear and 2SC equations of state. Inclusion of more relativistic effects to the nuclear equations of state partially reduces the mismatch. Another possible scenario is that 
baryonic three particle correlations are present in the 2SC phase, so that reactions to changes in $(\mu_Q,T)$ become similar to the nuclear's.. 
Such phenomenological corrections are introduced through evolving couplings, and we call the CSC with such contributions ``CSCX''.

The unified equations of state, which cover from the nuclear to quark matter domains, are constructed by connecting the nuclear and 2SCX phases. 
At higher density, the 2SCX phase turns into the CFLX phase. 

Unlike the previous crossover constructions, equations of state in this work include the first order phase transition, 
but it is not a hadron-quark phase transition but the 2SC-CFL transition within quark matter. 
The first order nature is associated with radical appearance of the strangeness. 
We suspect that the strangeness appears more smoothly if we manifestly treat hyperonic baryons,
and would temper the softening associated with the first order phase transition.
We leave this issue as a future problem.

The strangeness fraction has important impacts on the structure of neutrino trapped, hot neutron stars.
The abundance of neutrinos and thermal effects reduces the strangeness fraction and stiffens equations of state. 
For a neutrino trapped neutron star at $T\simeq 30 $ MeV with a lepton fraction $Y_L\simeq 0.05$, the mass is larger than its cold static counterpart by  $\sim 0.1M_\odot$. 
This should affect theoretical estimates on the lifetime of neutron star mergers. 
More detailed studies are called for.

Clearly, this work leaves a lot of of room for improvements. 
We close this paper by mentioning several possible extensions.

First, we need to make explicit what the X is. 
We suspect it to be a baryonic object; diquarks near the Fermi surface would further pick up another quark, developing three particle correlations. 
If such three particle correlations are sufficiently strong, this likely leads to quarkyonic matter proposed by McLerran and Pisarski \cite{McLerran:2007qj}. 
Recently, the picture was also discussed in the language of quantum percolation \cite{Fukushima:2020cmk}. 
Several schematic quarkyonic equations of state have been constructed \cite{McLerran:2018hbz,Jeong:2019lhv,Duarte:2020xsp,Duarte:2020kvi,Zhao:2020dvu,Sen:2020peq,Cao:2020byn}, 
leading to ``soft-to-stiff'' type equations compatible with observations for static neutron stars.
In the context of the quark-hadron continuity, this description is probably even more powerful at finite temperature and lepton fraction, 
as the Fermi surface is made of baryons in quarkyonic matter.

Other candidates for the X are additional pairings to usual diquark pairs. In fact, in the CFL domain, we have already checked that our quark model leads to charged meson condensations around $\mu_Q \gtrsim 20$ MeV and $\mu_Q \lesssim -100$ MeV \cite{Kojo:2016dhh}, as the CFL mesons have excitation energies much smaller than in the vacuum case. 
These charged mesons change the response to $\mu_Q$ already at $T=0$. The equations of state with these exotic phases will be reported elsewhere.

In this paper, we have omitted discussions on the inhomogeneous phases such as crystalline CSCs \cite{Bowers:2002xr,Casalbuoni:2003wh,Anglani:2013gfu} and chiral spirals (or chiral density waves) \cite{Kojo:2009ha,Andersen:2018osr,Ferrer:2019zfp,Buballa:2014tba,Nakano:2004cd,Kojo:2014fxa,Kojo:2010fe,Kojo:2011cn}. These phases have not been discussed in detail in light of recent neutron star observations and further studies are called for.

We plan to work out more systematic analyses, examining the sensitivity to the choice of nuclear equations of state, other choices of $(g_V,H)$, and so on. The results will be presented elsewhere.

\section*{Acknowledgments}
 T.K. is supported by NSFC Grant No. 11875144; 
 D.H. by NSFC Grant Nos. 11735007, 11890711;
 H.T. by JSPS KAKENHI No.18K13551.
   T.K. and H.T. were supported in part by the Aspen Center for Physics in 2018. They also thank G. Baym, T. Hatsuda, and S. Furusawa for discussions related to this work.

\appendix

\section{Extrapolating nuclear tables to proton rich domain}\label{app:proton-rich}

Nuclear equations of state are written as ($Y_p + Y_n =1$)
\beq
\varepsilon(Y_p) 
= \varepsilon_{\rm NR} (Y_p) + \big( m_p Y_p + m_n Y_n \big) n_B\,.
\label{eq:Yp}
\eeq
(We suppress $n_B$ and $T$ in the thermodynamic quantities for notational simplicity.)
We explicitly separated the mass contributions, 
as they should be most relevant isospin breaking terms coming from $m_d-m_u$. 
For the other parts, the small mass difference is suppressed by the Fermi momentum $p_F$ or the dynamical mass scale $\lqcd$, so we neglect the isospin breaking effects,
\beq
\varepsilon_{\rm NR} (Y_p) \simeq \varepsilon_{\rm NR} (Y_n) \,,
\eeq
then
\beq
\varepsilon(Y_n) 
\simeq \varepsilon_{\rm NR} (Y_p) + \big( m_p Y_n + m_n Y_p \big) n_B\,.
\label{eq:Yn}
\eeq
Eliminating $\varepsilon_{\rm NR} (Y_p)$ from Eqs.(\ref{eq:Yp}) and (\ref{eq:Yn}), we get an approximate relation,
\beq
\varepsilon(Y_n)
\simeq
\varepsilon(Y_p) 
+ ( m_p -  m_n ) (Y_n -Y_p) n_B \,.
\eeq
For the entropy we do not expect significant isospin breaking effects and assume
\beq
s(Y_n) \simeq s(Y_p) \,.
\eeq
The other thermodynamic quantities are derived from these approximate relations. 
The charge chemical potential is obtained from $\partial \varepsilon/\partial n_Q |_{n_B} = n_B^{-1} \partial  \varepsilon/\partial Y_Q |_{n_B} $,
\beq
\mu_Q (Y_{n} ) \simeq - \mu_Q ( Y_{p} ) - 2 \big( m_n - m_p \big) \,,
\eeq
and the baryon chemical potential is $\partial \varepsilon/\partial n_B |_{n_Q}$,
\beq
\mu_B (Y_{n} ) 
\simeq  \mu_B (Y_{p} ) + \mu_Q (Y_{p} ) + m_n -  m_p 
 \,.
\eeq
The pressure is $P=\mu_B n_B+ \mu_Q n_Q + T s - \varepsilon$, so the above relations lead to
\beq
P (Y_{n} ) \simeq  P (Y_{p} )  \,.
\eeq
\bibliographystyle{apsrev4-2}
\bibliography{Ref_running_g}

\begin{thebibliography}{103}%
\makeatletter
\providecommand \@ifxundefined [1]{%
 \@ifx{#1\undefined}
}%
\providecommand \@ifnum [1]{%
 \ifnum #1\expandafter \@firstoftwo
 \else \expandafter \@secondoftwo
 \fi
}%
\providecommand \@ifx [1]{%
 \ifx #1\expandafter \@firstoftwo
 \else \expandafter \@secondoftwo
 \fi
}%
\providecommand \natexlab [1]{#1}%
\providecommand \enquote  [1]{``#1''}%
\providecommand \bibnamefont  [1]{#1}%
\providecommand \bibfnamefont [1]{#1}%
\providecommand \citenamefont [1]{#1}%
\providecommand \href@noop [0]{\@secondoftwo}%
\providecommand \href [0]{\begingroup \@sanitize@url \@href}%
\providecommand \@href[1]{\@@startlink{#1}\@@href}%
\providecommand \@@href[1]{\endgroup#1\@@endlink}%
\providecommand \@sanitize@url [0]{\catcode `\\12\catcode `\$12\catcode
  `\&12\catcode `\#12\catcode `\^12\catcode `\_12\catcode `\%12\relax}%
\providecommand \@@startlink[1]{}%
\providecommand \@@endlink[0]{}%
\providecommand \url  [0]{\begingroup\@sanitize@url \@url }%
\providecommand \@url [1]{\endgroup\@href {#1}{\urlprefix }}%
\providecommand \urlprefix  [0]{URL }%
\providecommand \Eprint [0]{\href }%
\providecommand \doibase [0]{https://doi.org/}%
\providecommand \selectlanguage [0]{\@gobble}%
\providecommand \bibinfo  [0]{\@secondoftwo}%
\providecommand \bibfield  [0]{\@secondoftwo}%
\providecommand \translation [1]{[#1]}%
\providecommand \BibitemOpen [0]{}%
\providecommand \bibitemStop [0]{}%
\providecommand \bibitemNoStop [0]{.\EOS\space}%
\providecommand \EOS [0]{\spacefactor3000\relax}%
\providecommand \BibitemShut  [1]{\csname bibitem#1\endcsname}%
\let\auto@bib@innerbib\@empty
\bibitem [{\citenamefont {Fukushima}\ and\ \citenamefont
  {Hatsuda}(2011)}]{Fukushima:2010bq}%
  \BibitemOpen
  \bibfield  {author} {\bibinfo {author} {\bibfnamefont {K.}~\bibnamefont
  {Fukushima}}\ and\ \bibinfo {author} {\bibfnamefont {T.}~\bibnamefont
  {Hatsuda}},\ }\href {https://doi.org/10.1088/0034-4885/74/1/014001}
  {\bibfield  {journal} {\bibinfo  {journal} {Rept. Prog. Phys.}\ }\textbf
  {\bibinfo {volume} {74}},\ \bibinfo {pages} {014001} (\bibinfo {year}
  {2011})},\ \Eprint {https://arxiv.org/abs/1005.4814} {arXiv:1005.4814
  [hep-ph]} \BibitemShut {NoStop}%
\bibitem [{\citenamefont {Fukushima}\ and\ \citenamefont
  {Sasaki}(2013)}]{Fukushima:2013rx}%
  \BibitemOpen
  \bibfield  {author} {\bibinfo {author} {\bibfnamefont {K.}~\bibnamefont
  {Fukushima}}\ and\ \bibinfo {author} {\bibfnamefont {C.}~\bibnamefont
  {Sasaki}},\ }\href {https://doi.org/10.1016/j.ppnp.2013.05.003} {\bibfield
  {journal} {\bibinfo  {journal} {Prog. Part. Nucl. Phys.}\ }\textbf {\bibinfo
  {volume} {72}},\ \bibinfo {pages} {99} (\bibinfo {year} {2013})},\ \Eprint
  {https://arxiv.org/abs/1301.6377} {arXiv:1301.6377 [hep-ph]} \BibitemShut
  {NoStop}%
\bibitem [{\citenamefont {Arzoumanian}\ \emph {et~al.}(2018)\citenamefont
  {Arzoumanian} \emph {et~al.}}]{Arzoumanian:2017puf}%
  \BibitemOpen
  \bibfield  {author} {\bibinfo {author} {\bibfnamefont {Z.}~\bibnamefont
  {Arzoumanian}} \emph {et~al.} (\bibinfo {collaboration} {NANOGrav}),\ }\href
  {https://doi.org/10.3847/1538-4365/aab5b0} {\bibfield  {journal} {\bibinfo
  {journal} {Astrophys. J. Suppl.}\ }\textbf {\bibinfo {volume} {235}},\
  \bibinfo {pages} {37} (\bibinfo {year} {2018})},\ \Eprint
  {https://arxiv.org/abs/1801.01837} {arXiv:1801.01837 [astro-ph.HE]}
  \BibitemShut {NoStop}%
\bibitem [{\citenamefont {Antoniadis}\ \emph {et~al.}(2013)\citenamefont
  {Antoniadis} \emph {et~al.}}]{Antoniadis:2013pzd}%
  \BibitemOpen
  \bibfield  {author} {\bibinfo {author} {\bibfnamefont {J.}~\bibnamefont
  {Antoniadis}} \emph {et~al.},\ }\href
  {https://doi.org/10.1126/science.1233232} {\bibfield  {journal} {\bibinfo
  {journal} {Science}\ }\textbf {\bibinfo {volume} {340}},\ \bibinfo {pages}
  {6131} (\bibinfo {year} {2013})},\ \Eprint {https://arxiv.org/abs/1304.6875}
  {arXiv:1304.6875 [astro-ph.HE]} \BibitemShut {NoStop}%
\bibitem [{\citenamefont {Cromartie}\ \emph {et~al.}(2019)\citenamefont
  {Cromartie} \emph {et~al.}}]{Cromartie:2019kug}%
  \BibitemOpen
  \bibfield  {author} {\bibinfo {author} {\bibfnamefont {H.~T.}\ \bibnamefont
  {Cromartie}} \emph {et~al.},\ }\href
  {https://doi.org/10.1038/s41550-019-0880-2} {\bibfield  {journal} {\bibinfo
  {journal} {Nature Astron.}\ }\textbf {\bibinfo {volume} {4}},\ \bibinfo
  {pages} {72} (\bibinfo {year} {2019})},\ \Eprint
  {https://arxiv.org/abs/1904.06759} {arXiv:1904.06759 [astro-ph.HE]}
  \BibitemShut {NoStop}%
\bibitem [{\citenamefont {Watts}\ \emph {et~al.}(2016)\citenamefont {Watts}
  \emph {et~al.}}]{Watts:2016uzu}%
  \BibitemOpen
  \bibfield  {author} {\bibinfo {author} {\bibfnamefont {A.~L.}\ \bibnamefont
  {Watts}} \emph {et~al.},\ }\href
  {https://doi.org/10.1103/RevModPhys.88.021001} {\bibfield  {journal}
  {\bibinfo  {journal} {Rev. Mod. Phys.}\ }\textbf {\bibinfo {volume} {88}},\
  \bibinfo {pages} {021001} (\bibinfo {year} {2016})},\ \Eprint
  {https://arxiv.org/abs/1602.01081} {arXiv:1602.01081 [astro-ph.HE]}
  \BibitemShut {NoStop}%
\bibitem [{\citenamefont {Miller}\ \emph {et~al.}(2019)\citenamefont {Miller}
  \emph {et~al.}}]{Miller:2019cac}%
  \BibitemOpen
  \bibfield  {author} {\bibinfo {author} {\bibfnamefont {M.}~\bibnamefont
  {Miller}} \emph {et~al.},\ }\href {https://doi.org/10.3847/2041-8213/ab50c5}
  {\bibfield  {journal} {\bibinfo  {journal} {Astrophys. J. Lett.}\ }\textbf
  {\bibinfo {volume} {887}},\ \bibinfo {pages} {L24} (\bibinfo {year}
  {2019})},\ \Eprint {https://arxiv.org/abs/1912.05705} {arXiv:1912.05705
  [astro-ph.HE]} \BibitemShut {NoStop}%
\bibitem [{\citenamefont {Riley}\ \emph {et~al.}(2019)\citenamefont {Riley}
  \emph {et~al.}}]{Riley:2019yda}%
  \BibitemOpen
  \bibfield  {author} {\bibinfo {author} {\bibfnamefont {T.~E.}\ \bibnamefont
  {Riley}} \emph {et~al.},\ }\href {https://doi.org/10.3847/2041-8213/ab481c}
  {\bibfield  {journal} {\bibinfo  {journal} {Astrophys. J. Lett.}\ }\textbf
  {\bibinfo {volume} {887}},\ \bibinfo {pages} {L21} (\bibinfo {year}
  {2019})},\ \Eprint {https://arxiv.org/abs/1912.05702} {arXiv:1912.05702
  [astro-ph.HE]} \BibitemShut {NoStop}%
\bibitem [{\citenamefont {Abbott}\ \emph {et~al.}(2019)\citenamefont {Abbott}
  \emph {et~al.}}]{Abbott:2018wiz}%
  \BibitemOpen
  \bibfield  {author} {\bibinfo {author} {\bibfnamefont {B.}~\bibnamefont
  {Abbott}} \emph {et~al.} (\bibinfo {collaboration} {LIGO Scientific,
  Virgo}),\ }\href {https://doi.org/10.1103/PhysRevX.9.011001} {\bibfield
  {journal} {\bibinfo  {journal} {Phys. Rev. X}\ }\textbf {\bibinfo {volume}
  {9}},\ \bibinfo {pages} {011001} (\bibinfo {year} {2019})},\ \Eprint
  {https://arxiv.org/abs/1805.11579} {arXiv:1805.11579 [gr-qc]} \BibitemShut
  {NoStop}%
\bibitem [{\citenamefont {Annala}\ \emph {et~al.}(2018)\citenamefont {Annala},
  \citenamefont {Gorda}, \citenamefont {Kurkela},\ and\ \citenamefont
  {Vuorinen}}]{Annala:2017llu}%
  \BibitemOpen
  \bibfield  {author} {\bibinfo {author} {\bibfnamefont {E.}~\bibnamefont
  {Annala}}, \bibinfo {author} {\bibfnamefont {T.}~\bibnamefont {Gorda}},
  \bibinfo {author} {\bibfnamefont {A.}~\bibnamefont {Kurkela}},\ and\ \bibinfo
  {author} {\bibfnamefont {A.}~\bibnamefont {Vuorinen}},\ }\href
  {https://doi.org/10.1103/PhysRevLett.120.172703} {\bibfield  {journal}
  {\bibinfo  {journal} {Phys. Rev. Lett.}\ }\textbf {\bibinfo {volume} {120}},\
  \bibinfo {pages} {172703} (\bibinfo {year} {2018})},\ \Eprint
  {https://arxiv.org/abs/1711.02644} {arXiv:1711.02644 [astro-ph.HE]}
  \BibitemShut {NoStop}%
\bibitem [{\citenamefont {De}\ \emph {et~al.}(2018)\citenamefont {De},
  \citenamefont {Finstad}, \citenamefont {Lattimer}, \citenamefont {Brown},
  \citenamefont {Berger},\ and\ \citenamefont {Biwer}}]{De:2018uhw}%
  \BibitemOpen
  \bibfield  {author} {\bibinfo {author} {\bibfnamefont {S.}~\bibnamefont
  {De}}, \bibinfo {author} {\bibfnamefont {D.}~\bibnamefont {Finstad}},
  \bibinfo {author} {\bibfnamefont {J.~M.}\ \bibnamefont {Lattimer}}, \bibinfo
  {author} {\bibfnamefont {D.~A.}\ \bibnamefont {Brown}}, \bibinfo {author}
  {\bibfnamefont {E.}~\bibnamefont {Berger}},\ and\ \bibinfo {author}
  {\bibfnamefont {C.~M.}\ \bibnamefont {Biwer}},\ }\href
  {https://doi.org/10.1103/PhysRevLett.121.091102} {\bibfield  {journal}
  {\bibinfo  {journal} {Phys. Rev. Lett.}\ }\textbf {\bibinfo {volume} {121}},\
  \bibinfo {pages} {091102} (\bibinfo {year} {2018})},\ \bibinfo {note}
  {[Erratum: Phys.Rev.Lett. 121, 259902 (2018)]},\ \Eprint
  {https://arxiv.org/abs/1804.08583} {arXiv:1804.08583 [astro-ph.HE]}
  \BibitemShut {NoStop}%
\bibitem [{\citenamefont {Bedaque}\ and\ \citenamefont
  {Steiner}(2015)}]{Bedaque:2014sqa}%
  \BibitemOpen
  \bibfield  {author} {\bibinfo {author} {\bibfnamefont {P.}~\bibnamefont
  {Bedaque}}\ and\ \bibinfo {author} {\bibfnamefont {A.~W.}\ \bibnamefont
  {Steiner}},\ }\href {https://doi.org/10.1103/PhysRevLett.114.031103}
  {\bibfield  {journal} {\bibinfo  {journal} {Phys. Rev. Lett.}\ }\textbf
  {\bibinfo {volume} {114}},\ \bibinfo {pages} {031103} (\bibinfo {year}
  {2015})},\ \Eprint {https://arxiv.org/abs/1408.5116} {arXiv:1408.5116
  [nucl-th]} \BibitemShut {NoStop}%
\bibitem [{\citenamefont {Tews}\ \emph {et~al.}(2018)\citenamefont {Tews},
  \citenamefont {Carlson}, \citenamefont {Gandolfi},\ and\ \citenamefont
  {Reddy}}]{Tews:2018kmu}%
  \BibitemOpen
  \bibfield  {author} {\bibinfo {author} {\bibfnamefont {I.}~\bibnamefont
  {Tews}}, \bibinfo {author} {\bibfnamefont {J.}~\bibnamefont {Carlson}},
  \bibinfo {author} {\bibfnamefont {S.}~\bibnamefont {Gandolfi}},\ and\
  \bibinfo {author} {\bibfnamefont {S.}~\bibnamefont {Reddy}},\ }\href
  {https://doi.org/10.3847/1538-4357/aac267} {\bibfield  {journal} {\bibinfo
  {journal} {Astrophys. J.}\ }\textbf {\bibinfo {volume} {860}},\ \bibinfo
  {pages} {149} (\bibinfo {year} {2018})},\ \Eprint
  {https://arxiv.org/abs/1801.01923} {arXiv:1801.01923 [nucl-th]} \BibitemShut
  {NoStop}%
\bibitem [{\citenamefont {Drischler}\ \emph {et~al.}(2021)\citenamefont
  {Drischler}, \citenamefont {Han}, \citenamefont {Lattimer}, \citenamefont
  {Prakash}, \citenamefont {Reddy},\ and\ \citenamefont
  {Zhao}}]{Drischler:2020fvz}%
  \BibitemOpen
  \bibfield  {author} {\bibinfo {author} {\bibfnamefont {C.}~\bibnamefont
  {Drischler}}, \bibinfo {author} {\bibfnamefont {S.}~\bibnamefont {Han}},
  \bibinfo {author} {\bibfnamefont {J.~M.}\ \bibnamefont {Lattimer}}, \bibinfo
  {author} {\bibfnamefont {M.}~\bibnamefont {Prakash}}, \bibinfo {author}
  {\bibfnamefont {S.}~\bibnamefont {Reddy}},\ and\ \bibinfo {author}
  {\bibfnamefont {T.}~\bibnamefont {Zhao}},\ }\href
  {https://doi.org/10.1103/PhysRevC.103.045808} {\bibfield  {journal} {\bibinfo
   {journal} {Phys. Rev. C}\ }\textbf {\bibinfo {volume} {103}},\ \bibinfo
  {pages} {045808} (\bibinfo {year} {2021})},\ \Eprint
  {https://arxiv.org/abs/2009.06441} {arXiv:2009.06441 [nucl-th]} \BibitemShut
  {NoStop}%
\bibitem [{\citenamefont {Itoh}(1970)}]{Itoh:1970uw}%
  \BibitemOpen
  \bibfield  {author} {\bibinfo {author} {\bibfnamefont {N.}~\bibnamefont
  {Itoh}},\ }\href {https://doi.org/10.1143/PTP.44.291} {\bibfield  {journal}
  {\bibinfo  {journal} {Prog. Theor. Phys.}\ }\textbf {\bibinfo {volume}
  {44}},\ \bibinfo {pages} {291} (\bibinfo {year} {1970})}\BibitemShut
  {NoStop}%
\bibitem [{\citenamefont {Collins}\ and\ \citenamefont
  {Perry}(1975)}]{Collins:1974ky}%
  \BibitemOpen
  \bibfield  {author} {\bibinfo {author} {\bibfnamefont {J.~C.}\ \bibnamefont
  {Collins}}\ and\ \bibinfo {author} {\bibfnamefont {M.}~\bibnamefont
  {Perry}},\ }\href {https://doi.org/10.1103/PhysRevLett.34.1353} {\bibfield
  {journal} {\bibinfo  {journal} {Phys. Rev. Lett.}\ }\textbf {\bibinfo
  {volume} {34}},\ \bibinfo {pages} {1353} (\bibinfo {year}
  {1975})}\BibitemShut {NoStop}%
\bibitem [{\citenamefont {Kojo}(2021)}]{Kojo:2020krb}%
  \BibitemOpen
  \bibfield  {author} {\bibinfo {author} {\bibfnamefont {T.}~\bibnamefont
  {Kojo}},\ }\href {https://doi.org/10.1007/s43673-021-00011-6} {\bibfield
  {journal} {\bibinfo  {journal} {AAPPS Bull.}\ }\textbf {\bibinfo {volume}
  {31}},\ \bibinfo {pages} {11} (\bibinfo {year} {2021})},\ \Eprint
  {https://arxiv.org/abs/2011.10940} {arXiv:2011.10940 [nucl-th]} \BibitemShut
  {NoStop}%
\bibitem [{\citenamefont {Alford}\ \emph {et~al.}(2013)\citenamefont {Alford},
  \citenamefont {Han},\ and\ \citenamefont {Prakash}}]{Alford:2013aca}%
  \BibitemOpen
  \bibfield  {author} {\bibinfo {author} {\bibfnamefont {M.~G.}\ \bibnamefont
  {Alford}}, \bibinfo {author} {\bibfnamefont {S.}~\bibnamefont {Han}},\ and\
  \bibinfo {author} {\bibfnamefont {M.}~\bibnamefont {Prakash}},\ }\href
  {https://doi.org/10.1103/PhysRevD.88.083013} {\bibfield  {journal} {\bibinfo
  {journal} {Phys. Rev. D}\ }\textbf {\bibinfo {volume} {88}},\ \bibinfo
  {pages} {083013} (\bibinfo {year} {2013})},\ \Eprint
  {https://arxiv.org/abs/1302.4732} {arXiv:1302.4732 [astro-ph.SR]}
  \BibitemShut {NoStop}%
\bibitem [{\citenamefont {Benic}\ \emph {et~al.}(2015)\citenamefont {Benic},
  \citenamefont {Blaschke}, \citenamefont {Alvarez-Castillo}, \citenamefont
  {Fischer},\ and\ \citenamefont {Typel}}]{Benic:2014jia}%
  \BibitemOpen
  \bibfield  {author} {\bibinfo {author} {\bibfnamefont {S.}~\bibnamefont
  {Benic}}, \bibinfo {author} {\bibfnamefont {D.}~\bibnamefont {Blaschke}},
  \bibinfo {author} {\bibfnamefont {D.~E.}\ \bibnamefont {Alvarez-Castillo}},
  \bibinfo {author} {\bibfnamefont {T.}~\bibnamefont {Fischer}},\ and\ \bibinfo
  {author} {\bibfnamefont {S.}~\bibnamefont {Typel}},\ }\href
  {https://doi.org/10.1051/0004-6361/201425318} {\bibfield  {journal} {\bibinfo
   {journal} {Astron. Astrophys.}\ }\textbf {\bibinfo {volume} {577}},\
  \bibinfo {pages} {A40} (\bibinfo {year} {2015})},\ \Eprint
  {https://arxiv.org/abs/1411.2856} {arXiv:1411.2856 [astro-ph.HE]}
  \BibitemShut {NoStop}%
\bibitem [{\citenamefont {Alvarez-Castillo}\ \emph {et~al.}(2016)\citenamefont
  {Alvarez-Castillo}, \citenamefont {Ayriyan}, \citenamefont {Benic},
  \citenamefont {Blaschke}, \citenamefont {Grigorian},\ and\ \citenamefont
  {Typel}}]{Alvarez-Castillo:2016oln}%
  \BibitemOpen
  \bibfield  {author} {\bibinfo {author} {\bibfnamefont {D.}~\bibnamefont
  {Alvarez-Castillo}}, \bibinfo {author} {\bibfnamefont {A.}~\bibnamefont
  {Ayriyan}}, \bibinfo {author} {\bibfnamefont {S.}~\bibnamefont {Benic}},
  \bibinfo {author} {\bibfnamefont {D.}~\bibnamefont {Blaschke}}, \bibinfo
  {author} {\bibfnamefont {H.}~\bibnamefont {Grigorian}},\ and\ \bibinfo
  {author} {\bibfnamefont {S.}~\bibnamefont {Typel}},\ }\href
  {https://doi.org/10.1140/epja/i2016-16069-2} {\bibfield  {journal} {\bibinfo
  {journal} {Eur. Phys. J. A}\ }\textbf {\bibinfo {volume} {52}},\ \bibinfo
  {pages} {69} (\bibinfo {year} {2016})},\ \Eprint
  {https://arxiv.org/abs/1603.03457} {arXiv:1603.03457 [nucl-th]} \BibitemShut
  {NoStop}%
\bibitem [{\citenamefont {Freedman}\ and\ \citenamefont
  {McLerran}(1978)}]{Freedman:1977gz}%
  \BibitemOpen
  \bibfield  {author} {\bibinfo {author} {\bibfnamefont {B.}~\bibnamefont
  {Freedman}}\ and\ \bibinfo {author} {\bibfnamefont {L.~D.}\ \bibnamefont
  {McLerran}},\ }\href {https://doi.org/10.1103/PhysRevD.17.1109} {\bibfield
  {journal} {\bibinfo  {journal} {Phys. Rev. D}\ }\textbf {\bibinfo {volume}
  {17}},\ \bibinfo {pages} {1109} (\bibinfo {year} {1978})}\BibitemShut
  {NoStop}%
\bibitem [{\citenamefont {Witten}(1984)}]{Witten:1984rs}%
  \BibitemOpen
  \bibfield  {author} {\bibinfo {author} {\bibfnamefont {E.}~\bibnamefont
  {Witten}},\ }\href {https://doi.org/10.1103/PhysRevD.30.272} {\bibfield
  {journal} {\bibinfo  {journal} {Phys. Rev. D}\ }\textbf {\bibinfo {volume}
  {30}},\ \bibinfo {pages} {272} (\bibinfo {year} {1984})}\BibitemShut
  {NoStop}%
\bibitem [{\citenamefont {Annala}\ \emph {et~al.}(2020)\citenamefont {Annala},
  \citenamefont {Gorda}, \citenamefont {Kurkela}, \citenamefont {N\"attil\"a},\
  and\ \citenamefont {Vuorinen}}]{Annala:2019puf}%
  \BibitemOpen
  \bibfield  {author} {\bibinfo {author} {\bibfnamefont {E.}~\bibnamefont
  {Annala}}, \bibinfo {author} {\bibfnamefont {T.}~\bibnamefont {Gorda}},
  \bibinfo {author} {\bibfnamefont {A.}~\bibnamefont {Kurkela}}, \bibinfo
  {author} {\bibfnamefont {J.}~\bibnamefont {N\"attil\"a}},\ and\ \bibinfo
  {author} {\bibfnamefont {A.}~\bibnamefont {Vuorinen}},\ }\bibfield  {journal}
  {\bibinfo  {journal} {Nature Phys.}\ }\href
  {https://doi.org/10.1038/s41567-020-0914-9} {10.1038/s41567-020-0914-9}
  (\bibinfo {year} {2020}),\ \Eprint {https://arxiv.org/abs/1903.09121}
  {arXiv:1903.09121 [astro-ph.HE]} \BibitemShut {NoStop}%
\bibitem [{\citenamefont {Masuda}\ \emph
  {et~al.}(2013{\natexlab{a}})\citenamefont {Masuda}, \citenamefont {Hatsuda},\
  and\ \citenamefont {Takatsuka}}]{Masuda:2012kf}%
  \BibitemOpen
  \bibfield  {author} {\bibinfo {author} {\bibfnamefont {K.}~\bibnamefont
  {Masuda}}, \bibinfo {author} {\bibfnamefont {T.}~\bibnamefont {Hatsuda}},\
  and\ \bibinfo {author} {\bibfnamefont {T.}~\bibnamefont {Takatsuka}},\ }\href
  {https://doi.org/10.1088/0004-637X/764/1/12} {\bibfield  {journal} {\bibinfo
  {journal} {Astrophys. J.}\ }\textbf {\bibinfo {volume} {764}},\ \bibinfo
  {pages} {12} (\bibinfo {year} {2013}{\natexlab{a}})},\ \Eprint
  {https://arxiv.org/abs/1205.3621} {arXiv:1205.3621 [nucl-th]} \BibitemShut
  {NoStop}%
\bibitem [{\citenamefont {Kojo}\ \emph {et~al.}(2015)\citenamefont {Kojo},
  \citenamefont {Powell}, \citenamefont {Song},\ and\ \citenamefont
  {Baym}}]{Kojo:2014rca}%
  \BibitemOpen
  \bibfield  {author} {\bibinfo {author} {\bibfnamefont {T.}~\bibnamefont
  {Kojo}}, \bibinfo {author} {\bibfnamefont {P.~D.}\ \bibnamefont {Powell}},
  \bibinfo {author} {\bibfnamefont {Y.}~\bibnamefont {Song}},\ and\ \bibinfo
  {author} {\bibfnamefont {G.}~\bibnamefont {Baym}},\ }\href
  {https://doi.org/10.1103/PhysRevD.91.045003} {\bibfield  {journal} {\bibinfo
  {journal} {Phys. Rev. D}\ }\textbf {\bibinfo {volume} {91}},\ \bibinfo
  {pages} {045003} (\bibinfo {year} {2015})},\ \Eprint
  {https://arxiv.org/abs/1412.1108} {arXiv:1412.1108 [hep-ph]} \BibitemShut
  {NoStop}%
\bibitem [{\citenamefont {McLerran}\ and\ \citenamefont
  {Reddy}(2019)}]{McLerran:2018hbz}%
  \BibitemOpen
  \bibfield  {author} {\bibinfo {author} {\bibfnamefont {L.}~\bibnamefont
  {McLerran}}\ and\ \bibinfo {author} {\bibfnamefont {S.}~\bibnamefont
  {Reddy}},\ }\href {https://doi.org/10.1103/PhysRevLett.122.122701} {\bibfield
   {journal} {\bibinfo  {journal} {Phys. Rev. Lett.}\ }\textbf {\bibinfo
  {volume} {122}},\ \bibinfo {pages} {122701} (\bibinfo {year} {2019})},\
  \Eprint {https://arxiv.org/abs/1811.12503} {arXiv:1811.12503 [nucl-th]}
  \BibitemShut {NoStop}%
\bibitem [{\citenamefont {Baym}\ and\ \citenamefont
  {Chin}(1976)}]{Baym:1976yu}%
  \BibitemOpen
  \bibfield  {author} {\bibinfo {author} {\bibfnamefont {G.}~\bibnamefont
  {Baym}}\ and\ \bibinfo {author} {\bibfnamefont {S.}~\bibnamefont {Chin}},\
  }\href {https://doi.org/10.1016/0370-2693(76)90517-7} {\bibfield  {journal}
  {\bibinfo  {journal} {Phys. Lett. B}\ }\textbf {\bibinfo {volume} {62}},\
  \bibinfo {pages} {241} (\bibinfo {year} {1976})}\BibitemShut {NoStop}%
\bibitem [{\citenamefont {Schäfer}\ and\ \citenamefont
  {Wilczek}(1999)}]{Schafer:1998ef}%
  \BibitemOpen
  \bibfield  {author} {\bibinfo {author} {\bibfnamefont {T.}~\bibnamefont
  {Schäfer}}\ and\ \bibinfo {author} {\bibfnamefont {F.}~\bibnamefont
  {Wilczek}},\ }\href {https://doi.org/10.1103/PhysRevLett.82.3956} {\bibfield
  {journal} {\bibinfo  {journal} {Phys. Rev. Lett.}\ }\textbf {\bibinfo
  {volume} {82}},\ \bibinfo {pages} {3956} (\bibinfo {year} {1999})},\ \Eprint
  {https://arxiv.org/abs/hep-ph/9811473} {arXiv:hep-ph/9811473} \BibitemShut
  {NoStop}%
\bibitem [{\citenamefont {Hatsuda}\ \emph {et~al.}(2006)\citenamefont
  {Hatsuda}, \citenamefont {Tachibana}, \citenamefont {Yamamoto},\ and\
  \citenamefont {Baym}}]{Hatsuda:2006ps}%
  \BibitemOpen
  \bibfield  {author} {\bibinfo {author} {\bibfnamefont {T.}~\bibnamefont
  {Hatsuda}}, \bibinfo {author} {\bibfnamefont {M.}~\bibnamefont {Tachibana}},
  \bibinfo {author} {\bibfnamefont {N.}~\bibnamefont {Yamamoto}},\ and\
  \bibinfo {author} {\bibfnamefont {G.}~\bibnamefont {Baym}},\ }\href
  {https://doi.org/10.1103/PhysRevLett.97.122001} {\bibfield  {journal}
  {\bibinfo  {journal} {Phys. Rev. Lett.}\ }\textbf {\bibinfo {volume} {97}},\
  \bibinfo {pages} {122001} (\bibinfo {year} {2006})},\ \Eprint
  {https://arxiv.org/abs/hep-ph/0605018} {arXiv:hep-ph/0605018} \BibitemShut
  {NoStop}%
\bibitem [{\citenamefont {Yamamoto}\ \emph {et~al.}(2007)\citenamefont
  {Yamamoto}, \citenamefont {Tachibana}, \citenamefont {Hatsuda},\ and\
  \citenamefont {Baym}}]{Yamamoto:2007ah}%
  \BibitemOpen
  \bibfield  {author} {\bibinfo {author} {\bibfnamefont {N.}~\bibnamefont
  {Yamamoto}}, \bibinfo {author} {\bibfnamefont {M.}~\bibnamefont {Tachibana}},
  \bibinfo {author} {\bibfnamefont {T.}~\bibnamefont {Hatsuda}},\ and\ \bibinfo
  {author} {\bibfnamefont {G.}~\bibnamefont {Baym}},\ }\href
  {https://doi.org/10.1103/PhysRevD.76.074001} {\bibfield  {journal} {\bibinfo
  {journal} {Phys. Rev. D}\ }\textbf {\bibinfo {volume} {76}},\ \bibinfo
  {pages} {074001} (\bibinfo {year} {2007})},\ \Eprint
  {https://arxiv.org/abs/0704.2654} {arXiv:0704.2654 [hep-ph]} \BibitemShut
  {NoStop}%
\bibitem [{\citenamefont {Masuda}\ \emph
  {et~al.}(2013{\natexlab{b}})\citenamefont {Masuda}, \citenamefont {Hatsuda},\
  and\ \citenamefont {Takatsuka}}]{Masuda:2012ed}%
  \BibitemOpen
  \bibfield  {author} {\bibinfo {author} {\bibfnamefont {K.}~\bibnamefont
  {Masuda}}, \bibinfo {author} {\bibfnamefont {T.}~\bibnamefont {Hatsuda}},\
  and\ \bibinfo {author} {\bibfnamefont {T.}~\bibnamefont {Takatsuka}},\ }\href
  {https://doi.org/10.1093/ptep/ptt045} {\bibfield  {journal} {\bibinfo
  {journal} {PTEP}\ }\textbf {\bibinfo {volume} {2013}},\ \bibinfo {pages}
  {073D01} (\bibinfo {year} {2013}{\natexlab{b}})},\ \Eprint
  {https://arxiv.org/abs/1212.6803} {arXiv:1212.6803 [nucl-th]} \BibitemShut
  {NoStop}%
\bibitem [{\citenamefont {Masuda}\ \emph
  {et~al.}(2016{\natexlab{a}})\citenamefont {Masuda}, \citenamefont {Hatsuda},\
  and\ \citenamefont {Takatsuka}}]{Masuda:2015kha}%
  \BibitemOpen
  \bibfield  {author} {\bibinfo {author} {\bibfnamefont {K.}~\bibnamefont
  {Masuda}}, \bibinfo {author} {\bibfnamefont {T.}~\bibnamefont {Hatsuda}},\
  and\ \bibinfo {author} {\bibfnamefont {T.}~\bibnamefont {Takatsuka}},\ }\href
  {https://doi.org/10.1140/epja/i2016-16065-6} {\bibfield  {journal} {\bibinfo
  {journal} {Eur. Phys. J. A}\ }\textbf {\bibinfo {volume} {52}},\ \bibinfo
  {pages} {65} (\bibinfo {year} {2016}{\natexlab{a}})},\ \Eprint
  {https://arxiv.org/abs/1508.04861} {arXiv:1508.04861 [nucl-th]} \BibitemShut
  {NoStop}%
\bibitem [{\citenamefont {Baym}\ \emph {et~al.}(2018)\citenamefont {Baym},
  \citenamefont {Hatsuda}, \citenamefont {Kojo}, \citenamefont {Powell},
  \citenamefont {Song},\ and\ \citenamefont {Takatsuka}}]{Baym:2017whm}%
  \BibitemOpen
  \bibfield  {author} {\bibinfo {author} {\bibfnamefont {G.}~\bibnamefont
  {Baym}}, \bibinfo {author} {\bibfnamefont {T.}~\bibnamefont {Hatsuda}},
  \bibinfo {author} {\bibfnamefont {T.}~\bibnamefont {Kojo}}, \bibinfo {author}
  {\bibfnamefont {P.~D.}\ \bibnamefont {Powell}}, \bibinfo {author}
  {\bibfnamefont {Y.}~\bibnamefont {Song}},\ and\ \bibinfo {author}
  {\bibfnamefont {T.}~\bibnamefont {Takatsuka}},\ }\href
  {https://doi.org/10.1088/1361-6633/aaae14} {\bibfield  {journal} {\bibinfo
  {journal} {Rept. Prog. Phys.}\ }\textbf {\bibinfo {volume} {81}},\ \bibinfo
  {pages} {056902} (\bibinfo {year} {2018})},\ \Eprint
  {https://arxiv.org/abs/1707.04966} {arXiv:1707.04966 [astro-ph.HE]}
  \BibitemShut {NoStop}%
\bibitem [{\citenamefont {Baym}\ \emph {et~al.}(2019)\citenamefont {Baym},
  \citenamefont {Furusawa}, \citenamefont {Hatsuda}, \citenamefont {Kojo},\
  and\ \citenamefont {Togashi}}]{Baym:2019iky}%
  \BibitemOpen
  \bibfield  {author} {\bibinfo {author} {\bibfnamefont {G.}~\bibnamefont
  {Baym}}, \bibinfo {author} {\bibfnamefont {S.}~\bibnamefont {Furusawa}},
  \bibinfo {author} {\bibfnamefont {T.}~\bibnamefont {Hatsuda}}, \bibinfo
  {author} {\bibfnamefont {T.}~\bibnamefont {Kojo}},\ and\ \bibinfo {author}
  {\bibfnamefont {H.}~\bibnamefont {Togashi}},\ }\href
  {https://doi.org/10.3847/1538-4357/ab441e} {\bibfield  {journal} {\bibinfo
  {journal} {Astrophys. J.}\ }\textbf {\bibinfo {volume} {885}},\ \bibinfo
  {pages} {42} (\bibinfo {year} {2019})},\ \Eprint
  {https://arxiv.org/abs/1903.08963} {arXiv:1903.08963 [astro-ph.HE]}
  \BibitemShut {NoStop}%
\bibitem [{\citenamefont {Alford}\ \emph {et~al.}(2008)\citenamefont {Alford},
  \citenamefont {Schmitt}, \citenamefont {Rajagopal},\ and\ \citenamefont
  {Schäfer}}]{Alford:2007xm}%
  \BibitemOpen
  \bibfield  {author} {\bibinfo {author} {\bibfnamefont {M.~G.}\ \bibnamefont
  {Alford}}, \bibinfo {author} {\bibfnamefont {A.}~\bibnamefont {Schmitt}},
  \bibinfo {author} {\bibfnamefont {K.}~\bibnamefont {Rajagopal}},\ and\
  \bibinfo {author} {\bibfnamefont {T.}~\bibnamefont {Schäfer}},\ }\href
  {https://doi.org/10.1103/RevModPhys.80.1455} {\bibfield  {journal} {\bibinfo
  {journal} {Rev. Mod. Phys.}\ }\textbf {\bibinfo {volume} {80}},\ \bibinfo
  {pages} {1455} (\bibinfo {year} {2008})},\ \Eprint
  {https://arxiv.org/abs/0709.4635} {arXiv:0709.4635 [hep-ph]} \BibitemShut
  {NoStop}%
\bibitem [{\citenamefont {Akmal}\ \emph {et~al.}(1998)\citenamefont {Akmal},
  \citenamefont {Pandharipande},\ and\ \citenamefont
  {Ravenhall}}]{Akmal:1998cf}%
  \BibitemOpen
  \bibfield  {author} {\bibinfo {author} {\bibfnamefont {A.}~\bibnamefont
  {Akmal}}, \bibinfo {author} {\bibfnamefont {V.}~\bibnamefont
  {Pandharipande}},\ and\ \bibinfo {author} {\bibfnamefont {D.}~\bibnamefont
  {Ravenhall}},\ }\href {https://doi.org/10.1103/PhysRevC.58.1804} {\bibfield
  {journal} {\bibinfo  {journal} {Phys. Rev. C}\ }\textbf {\bibinfo {volume}
  {58}},\ \bibinfo {pages} {1804} (\bibinfo {year} {1998})},\ \Eprint
  {https://arxiv.org/abs/nucl-th/9804027} {arXiv:nucl-th/9804027} \BibitemShut
  {NoStop}%
\bibitem [{\citenamefont {Togashi}\ \emph {et~al.}(2017)\citenamefont
  {Togashi}, \citenamefont {Nakazato}, \citenamefont {Takehara}, \citenamefont
  {Yamamuro}, \citenamefont {Suzuki},\ and\ \citenamefont
  {Takano}}]{Togashi:2017mjp}%
  \BibitemOpen
  \bibfield  {author} {\bibinfo {author} {\bibfnamefont {H.}~\bibnamefont
  {Togashi}}, \bibinfo {author} {\bibfnamefont {K.}~\bibnamefont {Nakazato}},
  \bibinfo {author} {\bibfnamefont {Y.}~\bibnamefont {Takehara}}, \bibinfo
  {author} {\bibfnamefont {S.}~\bibnamefont {Yamamuro}}, \bibinfo {author}
  {\bibfnamefont {H.}~\bibnamefont {Suzuki}},\ and\ \bibinfo {author}
  {\bibfnamefont {M.}~\bibnamefont {Takano}},\ }\href
  {https://doi.org/10.1016/j.nuclphysa.2017.02.010} {\bibfield  {journal}
  {\bibinfo  {journal} {Nucl. Phys. A}\ }\textbf {\bibinfo {volume} {961}},\
  \bibinfo {pages} {78} (\bibinfo {year} {2017})},\ \Eprint
  {https://arxiv.org/abs/1702.05324} {arXiv:1702.05324 [nucl-th]} \BibitemShut
  {NoStop}%
\bibitem [{\citenamefont {Oertel}\ \emph {et~al.}(2017)\citenamefont {Oertel},
  \citenamefont {Hempel}, \citenamefont {Kl\"ahn},\ and\ \citenamefont
  {Typel}}]{Oertel:2016bki}%
  \BibitemOpen
  \bibfield  {author} {\bibinfo {author} {\bibfnamefont {M.}~\bibnamefont
  {Oertel}}, \bibinfo {author} {\bibfnamefont {M.}~\bibnamefont {Hempel}},
  \bibinfo {author} {\bibfnamefont {T.}~\bibnamefont {Kl\"ahn}},\ and\ \bibinfo
  {author} {\bibfnamefont {S.}~\bibnamefont {Typel}},\ }\href
  {https://doi.org/10.1103/RevModPhys.89.015007} {\bibfield  {journal}
  {\bibinfo  {journal} {Rev. Mod. Phys.}\ }\textbf {\bibinfo {volume} {89}},\
  \bibinfo {pages} {015007} (\bibinfo {year} {2017})},\ \Eprint
  {https://arxiv.org/abs/1610.03361} {arXiv:1610.03361 [astro-ph.HE]}
  \BibitemShut {NoStop}%
\bibitem [{\citenamefont {Schmitt}\ and\ \citenamefont
  {Shternin}(2018)}]{Schmitt:2017efp}%
  \BibitemOpen
  \bibfield  {author} {\bibinfo {author} {\bibfnamefont {A.}~\bibnamefont
  {Schmitt}}\ and\ \bibinfo {author} {\bibfnamefont {P.}~\bibnamefont
  {Shternin}},\ }\href {https://doi.org/10.1007/978-3-319-97616-7_9} {\bibfield
   {journal} {\bibinfo  {journal} {Astrophys. Space Sci. Libr.}\ }\textbf
  {\bibinfo {volume} {457}},\ \bibinfo {pages} {455} (\bibinfo {year}
  {2018})},\ \Eprint {https://arxiv.org/abs/1711.06520} {arXiv:1711.06520
  [astro-ph.HE]} \BibitemShut {NoStop}%
\bibitem [{\citenamefont {Alford}\ \emph {et~al.}(2018)\citenamefont {Alford},
  \citenamefont {Bovard}, \citenamefont {Hanauske}, \citenamefont {Rezzolla},\
  and\ \citenamefont {Schwenzer}}]{Alford:2017rxf}%
  \BibitemOpen
  \bibfield  {author} {\bibinfo {author} {\bibfnamefont {M.~G.}\ \bibnamefont
  {Alford}}, \bibinfo {author} {\bibfnamefont {L.}~\bibnamefont {Bovard}},
  \bibinfo {author} {\bibfnamefont {M.}~\bibnamefont {Hanauske}}, \bibinfo
  {author} {\bibfnamefont {L.}~\bibnamefont {Rezzolla}},\ and\ \bibinfo
  {author} {\bibfnamefont {K.}~\bibnamefont {Schwenzer}},\ }\href
  {https://doi.org/10.1103/PhysRevLett.120.041101} {\bibfield  {journal}
  {\bibinfo  {journal} {Phys. Rev. Lett.}\ }\textbf {\bibinfo {volume} {120}},\
  \bibinfo {pages} {041101} (\bibinfo {year} {2018})},\ \Eprint
  {https://arxiv.org/abs/1707.09475} {arXiv:1707.09475 [gr-qc]} \BibitemShut
  {NoStop}%
\bibitem [{\citenamefont {Most}\ \emph {et~al.}(2019)\citenamefont {Most},
  \citenamefont {Papenfort}, \citenamefont {Dexheimer}, \citenamefont
  {Hanauske}, \citenamefont {Schramm}, \citenamefont {St\"ocker},\ and\
  \citenamefont {Rezzolla}}]{Most:2018eaw}%
  \BibitemOpen
  \bibfield  {author} {\bibinfo {author} {\bibfnamefont {E.~R.}\ \bibnamefont
  {Most}}, \bibinfo {author} {\bibfnamefont {L.~J.}\ \bibnamefont {Papenfort}},
  \bibinfo {author} {\bibfnamefont {V.}~\bibnamefont {Dexheimer}}, \bibinfo
  {author} {\bibfnamefont {M.}~\bibnamefont {Hanauske}}, \bibinfo {author}
  {\bibfnamefont {S.}~\bibnamefont {Schramm}}, \bibinfo {author} {\bibfnamefont
  {H.}~\bibnamefont {St\"ocker}},\ and\ \bibinfo {author} {\bibfnamefont
  {L.}~\bibnamefont {Rezzolla}},\ }\href
  {https://doi.org/10.1103/PhysRevLett.122.061101} {\bibfield  {journal}
  {\bibinfo  {journal} {Phys. Rev. Lett.}\ }\textbf {\bibinfo {volume} {122}},\
  \bibinfo {pages} {061101} (\bibinfo {year} {2019})},\ \Eprint
  {https://arxiv.org/abs/1807.03684} {arXiv:1807.03684 [astro-ph.HE]}
  \BibitemShut {NoStop}%
\bibitem [{\citenamefont {Bauswein}\ \emph {et~al.}(2020)\citenamefont
  {Bauswein}, \citenamefont {Blacker}, \citenamefont {Vijayan}, \citenamefont
  {Stergioulas}, \citenamefont {Chatziioannou}, \citenamefont {Clark},
  \citenamefont {Bastian}, \citenamefont {Blaschke}, \citenamefont {Cierniak},\
  and\ \citenamefont {Fischer}}]{Bauswein:2020aag}%
  \BibitemOpen
  \bibfield  {author} {\bibinfo {author} {\bibfnamefont {A.}~\bibnamefont
  {Bauswein}}, \bibinfo {author} {\bibfnamefont {S.}~\bibnamefont {Blacker}},
  \bibinfo {author} {\bibfnamefont {V.}~\bibnamefont {Vijayan}}, \bibinfo
  {author} {\bibfnamefont {N.}~\bibnamefont {Stergioulas}}, \bibinfo {author}
  {\bibfnamefont {K.}~\bibnamefont {Chatziioannou}}, \bibinfo {author}
  {\bibfnamefont {J.~A.}\ \bibnamefont {Clark}}, \bibinfo {author}
  {\bibfnamefont {N.-U.~F.}\ \bibnamefont {Bastian}}, \bibinfo {author}
  {\bibfnamefont {D.~B.}\ \bibnamefont {Blaschke}}, \bibinfo {author}
  {\bibfnamefont {M.}~\bibnamefont {Cierniak}},\ and\ \bibinfo {author}
  {\bibfnamefont {T.}~\bibnamefont {Fischer}},\ }\href
  {https://doi.org/10.1103/PhysRevLett.125.141103} {\bibfield  {journal}
  {\bibinfo  {journal} {Phys. Rev. Lett.}\ }\textbf {\bibinfo {volume} {125}},\
  \bibinfo {pages} {141103} (\bibinfo {year} {2020})},\ \Eprint
  {https://arxiv.org/abs/2004.00846} {arXiv:2004.00846 [astro-ph.HE]}
  \BibitemShut {NoStop}%
\bibitem [{\citenamefont {Yakovlev}\ and\ \citenamefont
  {Pethick}(2004)}]{Yakovlev:2004iq}%
  \BibitemOpen
  \bibfield  {author} {\bibinfo {author} {\bibfnamefont {D.~G.}\ \bibnamefont
  {Yakovlev}}\ and\ \bibinfo {author} {\bibfnamefont {C.}~\bibnamefont
  {Pethick}},\ }\href {https://doi.org/10.1146/annurev.astro.42.053102.134013}
  {\bibfield  {journal} {\bibinfo  {journal} {Ann. Rev. Astron. Astrophys.}\
  }\textbf {\bibinfo {volume} {42}},\ \bibinfo {pages} {169} (\bibinfo {year}
  {2004})},\ \Eprint {https://arxiv.org/abs/astro-ph/0402143}
  {arXiv:astro-ph/0402143} \BibitemShut {NoStop}%
\bibitem [{\citenamefont {Page}\ \emph {et~al.}(2004)\citenamefont {Page},
  \citenamefont {Lattimer}, \citenamefont {Prakash},\ and\ \citenamefont
  {Steiner}}]{Page:2004fy}%
  \BibitemOpen
  \bibfield  {author} {\bibinfo {author} {\bibfnamefont {D.}~\bibnamefont
  {Page}}, \bibinfo {author} {\bibfnamefont {J.~M.}\ \bibnamefont {Lattimer}},
  \bibinfo {author} {\bibfnamefont {M.}~\bibnamefont {Prakash}},\ and\ \bibinfo
  {author} {\bibfnamefont {A.~W.}\ \bibnamefont {Steiner}},\ }\href
  {https://doi.org/10.1086/424844} {\bibfield  {journal} {\bibinfo  {journal}
  {Astrophys. J. Suppl.}\ }\textbf {\bibinfo {volume} {155}},\ \bibinfo {pages}
  {623} (\bibinfo {year} {2004})},\ \Eprint
  {https://arxiv.org/abs/astro-ph/0403657} {arXiv:astro-ph/0403657}
  \BibitemShut {NoStop}%
\bibitem [{\citenamefont {Lattimer}\ and\ \citenamefont
  {Swesty}(1991)}]{Lattimer:1991nc}%
  \BibitemOpen
  \bibfield  {author} {\bibinfo {author} {\bibfnamefont {J.~M.}\ \bibnamefont
  {Lattimer}}\ and\ \bibinfo {author} {\bibfnamefont {F.}~\bibnamefont
  {Swesty}},\ }\href {https://doi.org/10.1016/0375-9474(91)90452-C} {\bibfield
  {journal} {\bibinfo  {journal} {Nucl. Phys. A}\ }\textbf {\bibinfo {volume}
  {535}},\ \bibinfo {pages} {331} (\bibinfo {year} {1991})}\BibitemShut
  {NoStop}%
\bibitem [{\citenamefont {Shen}\ \emph {et~al.}(1998)\citenamefont {Shen},
  \citenamefont {Toki}, \citenamefont {Oyamatsu},\ and\ \citenamefont
  {Sumiyoshi}}]{Shen:1998gq}%
  \BibitemOpen
  \bibfield  {author} {\bibinfo {author} {\bibfnamefont {H.}~\bibnamefont
  {Shen}}, \bibinfo {author} {\bibfnamefont {H.}~\bibnamefont {Toki}}, \bibinfo
  {author} {\bibfnamefont {K.}~\bibnamefont {Oyamatsu}},\ and\ \bibinfo
  {author} {\bibfnamefont {K.}~\bibnamefont {Sumiyoshi}},\ }\href
  {https://doi.org/10.1016/S0375-9474(98)00236-X} {\bibfield  {journal}
  {\bibinfo  {journal} {Nucl. Phys. A}\ }\textbf {\bibinfo {volume} {637}},\
  \bibinfo {pages} {435} (\bibinfo {year} {1998})},\ \Eprint
  {https://arxiv.org/abs/nucl-th/9805035} {arXiv:nucl-th/9805035} \BibitemShut
  {NoStop}%
\bibitem [{\citenamefont {Steiner}\ \emph {et~al.}(2013)\citenamefont
  {Steiner}, \citenamefont {Hempel},\ and\ \citenamefont
  {Fischer}}]{Steiner:2012rk}%
  \BibitemOpen
  \bibfield  {author} {\bibinfo {author} {\bibfnamefont {A.~W.}\ \bibnamefont
  {Steiner}}, \bibinfo {author} {\bibfnamefont {M.}~\bibnamefont {Hempel}},\
  and\ \bibinfo {author} {\bibfnamefont {T.}~\bibnamefont {Fischer}},\ }\href
  {https://doi.org/10.1088/0004-637X/774/1/17} {\bibfield  {journal} {\bibinfo
  {journal} {Astrophys. J.}\ }\textbf {\bibinfo {volume} {774}},\ \bibinfo
  {pages} {17} (\bibinfo {year} {2013})},\ \Eprint
  {https://arxiv.org/abs/1207.2184} {arXiv:1207.2184 [astro-ph.SR]}
  \BibitemShut {NoStop}%
\bibitem [{\citenamefont {Typel}\ \emph {et~al.}(2010)\citenamefont {Typel},
  \citenamefont {Ropke}, \citenamefont {Klahn}, \citenamefont {Blaschke},\ and\
  \citenamefont {Wolter}}]{Typel:2009sy}%
  \BibitemOpen
  \bibfield  {author} {\bibinfo {author} {\bibfnamefont {S.}~\bibnamefont
  {Typel}}, \bibinfo {author} {\bibfnamefont {G.}~\bibnamefont {Ropke}},
  \bibinfo {author} {\bibfnamefont {T.}~\bibnamefont {Klahn}}, \bibinfo
  {author} {\bibfnamefont {D.}~\bibnamefont {Blaschke}},\ and\ \bibinfo
  {author} {\bibfnamefont {H.}~\bibnamefont {Wolter}},\ }\href
  {https://doi.org/10.1103/PhysRevC.81.015803} {\bibfield  {journal} {\bibinfo
  {journal} {Phys. Rev. C}\ }\textbf {\bibinfo {volume} {81}},\ \bibinfo
  {pages} {015803} (\bibinfo {year} {2010})},\ \Eprint
  {https://arxiv.org/abs/0908.2344} {arXiv:0908.2344 [nucl-th]} \BibitemShut
  {NoStop}%
\bibitem [{\citenamefont {Fukushima}\ and\ \citenamefont
  {Kojo}(2016)}]{Fukushima:2015bda}%
  \BibitemOpen
  \bibfield  {author} {\bibinfo {author} {\bibfnamefont {K.}~\bibnamefont
  {Fukushima}}\ and\ \bibinfo {author} {\bibfnamefont {T.}~\bibnamefont
  {Kojo}},\ }\href {https://doi.org/10.3847/0004-637X/817/2/180} {\bibfield
  {journal} {\bibinfo  {journal} {Astrophys. J.}\ }\textbf {\bibinfo {volume}
  {817}},\ \bibinfo {pages} {180} (\bibinfo {year} {2016})},\ \Eprint
  {https://arxiv.org/abs/1509.00356} {arXiv:1509.00356 [nucl-th]} \BibitemShut
  {NoStop}%
\bibitem [{\citenamefont {Fujimoto}\ \emph {et~al.}(2020)\citenamefont
  {Fujimoto}, \citenamefont {Fukushima},\ and\ \citenamefont
  {Weise}}]{Fujimoto:2019sxg}%
  \BibitemOpen
  \bibfield  {author} {\bibinfo {author} {\bibfnamefont {Y.}~\bibnamefont
  {Fujimoto}}, \bibinfo {author} {\bibfnamefont {K.}~\bibnamefont
  {Fukushima}},\ and\ \bibinfo {author} {\bibfnamefont {W.}~\bibnamefont
  {Weise}},\ }\href {https://doi.org/10.1103/PhysRevD.101.094009} {\bibfield
  {journal} {\bibinfo  {journal} {Phys. Rev. D}\ }\textbf {\bibinfo {volume}
  {101}},\ \bibinfo {pages} {094009} (\bibinfo {year} {2020})},\ \Eprint
  {https://arxiv.org/abs/1908.09360} {arXiv:1908.09360 [hep-ph]} \BibitemShut
  {NoStop}%
\bibitem [{\citenamefont {Ishii}\ \emph {et~al.}(2007)\citenamefont {Ishii},
  \citenamefont {Aoki},\ and\ \citenamefont {Hatsuda}}]{Ishii:2006ec}%
  \BibitemOpen
  \bibfield  {author} {\bibinfo {author} {\bibfnamefont {N.}~\bibnamefont
  {Ishii}}, \bibinfo {author} {\bibfnamefont {S.}~\bibnamefont {Aoki}},\ and\
  \bibinfo {author} {\bibfnamefont {T.}~\bibnamefont {Hatsuda}},\ }\href
  {https://doi.org/10.1103/PhysRevLett.99.022001} {\bibfield  {journal}
  {\bibinfo  {journal} {Phys. Rev. Lett.}\ }\textbf {\bibinfo {volume} {99}},\
  \bibinfo {pages} {022001} (\bibinfo {year} {2007})},\ \Eprint
  {https://arxiv.org/abs/nucl-th/0611096} {arXiv:nucl-th/0611096} \BibitemShut
  {NoStop}%
\bibitem [{\citenamefont {Iritani}\ \emph {et~al.}(2019)\citenamefont {Iritani}
  \emph {et~al.}}]{Iritani:2018sra}%
  \BibitemOpen
  \bibfield  {author} {\bibinfo {author} {\bibfnamefont {T.}~\bibnamefont
  {Iritani}} \emph {et~al.} (\bibinfo {collaboration} {HAL QCD}),\ }\href
  {https://doi.org/10.1016/j.physletb.2019.03.050} {\bibfield  {journal}
  {\bibinfo  {journal} {Phys. Lett. B}\ }\textbf {\bibinfo {volume} {792}},\
  \bibinfo {pages} {284} (\bibinfo {year} {2019})},\ \Eprint
  {https://arxiv.org/abs/1810.03416} {arXiv:1810.03416 [hep-lat]} \BibitemShut
  {NoStop}%
\bibitem [{\citenamefont {Oka}\ and\ \citenamefont
  {Yazaki}(1980)}]{Oka:1980ax}%
  \BibitemOpen
  \bibfield  {author} {\bibinfo {author} {\bibfnamefont {M.}~\bibnamefont
  {Oka}}\ and\ \bibinfo {author} {\bibfnamefont {K.}~\bibnamefont {Yazaki}},\
  }\href {https://doi.org/10.1016/0370-2693(80)90046-5} {\bibfield  {journal}
  {\bibinfo  {journal} {Phys. Lett. B}\ }\textbf {\bibinfo {volume} {90}},\
  \bibinfo {pages} {41} (\bibinfo {year} {1980})}\BibitemShut {NoStop}%
\bibitem [{\citenamefont {Oka}\ and\ \citenamefont
  {Yazaki}(1983)}]{Oka:1982qa}%
  \BibitemOpen
  \bibfield  {author} {\bibinfo {author} {\bibfnamefont {M.}~\bibnamefont
  {Oka}}\ and\ \bibinfo {author} {\bibfnamefont {K.}~\bibnamefont {Yazaki}},\
  }\href {https://doi.org/10.1016/0375-9474(86)90199-5} {\bibfield  {journal}
  {\bibinfo  {journal} {Nucl. Phys. A}\ }\textbf {\bibinfo {volume} {402}},\
  \bibinfo {pages} {477} (\bibinfo {year} {1983})},\ \bibinfo {note} {[Erratum:
  Nucl.Phys.A 458, 773--773 (1986)]}\BibitemShut {NoStop}%
\bibitem [{\citenamefont {Nambu}\ and\ \citenamefont
  {Jona-Lasinio}(1961)}]{Nambu:1961tp}%
  \BibitemOpen
  \bibfield  {author} {\bibinfo {author} {\bibfnamefont {Y.}~\bibnamefont
  {Nambu}}\ and\ \bibinfo {author} {\bibfnamefont {G.}~\bibnamefont
  {Jona-Lasinio}},\ }\href {https://doi.org/10.1103/PhysRev.122.345} {\bibfield
   {journal} {\bibinfo  {journal} {Phys. Rev.}\ }\textbf {\bibinfo {volume}
  {122}},\ \bibinfo {pages} {345} (\bibinfo {year} {1961})}\BibitemShut
  {NoStop}%
\bibitem [{\citenamefont {Manohar}\ and\ \citenamefont
  {Georgi}(1984)}]{Manohar:1983md}%
  \BibitemOpen
  \bibfield  {author} {\bibinfo {author} {\bibfnamefont {A.}~\bibnamefont
  {Manohar}}\ and\ \bibinfo {author} {\bibfnamefont {H.}~\bibnamefont
  {Georgi}},\ }\href {https://doi.org/10.1016/0550-3213(84)90231-1} {\bibfield
  {journal} {\bibinfo  {journal} {Nucl. Phys. B}\ }\textbf {\bibinfo {volume}
  {234}},\ \bibinfo {pages} {189} (\bibinfo {year} {1984})}\BibitemShut
  {NoStop}%
\bibitem [{\citenamefont {De~Rujula}\ \emph {et~al.}(1975)\citenamefont
  {De~Rujula}, \citenamefont {Georgi},\ and\ \citenamefont
  {Glashow}}]{DeRujula:1975qlm}%
  \BibitemOpen
  \bibfield  {author} {\bibinfo {author} {\bibfnamefont {A.}~\bibnamefont
  {De~Rujula}}, \bibinfo {author} {\bibfnamefont {H.}~\bibnamefont {Georgi}},\
  and\ \bibinfo {author} {\bibfnamefont {S.}~\bibnamefont {Glashow}},\ }\href
  {https://doi.org/10.1103/PhysRevD.12.147} {\bibfield  {journal} {\bibinfo
  {journal} {Phys. Rev. D}\ }\textbf {\bibinfo {volume} {12}},\ \bibinfo
  {pages} {147} (\bibinfo {year} {1975})}\BibitemShut {NoStop}%
\bibitem [{\citenamefont {Park}\ \emph {et~al.}(2020)\citenamefont {Park},
  \citenamefont {Lee}, \citenamefont {Inoue},\ and\ \citenamefont
  {Hatsuda}}]{Park:2019bsz}%
  \BibitemOpen
  \bibfield  {author} {\bibinfo {author} {\bibfnamefont {A.}~\bibnamefont
  {Park}}, \bibinfo {author} {\bibfnamefont {S.~H.}\ \bibnamefont {Lee}},
  \bibinfo {author} {\bibfnamefont {T.}~\bibnamefont {Inoue}},\ and\ \bibinfo
  {author} {\bibfnamefont {T.}~\bibnamefont {Hatsuda}},\ }\href
  {https://doi.org/10.1140/epja/s10050-020-00078-z} {\bibfield  {journal}
  {\bibinfo  {journal} {Eur. Phys. J. A}\ }\textbf {\bibinfo {volume} {56}},\
  \bibinfo {pages} {93} (\bibinfo {year} {2020})},\ \Eprint
  {https://arxiv.org/abs/1907.06351} {arXiv:1907.06351 [hep-ph]} \BibitemShut
  {NoStop}%
\bibitem [{\citenamefont {Hatsuda}\ and\ \citenamefont
  {Kunihiro}(1994)}]{Hatsuda:1994pi}%
  \BibitemOpen
  \bibfield  {author} {\bibinfo {author} {\bibfnamefont {T.}~\bibnamefont
  {Hatsuda}}\ and\ \bibinfo {author} {\bibfnamefont {T.}~\bibnamefont
  {Kunihiro}},\ }\href {https://doi.org/10.1016/0370-1573(94)90022-1}
  {\bibfield  {journal} {\bibinfo  {journal} {Phys. Rept.}\ }\textbf {\bibinfo
  {volume} {247}},\ \bibinfo {pages} {221} (\bibinfo {year} {1994})},\ \Eprint
  {https://arxiv.org/abs/hep-ph/9401310} {arXiv:hep-ph/9401310} \BibitemShut
  {NoStop}%
\bibitem [{\citenamefont {Iida}\ and\ \citenamefont
  {Baym}(2001)}]{Iida:2000ha}%
  \BibitemOpen
  \bibfield  {author} {\bibinfo {author} {\bibfnamefont {K.}~\bibnamefont
  {Iida}}\ and\ \bibinfo {author} {\bibfnamefont {G.}~\bibnamefont {Baym}},\
  }\href {https://doi.org/10.1103/PhysRevD.63.074018} {\bibfield  {journal}
  {\bibinfo  {journal} {Phys. Rev. D}\ }\textbf {\bibinfo {volume} {63}},\
  \bibinfo {pages} {074018} (\bibinfo {year} {2001})},\ \bibinfo {note}
  {[Erratum: Phys.Rev.D 66, 059903 (2002)]},\ \Eprint
  {https://arxiv.org/abs/hep-ph/0011229} {arXiv:hep-ph/0011229} \BibitemShut
  {NoStop}%
\bibitem [{\citenamefont {Zhang}\ \emph {et~al.}(2017)\citenamefont {Zhang},
  \citenamefont {Hou}, \citenamefont {Kojo},\ and\ \citenamefont
  {Qin}}]{Zhang:2017icm}%
  \BibitemOpen
  \bibfield  {author} {\bibinfo {author} {\bibfnamefont {H.}~\bibnamefont
  {Zhang}}, \bibinfo {author} {\bibfnamefont {D.}~\bibnamefont {Hou}}, \bibinfo
  {author} {\bibfnamefont {T.}~\bibnamefont {Kojo}},\ and\ \bibinfo {author}
  {\bibfnamefont {B.}~\bibnamefont {Qin}},\ }\href
  {https://doi.org/10.1103/PhysRevD.96.114029} {\bibfield  {journal} {\bibinfo
  {journal} {Phys. Rev. D}\ }\textbf {\bibinfo {volume} {96}},\ \bibinfo
  {pages} {114029} (\bibinfo {year} {2017})},\ \Eprint
  {https://arxiv.org/abs/1709.05654} {arXiv:1709.05654 [hep-ph]} \BibitemShut
  {NoStop}%
\bibitem [{\citenamefont {Nakazato}\ \emph
  {et~al.}(2010{\natexlab{a}})\citenamefont {Nakazato}, \citenamefont
  {Sumiyoshi},\ and\ \citenamefont {Yamada}}]{Nakazato:2010ue}%
  \BibitemOpen
  \bibfield  {author} {\bibinfo {author} {\bibfnamefont {K.}~\bibnamefont
  {Nakazato}}, \bibinfo {author} {\bibfnamefont {K.}~\bibnamefont
  {Sumiyoshi}},\ and\ \bibinfo {author} {\bibfnamefont {S.}~\bibnamefont
  {Yamada}},\ }\href {https://doi.org/10.1088/0004-637X/721/2/1284} {\bibfield
  {journal} {\bibinfo  {journal} {Astrophys. J.}\ }\textbf {\bibinfo {volume}
  {721}},\ \bibinfo {pages} {1284} (\bibinfo {year} {2010}{\natexlab{a}})},\
  \Eprint {https://arxiv.org/abs/1001.5084} {arXiv:1001.5084 [astro-ph.HE]}
  \BibitemShut {NoStop}%
\bibitem [{\citenamefont {Nakazato}\ \emph
  {et~al.}(2010{\natexlab{b}})\citenamefont {Nakazato}, \citenamefont
  {Sumiyoshi}, \citenamefont {Suzuki},\ and\ \citenamefont
  {Yamada}}]{Nakazato:2010qy}%
  \BibitemOpen
  \bibfield  {author} {\bibinfo {author} {\bibfnamefont {K.}~\bibnamefont
  {Nakazato}}, \bibinfo {author} {\bibfnamefont {K.}~\bibnamefont {Sumiyoshi}},
  \bibinfo {author} {\bibfnamefont {H.}~\bibnamefont {Suzuki}},\ and\ \bibinfo
  {author} {\bibfnamefont {S.}~\bibnamefont {Yamada}},\ }\href
  {https://doi.org/10.1103/PhysRevD.81.083009} {\bibfield  {journal} {\bibinfo
  {journal} {Phys. Rev. D}\ }\textbf {\bibinfo {volume} {81}},\ \bibinfo
  {pages} {083009} (\bibinfo {year} {2010}{\natexlab{b}})},\ \Eprint
  {https://arxiv.org/abs/1004.0291} {arXiv:1004.0291 [astro-ph.HE]}
  \BibitemShut {NoStop}%
\bibitem [{\citenamefont {Fischer}\ \emph {et~al.}(2011)\citenamefont
  {Fischer}, \citenamefont {Sagert}, \citenamefont {Pagliara}, \citenamefont
  {Hempel}, \citenamefont {Schaffner-Bielich}, \citenamefont {Rauscher},
  \citenamefont {Thielemann}, \citenamefont {Kappeli}, \citenamefont
  {Martinez-Pinedo},\ and\ \citenamefont {Liebendorfer}}]{Fischer:2010wp}%
  \BibitemOpen
  \bibfield  {author} {\bibinfo {author} {\bibfnamefont {T.}~\bibnamefont
  {Fischer}}, \bibinfo {author} {\bibfnamefont {I.}~\bibnamefont {Sagert}},
  \bibinfo {author} {\bibfnamefont {G.}~\bibnamefont {Pagliara}}, \bibinfo
  {author} {\bibfnamefont {M.}~\bibnamefont {Hempel}}, \bibinfo {author}
  {\bibfnamefont {J.}~\bibnamefont {Schaffner-Bielich}}, \bibinfo {author}
  {\bibfnamefont {T.}~\bibnamefont {Rauscher}}, \bibinfo {author}
  {\bibfnamefont {F.}~\bibnamefont {Thielemann}}, \bibinfo {author}
  {\bibfnamefont {R.}~\bibnamefont {Kappeli}}, \bibinfo {author} {\bibfnamefont
  {G.}~\bibnamefont {Martinez-Pinedo}},\ and\ \bibinfo {author} {\bibfnamefont
  {M.}~\bibnamefont {Liebendorfer}},\ }\href
  {https://doi.org/10.1088/0067-0049/194/2/39} {\bibfield  {journal} {\bibinfo
  {journal} {Astrophys. J. Suppl.}\ }\textbf {\bibinfo {volume} {194}},\
  \bibinfo {pages} {39} (\bibinfo {year} {2011})},\ \Eprint
  {https://arxiv.org/abs/1011.3409} {arXiv:1011.3409 [astro-ph.HE]}
  \BibitemShut {NoStop}%
\bibitem [{\citenamefont {Fischer}\ \emph {et~al.}(2018)\citenamefont
  {Fischer}, \citenamefont {Bastian}, \citenamefont {Wu}, \citenamefont
  {Baklanov}, \citenamefont {Sorokina}, \citenamefont {Blinnikov},
  \citenamefont {Typel}, \citenamefont {Kl\"ahn},\ and\ \citenamefont
  {Blaschke}}]{Fischer:2017lag}%
  \BibitemOpen
  \bibfield  {author} {\bibinfo {author} {\bibfnamefont {T.}~\bibnamefont
  {Fischer}}, \bibinfo {author} {\bibfnamefont {N.-U.~F.}\ \bibnamefont
  {Bastian}}, \bibinfo {author} {\bibfnamefont {M.-R.}\ \bibnamefont {Wu}},
  \bibinfo {author} {\bibfnamefont {P.}~\bibnamefont {Baklanov}}, \bibinfo
  {author} {\bibfnamefont {E.}~\bibnamefont {Sorokina}}, \bibinfo {author}
  {\bibfnamefont {S.}~\bibnamefont {Blinnikov}}, \bibinfo {author}
  {\bibfnamefont {S.}~\bibnamefont {Typel}}, \bibinfo {author} {\bibfnamefont
  {T.}~\bibnamefont {Kl\"ahn}},\ and\ \bibinfo {author} {\bibfnamefont {D.~B.}\
  \bibnamefont {Blaschke}},\ }\href {https://doi.org/10.1038/s41550-018-0583-0}
  {\bibfield  {journal} {\bibinfo  {journal} {Nature Astron.}\ }\textbf
  {\bibinfo {volume} {2}},\ \bibinfo {pages} {980} (\bibinfo {year} {2018})},\
  \Eprint {https://arxiv.org/abs/1712.08788} {arXiv:1712.08788 [astro-ph.HE]}
  \BibitemShut {NoStop}%
\bibitem [{\citenamefont {Pons}\ \emph {et~al.}(2001)\citenamefont {Pons},
  \citenamefont {Steiner}, \citenamefont {Prakash},\ and\ \citenamefont
  {Lattimer}}]{PhysRevLett.86.5223}%
  \BibitemOpen
  \bibfield  {author} {\bibinfo {author} {\bibfnamefont {J.~A.}\ \bibnamefont
  {Pons}}, \bibinfo {author} {\bibfnamefont {A.~W.}\ \bibnamefont {Steiner}},
  \bibinfo {author} {\bibfnamefont {M.}~\bibnamefont {Prakash}},\ and\ \bibinfo
  {author} {\bibfnamefont {J.~M.}\ \bibnamefont {Lattimer}},\ }\href
  {https://doi.org/10.1103/PhysRevLett.86.5223} {\bibfield  {journal} {\bibinfo
   {journal} {Phys. Rev. Lett.}\ }\textbf {\bibinfo {volume} {86}},\ \bibinfo
  {pages} {5223} (\bibinfo {year} {2001})}\BibitemShut {NoStop}%
\bibitem [{\citenamefont {Camelio}\ \emph {et~al.}(2017)\citenamefont
  {Camelio}, \citenamefont {Lovato}, \citenamefont {Gualtieri}, \citenamefont
  {Benhar}, \citenamefont {Pons},\ and\ \citenamefont
  {Ferrari}}]{Camelio:2017nka}%
  \BibitemOpen
  \bibfield  {author} {\bibinfo {author} {\bibfnamefont {G.}~\bibnamefont
  {Camelio}}, \bibinfo {author} {\bibfnamefont {A.}~\bibnamefont {Lovato}},
  \bibinfo {author} {\bibfnamefont {L.}~\bibnamefont {Gualtieri}}, \bibinfo
  {author} {\bibfnamefont {O.}~\bibnamefont {Benhar}}, \bibinfo {author}
  {\bibfnamefont {J.~A.}\ \bibnamefont {Pons}},\ and\ \bibinfo {author}
  {\bibfnamefont {V.}~\bibnamefont {Ferrari}},\ }\href
  {https://doi.org/10.1103/PhysRevD.96.043015} {\bibfield  {journal} {\bibinfo
  {journal} {Phys. Rev. D}\ }\textbf {\bibinfo {volume} {96}},\ \bibinfo
  {pages} {043015} (\bibinfo {year} {2017})},\ \Eprint
  {https://arxiv.org/abs/1704.01923} {arXiv:1704.01923 [astro-ph.HE]}
  \BibitemShut {NoStop}%
\bibitem [{\citenamefont {Vincent}\ \emph {et~al.}(2020)\citenamefont
  {Vincent}, \citenamefont {Foucart}, \citenamefont {Duez}, \citenamefont
  {Haas}, \citenamefont {Kidder}, \citenamefont {Pfeiffer},\ and\ \citenamefont
  {Scheel}}]{Vincent:2019kor}%
  \BibitemOpen
  \bibfield  {author} {\bibinfo {author} {\bibfnamefont {T.}~\bibnamefont
  {Vincent}}, \bibinfo {author} {\bibfnamefont {F.}~\bibnamefont {Foucart}},
  \bibinfo {author} {\bibfnamefont {M.~D.}\ \bibnamefont {Duez}}, \bibinfo
  {author} {\bibfnamefont {R.}~\bibnamefont {Haas}}, \bibinfo {author}
  {\bibfnamefont {L.~E.}\ \bibnamefont {Kidder}}, \bibinfo {author}
  {\bibfnamefont {H.~P.}\ \bibnamefont {Pfeiffer}},\ and\ \bibinfo {author}
  {\bibfnamefont {M.~A.}\ \bibnamefont {Scheel}},\ }\href
  {https://doi.org/10.1103/PhysRevD.101.044053} {\bibfield  {journal} {\bibinfo
   {journal} {Phys. Rev. D}\ }\textbf {\bibinfo {volume} {101}},\ \bibinfo
  {pages} {044053} (\bibinfo {year} {2020})},\ \Eprint
  {https://arxiv.org/abs/1908.00655} {arXiv:1908.00655 [gr-qc]} \BibitemShut
  {NoStop}%
\bibitem [{\citenamefont {Jim\'enez}\ and\ \citenamefont
  {Fraga}(2018)}]{PhysRevD.97.094023}%
  \BibitemOpen
  \bibfield  {author} {\bibinfo {author} {\bibfnamefont {J.~C.}\ \bibnamefont
  {Jim\'enez}}\ and\ \bibinfo {author} {\bibfnamefont {E.~S.}\ \bibnamefont
  {Fraga}},\ }\href {https://doi.org/10.1103/PhysRevD.97.094023} {\bibfield
  {journal} {\bibinfo  {journal} {Phys. Rev. D}\ }\textbf {\bibinfo {volume}
  {97}},\ \bibinfo {pages} {094023} (\bibinfo {year} {2018})}\BibitemShut
  {NoStop}%
\bibitem [{\citenamefont {Masuda}\ \emph
  {et~al.}(2016{\natexlab{b}})\citenamefont {Masuda}, \citenamefont {Hatsuda},\
  and\ \citenamefont {Takatsuka}}]{Masuda:2015wva}%
  \BibitemOpen
  \bibfield  {author} {\bibinfo {author} {\bibfnamefont {K.}~\bibnamefont
  {Masuda}}, \bibinfo {author} {\bibfnamefont {T.}~\bibnamefont {Hatsuda}},\
  and\ \bibinfo {author} {\bibfnamefont {T.}~\bibnamefont {Takatsuka}},\ }\href
  {https://doi.org/10.1093/ptep/ptv187} {\bibfield  {journal} {\bibinfo
  {journal} {PTEP}\ }\textbf {\bibinfo {volume} {2016}},\ \bibinfo {pages}
  {021D01} (\bibinfo {year} {2016}{\natexlab{b}})},\ \Eprint
  {https://arxiv.org/abs/1506.00984} {arXiv:1506.00984 [nucl-th]} \BibitemShut
  {NoStop}%
\bibitem [{\citenamefont {Togashi}\ \emph {et~al.}(2016)\citenamefont
  {Togashi}, \citenamefont {Hiyama}, \citenamefont {Yamamoto},\ and\
  \citenamefont {Takano}}]{Togashi:2016fky}%
  \BibitemOpen
  \bibfield  {author} {\bibinfo {author} {\bibfnamefont {H.}~\bibnamefont
  {Togashi}}, \bibinfo {author} {\bibfnamefont {E.}~\bibnamefont {Hiyama}},
  \bibinfo {author} {\bibfnamefont {Y.}~\bibnamefont {Yamamoto}},\ and\
  \bibinfo {author} {\bibfnamefont {M.}~\bibnamefont {Takano}},\ }\href
  {https://doi.org/10.1103/PhysRevC.93.035808} {\bibfield  {journal} {\bibinfo
  {journal} {Phys. Rev. C}\ }\textbf {\bibinfo {volume} {93}},\ \bibinfo
  {pages} {035808} (\bibinfo {year} {2016})},\ \Eprint
  {https://arxiv.org/abs/1602.08106} {arXiv:1602.08106 [nucl-th]} \BibitemShut
  {NoStop}%
\bibitem [{\citenamefont {Ishizuka}\ \emph {et~al.}(2008)\citenamefont
  {Ishizuka}, \citenamefont {Ohnishi}, \citenamefont {Tsubakihara},
  \citenamefont {Sumiyoshi},\ and\ \citenamefont {Yamada}}]{Ishizuka:2008gr}%
  \BibitemOpen
  \bibfield  {author} {\bibinfo {author} {\bibfnamefont {C.}~\bibnamefont
  {Ishizuka}}, \bibinfo {author} {\bibfnamefont {A.}~\bibnamefont {Ohnishi}},
  \bibinfo {author} {\bibfnamefont {K.}~\bibnamefont {Tsubakihara}}, \bibinfo
  {author} {\bibfnamefont {K.}~\bibnamefont {Sumiyoshi}},\ and\ \bibinfo
  {author} {\bibfnamefont {S.}~\bibnamefont {Yamada}},\ }\href
  {https://doi.org/10.1088/0954-3899/35/8/085201} {\bibfield  {journal}
  {\bibinfo  {journal} {J. Phys. G}\ }\textbf {\bibinfo {volume} {35}},\
  \bibinfo {pages} {085201} (\bibinfo {year} {2008})},\ \Eprint
  {https://arxiv.org/abs/0802.2318} {arXiv:0802.2318 [nucl-th]} \BibitemShut
  {NoStop}%
\bibitem [{\citenamefont {Fortin}\ \emph {et~al.}(2018)\citenamefont {Fortin},
  \citenamefont {Oertel},\ and\ \citenamefont {Providência}}]{Fortin:2017dsj}%
  \BibitemOpen
  \bibfield  {author} {\bibinfo {author} {\bibfnamefont {M.}~\bibnamefont
  {Fortin}}, \bibinfo {author} {\bibfnamefont {M.}~\bibnamefont {Oertel}},\
  and\ \bibinfo {author} {\bibfnamefont {C.}~\bibnamefont {Providência}},\
  }\href {https://doi.org/10.1017/pasa.2018.32} {\bibfield  {journal} {\bibinfo
   {journal} {Publ. Astron. Soc. Austral.}\ }\textbf {\bibinfo {volume} {35}},\
  \bibinfo {pages} {44} (\bibinfo {year} {2018})},\ \Eprint
  {https://arxiv.org/abs/1711.09427} {arXiv:1711.09427 [astro-ph.HE]}
  \BibitemShut {NoStop}%
\bibitem [{\citenamefont {Marques}\ \emph {et~al.}(2017)\citenamefont
  {Marques}, \citenamefont {Oertel}, \citenamefont {Hempel},\ and\
  \citenamefont {Novak}}]{Marques:2017zju}%
  \BibitemOpen
  \bibfield  {author} {\bibinfo {author} {\bibfnamefont {M.}~\bibnamefont
  {Marques}}, \bibinfo {author} {\bibfnamefont {M.}~\bibnamefont {Oertel}},
  \bibinfo {author} {\bibfnamefont {M.}~\bibnamefont {Hempel}},\ and\ \bibinfo
  {author} {\bibfnamefont {J.}~\bibnamefont {Novak}},\ }\href
  {https://doi.org/10.1103/PhysRevC.96.045806} {\bibfield  {journal} {\bibinfo
  {journal} {Phys. Rev. C}\ }\textbf {\bibinfo {volume} {96}},\ \bibinfo
  {pages} {045806} (\bibinfo {year} {2017})},\ \Eprint
  {https://arxiv.org/abs/1706.02913} {arXiv:1706.02913 [nucl-th]} \BibitemShut
  {NoStop}%
\bibitem [{\citenamefont {Burgio}\ \emph {et~al.}(2011)\citenamefont {Burgio},
  \citenamefont {Schulze},\ and\ \citenamefont {Li}}]{Burgio:2011wt}%
  \BibitemOpen
  \bibfield  {author} {\bibinfo {author} {\bibfnamefont {G.}~\bibnamefont
  {Burgio}}, \bibinfo {author} {\bibfnamefont {H.-J.}\ \bibnamefont
  {Schulze}},\ and\ \bibinfo {author} {\bibfnamefont {A.}~\bibnamefont {Li}},\
  }\href {https://doi.org/10.1103/PhysRevC.83.025804} {\bibfield  {journal}
  {\bibinfo  {journal} {Phys. Rev. C}\ }\textbf {\bibinfo {volume} {83}},\
  \bibinfo {pages} {025804} (\bibinfo {year} {2011})},\ \Eprint
  {https://arxiv.org/abs/1101.0726} {arXiv:1101.0726 [astro-ph.SR]}
  \BibitemShut {NoStop}%
\bibitem [{\citenamefont {Song}\ \emph {et~al.}(2019)\citenamefont {Song},
  \citenamefont {Baym}, \citenamefont {Hatsuda},\ and\ \citenamefont
  {Kojo}}]{Song:2019qoh}%
  \BibitemOpen
  \bibfield  {author} {\bibinfo {author} {\bibfnamefont {Y.}~\bibnamefont
  {Song}}, \bibinfo {author} {\bibfnamefont {G.}~\bibnamefont {Baym}}, \bibinfo
  {author} {\bibfnamefont {T.}~\bibnamefont {Hatsuda}},\ and\ \bibinfo {author}
  {\bibfnamefont {T.}~\bibnamefont {Kojo}},\ }\href
  {https://doi.org/10.1103/PhysRevD.100.034018} {\bibfield  {journal} {\bibinfo
   {journal} {Phys. Rev. D}\ }\textbf {\bibinfo {volume} {100}},\ \bibinfo
  {pages} {034018} (\bibinfo {year} {2019})},\ \Eprint
  {https://arxiv.org/abs/1905.01005} {arXiv:1905.01005 [astro-ph.HE]}
  \BibitemShut {NoStop}%
\bibitem [{\citenamefont {Tissier}\ and\ \citenamefont
  {Wschebor}(2011)}]{PhysRevD.84.045018}%
  \BibitemOpen
  \bibfield  {author} {\bibinfo {author} {\bibfnamefont {M.}~\bibnamefont
  {Tissier}}\ and\ \bibinfo {author} {\bibfnamefont {N.}~\bibnamefont
  {Wschebor}},\ }\href {https://doi.org/10.1103/PhysRevD.84.045018} {\bibfield
  {journal} {\bibinfo  {journal} {Phys. Rev. D}\ }\textbf {\bibinfo {volume}
  {84}},\ \bibinfo {pages} {045018} (\bibinfo {year} {2011})}\BibitemShut
  {NoStop}%
\bibitem [{\citenamefont {Suenaga}\ and\ \citenamefont
  {Kojo}(2019)}]{Suenaga:2019jjv}%
  \BibitemOpen
  \bibfield  {author} {\bibinfo {author} {\bibfnamefont {D.}~\bibnamefont
  {Suenaga}}\ and\ \bibinfo {author} {\bibfnamefont {T.}~\bibnamefont {Kojo}},\
  }\href {https://doi.org/10.1103/PhysRevD.100.076017} {\bibfield  {journal}
  {\bibinfo  {journal} {Phys. Rev. D}\ }\textbf {\bibinfo {volume} {100}},\
  \bibinfo {pages} {076017} (\bibinfo {year} {2019})},\ \Eprint
  {https://arxiv.org/abs/1905.08751} {arXiv:1905.08751 [hep-ph]} \BibitemShut
  {NoStop}%
\bibitem [{\citenamefont {Kojo}\ \emph {et~al.}()\citenamefont {Kojo},
  \citenamefont {Hou}, \citenamefont {Okafor},\ and\ \citenamefont
  {Togashi}}]{preparation1}%
  \BibitemOpen
  \bibfield  {author} {\bibinfo {author} {\bibfnamefont {T.}~\bibnamefont
  {Kojo}}, \bibinfo {author} {\bibfnamefont {D.}~\bibnamefont {Hou}}, \bibinfo
  {author} {\bibfnamefont {J.}~\bibnamefont {Okafor}},\ and\ \bibinfo {author}
  {\bibfnamefont {H.}~\bibnamefont {Togashi}},\ }\href@noop {} {\bibinfo
  {journal} {in preparation}\ }\BibitemShut {NoStop}%
\bibitem [{\citenamefont {Kojo}(2016)}]{Kojo:2015fua}%
  \BibitemOpen
\bibfield  {journal} {  }\bibfield  {author} {\bibinfo {author} {\bibfnamefont
  {T.}~\bibnamefont {Kojo}},\ }\href
  {https://doi.org/10.1140/epja/i2016-16051-0} {\bibfield  {journal} {\bibinfo
  {journal} {Eur. Phys. J. A}\ }\textbf {\bibinfo {volume} {52}},\ \bibinfo
  {pages} {51} (\bibinfo {year} {2016})},\ \Eprint
  {https://arxiv.org/abs/1508.04408} {arXiv:1508.04408 [hep-ph]} \BibitemShut
  {NoStop}%
\bibitem [{\citenamefont {Fukushima}(2004)}]{Fukushima:2003fw}%
  \BibitemOpen
  \bibfield  {author} {\bibinfo {author} {\bibfnamefont {K.}~\bibnamefont
  {Fukushima}},\ }\href {https://doi.org/10.1016/j.physletb.2004.04.027}
  {\bibfield  {journal} {\bibinfo  {journal} {Phys. Lett. B}\ }\textbf
  {\bibinfo {volume} {591}},\ \bibinfo {pages} {277} (\bibinfo {year}
  {2004})},\ \Eprint {https://arxiv.org/abs/hep-ph/0310121}
  {arXiv:hep-ph/0310121} \BibitemShut {NoStop}%
\bibitem [{\citenamefont {Roessner}\ \emph {et~al.}(2007)\citenamefont
  {Roessner}, \citenamefont {Ratti},\ and\ \citenamefont
  {Weise}}]{Roessner:2006xn}%
  \BibitemOpen
  \bibfield  {author} {\bibinfo {author} {\bibfnamefont {S.}~\bibnamefont
  {Roessner}}, \bibinfo {author} {\bibfnamefont {C.}~\bibnamefont {Ratti}},\
  and\ \bibinfo {author} {\bibfnamefont {W.}~\bibnamefont {Weise}},\ }\href
  {https://doi.org/10.1103/PhysRevD.75.034007} {\bibfield  {journal} {\bibinfo
  {journal} {Phys. Rev. D}\ }\textbf {\bibinfo {volume} {75}},\ \bibinfo
  {pages} {034007} (\bibinfo {year} {2007})},\ \Eprint
  {https://arxiv.org/abs/hep-ph/0609281} {arXiv:hep-ph/0609281} \BibitemShut
  {NoStop}%
\bibitem [{\citenamefont {Weinberg}\ and\ \citenamefont
  {de~Campos}(1995)}]{weinberg1995quantum}%
  \BibitemOpen
  \bibfield  {author} {\bibinfo {author} {\bibfnamefont {S.}~\bibnamefont
  {Weinberg}}\ and\ \bibinfo {author} {\bibfnamefont {T.}~\bibnamefont
  {de~Campos}},\ }\href {https://books.google.co.jp/books?id=doeDB3\_WLvwC}
  {\emph {\bibinfo {title} {The Quantum Theory of Fields}}},\ Quantum Theory of
  Fields, Vol. 2: Modern Applications\ (\bibinfo  {publisher} {Cambridge
  University Press, Cambridge, England, 1995},\ \bibinfo {year}
  {1995})\BibitemShut {NoStop}%
\bibitem [{\citenamefont {McLerran}\ and\ \citenamefont
  {Pisarski}(2007)}]{McLerran:2007qj}%
  \BibitemOpen
  \bibfield  {author} {\bibinfo {author} {\bibfnamefont {L.}~\bibnamefont
  {McLerran}}\ and\ \bibinfo {author} {\bibfnamefont {R.~D.}\ \bibnamefont
  {Pisarski}},\ }\href {https://doi.org/10.1016/j.nuclphysa.2007.08.013}
  {\bibfield  {journal} {\bibinfo  {journal} {Nucl. Phys. A}\ }\textbf
  {\bibinfo {volume} {796}},\ \bibinfo {pages} {83} (\bibinfo {year} {2007})},\
  \Eprint {https://arxiv.org/abs/0706.2191} {arXiv:0706.2191 [hep-ph]}
  \BibitemShut {NoStop}%
\bibitem [{\citenamefont {Fukushima}\ \emph {et~al.}(2020)\citenamefont
  {Fukushima}, \citenamefont {Kojo},\ and\ \citenamefont
  {Weise}}]{Fukushima:2020cmk}%
  \BibitemOpen
  \bibfield  {author} {\bibinfo {author} {\bibfnamefont {K.}~\bibnamefont
  {Fukushima}}, \bibinfo {author} {\bibfnamefont {T.}~\bibnamefont {Kojo}},\
  and\ \bibinfo {author} {\bibfnamefont {W.}~\bibnamefont {Weise}},\ }\href
  {https://doi.org/10.1103/PhysRevD.102.096017} {\bibfield  {journal} {\bibinfo
   {journal} {Phys. Rev. D}\ }\textbf {\bibinfo {volume} {102}},\ \bibinfo
  {pages} {096017} (\bibinfo {year} {2020})},\ \Eprint
  {https://arxiv.org/abs/2008.08436} {arXiv:2008.08436 [hep-ph]} \BibitemShut
  {NoStop}%
\bibitem [{\citenamefont {Jeong}\ \emph {et~al.}(2020)\citenamefont {Jeong},
  \citenamefont {McLerran},\ and\ \citenamefont {Sen}}]{Jeong:2019lhv}%
  \BibitemOpen
  \bibfield  {author} {\bibinfo {author} {\bibfnamefont {K.~S.}\ \bibnamefont
  {Jeong}}, \bibinfo {author} {\bibfnamefont {L.}~\bibnamefont {McLerran}},\
  and\ \bibinfo {author} {\bibfnamefont {S.}~\bibnamefont {Sen}},\ }\href
  {https://doi.org/10.1103/PhysRevC.101.035201} {\bibfield  {journal} {\bibinfo
   {journal} {Phys. Rev. C}\ }\textbf {\bibinfo {volume} {101}},\ \bibinfo
  {pages} {035201} (\bibinfo {year} {2020})},\ \Eprint
  {https://arxiv.org/abs/1908.04799} {arXiv:1908.04799 [nucl-th]} \BibitemShut
  {NoStop}%
\bibitem [{\citenamefont {Duarte}\ \emph
  {et~al.}(2020{\natexlab{a}})\citenamefont {Duarte}, \citenamefont
  {Hernandez-Ortiz},\ and\ \citenamefont {Jeong}}]{Duarte:2020xsp}%
  \BibitemOpen
  \bibfield  {author} {\bibinfo {author} {\bibfnamefont {D.~C.}\ \bibnamefont
  {Duarte}}, \bibinfo {author} {\bibfnamefont {S.}~\bibnamefont
  {Hernandez-Ortiz}},\ and\ \bibinfo {author} {\bibfnamefont {K.~S.}\
  \bibnamefont {Jeong}},\ }\href {https://doi.org/10.1103/PhysRevC.102.025203}
  {\bibfield  {journal} {\bibinfo  {journal} {Phys. Rev. C}\ }\textbf {\bibinfo
  {volume} {102}},\ \bibinfo {pages} {025203} (\bibinfo {year}
  {2020}{\natexlab{a}})},\ \Eprint {https://arxiv.org/abs/2003.02362}
  {arXiv:2003.02362 [nucl-th]} \BibitemShut {NoStop}%
\bibitem [{\citenamefont {Duarte}\ \emph
  {et~al.}(2020{\natexlab{b}})\citenamefont {Duarte}, \citenamefont
  {Hernandez-Ortiz},\ and\ \citenamefont {Jeong}}]{Duarte:2020kvi}%
  \BibitemOpen
  \bibfield  {author} {\bibinfo {author} {\bibfnamefont {D.~C.}\ \bibnamefont
  {Duarte}}, \bibinfo {author} {\bibfnamefont {S.}~\bibnamefont
  {Hernandez-Ortiz}},\ and\ \bibinfo {author} {\bibfnamefont {K.~S.}\
  \bibnamefont {Jeong}},\ }\href@noop {} {\  (\bibinfo {year}
  {2020}{\natexlab{b}})},\ \Eprint {https://arxiv.org/abs/2007.08098}
  {arXiv:2007.08098 [nucl-th]} \BibitemShut {NoStop}%
\bibitem [{\citenamefont {Zhao}\ and\ \citenamefont
  {Lattimer}(2020)}]{Zhao:2020dvu}%
  \BibitemOpen
  \bibfield  {author} {\bibinfo {author} {\bibfnamefont {T.}~\bibnamefont
  {Zhao}}\ and\ \bibinfo {author} {\bibfnamefont {J.~M.}\ \bibnamefont
  {Lattimer}},\ }\href@noop {} {\  (\bibinfo {year} {2020})},\ \Eprint
  {https://arxiv.org/abs/2004.08293} {arXiv:2004.08293 [astro-ph.HE]}
  \BibitemShut {NoStop}%
\bibitem [{\citenamefont {Sen}\ and\ \citenamefont
  {Warrington}(2020)}]{Sen:2020peq}%
  \BibitemOpen
  \bibfield  {author} {\bibinfo {author} {\bibfnamefont {S.}~\bibnamefont
  {Sen}}\ and\ \bibinfo {author} {\bibfnamefont {N.~C.}\ \bibnamefont
  {Warrington}},\ }\href@noop {} {\  (\bibinfo {year} {2020})},\ \Eprint
  {https://arxiv.org/abs/2002.11133} {arXiv:2002.11133 [nucl-th]} \BibitemShut
  {NoStop}%
\bibitem [{\citenamefont {Cao}\ and\ \citenamefont {Liao}(2020)}]{Cao:2020byn}%
  \BibitemOpen
  \bibfield  {author} {\bibinfo {author} {\bibfnamefont {G.}~\bibnamefont
  {Cao}}\ and\ \bibinfo {author} {\bibfnamefont {J.}~\bibnamefont {Liao}},\
  }\href {https://doi.org/10.1007/JHEP10(2020)168} {\bibfield  {journal}
  {\bibinfo  {journal} {JHEP}\ }\textbf {\bibinfo {volume} {10}},\ \bibinfo
  {pages} {168}},\ \Eprint {https://arxiv.org/abs/2007.02028} {arXiv:2007.02028
  [nucl-th]} \BibitemShut {NoStop}%
\bibitem [{\citenamefont {Kojo}(2017)}]{Kojo:2016dhh}%
  \BibitemOpen
  \bibfield  {author} {\bibinfo {author} {\bibfnamefont {T.}~\bibnamefont
  {Kojo}},\ }\href {https://doi.org/10.1016/j.physletb.2017.03.023} {\bibfield
  {journal} {\bibinfo  {journal} {Phys. Lett. B}\ }\textbf {\bibinfo {volume}
  {769}},\ \bibinfo {pages} {14} (\bibinfo {year} {2017})},\ \Eprint
  {https://arxiv.org/abs/1610.05486} {arXiv:1610.05486 [hep-ph]} \BibitemShut
  {NoStop}%
\bibitem [{\citenamefont {Bowers}\ and\ \citenamefont
  {Rajagopal}(2002)}]{Bowers:2002xr}%
  \BibitemOpen
  \bibfield  {author} {\bibinfo {author} {\bibfnamefont {J.~A.}\ \bibnamefont
  {Bowers}}\ and\ \bibinfo {author} {\bibfnamefont {K.}~\bibnamefont
  {Rajagopal}},\ }\href {https://doi.org/10.1103/PhysRevD.66.065002} {\bibfield
   {journal} {\bibinfo  {journal} {Phys. Rev. D}\ }\textbf {\bibinfo {volume}
  {66}},\ \bibinfo {pages} {065002} (\bibinfo {year} {2002})},\ \Eprint
  {https://arxiv.org/abs/hep-ph/0204079} {arXiv:hep-ph/0204079} \BibitemShut
  {NoStop}%
\bibitem [{\citenamefont {Casalbuoni}\ and\ \citenamefont
  {Nardulli}(2004)}]{Casalbuoni:2003wh}%
  \BibitemOpen
  \bibfield  {author} {\bibinfo {author} {\bibfnamefont {R.}~\bibnamefont
  {Casalbuoni}}\ and\ \bibinfo {author} {\bibfnamefont {G.}~\bibnamefont
  {Nardulli}},\ }\href {https://doi.org/10.1103/RevModPhys.76.263} {\bibfield
  {journal} {\bibinfo  {journal} {Rev. Mod. Phys.}\ }\textbf {\bibinfo {volume}
  {76}},\ \bibinfo {pages} {263} (\bibinfo {year} {2004})},\ \Eprint
  {https://arxiv.org/abs/hep-ph/0305069} {arXiv:hep-ph/0305069} \BibitemShut
  {NoStop}%
\bibitem [{\citenamefont {Anglani}\ \emph {et~al.}(2014)\citenamefont
  {Anglani}, \citenamefont {Casalbuoni}, \citenamefont {Ciminale},
  \citenamefont {Ippolito}, \citenamefont {Gatto}, \citenamefont {Mannarelli},\
  and\ \citenamefont {Ruggieri}}]{Anglani:2013gfu}%
  \BibitemOpen
  \bibfield  {author} {\bibinfo {author} {\bibfnamefont {R.}~\bibnamefont
  {Anglani}}, \bibinfo {author} {\bibfnamefont {R.}~\bibnamefont {Casalbuoni}},
  \bibinfo {author} {\bibfnamefont {M.}~\bibnamefont {Ciminale}}, \bibinfo
  {author} {\bibfnamefont {N.}~\bibnamefont {Ippolito}}, \bibinfo {author}
  {\bibfnamefont {R.}~\bibnamefont {Gatto}}, \bibinfo {author} {\bibfnamefont
  {M.}~\bibnamefont {Mannarelli}},\ and\ \bibinfo {author} {\bibfnamefont
  {M.}~\bibnamefont {Ruggieri}},\ }\href
  {https://doi.org/10.1103/RevModPhys.86.509} {\bibfield  {journal} {\bibinfo
  {journal} {Rev. Mod. Phys.}\ }\textbf {\bibinfo {volume} {86}},\ \bibinfo
  {pages} {509} (\bibinfo {year} {2014})},\ \Eprint
  {https://arxiv.org/abs/1302.4264} {arXiv:1302.4264 [hep-ph]} \BibitemShut
  {NoStop}%
\bibitem [{\citenamefont {Kojo}\ \emph
  {et~al.}(2010{\natexlab{a}})\citenamefont {Kojo}, \citenamefont {Hidaka},
  \citenamefont {McLerran},\ and\ \citenamefont {Pisarski}}]{Kojo:2009ha}%
  \BibitemOpen
  \bibfield  {author} {\bibinfo {author} {\bibfnamefont {T.}~\bibnamefont
  {Kojo}}, \bibinfo {author} {\bibfnamefont {Y.}~\bibnamefont {Hidaka}},
  \bibinfo {author} {\bibfnamefont {L.}~\bibnamefont {McLerran}},\ and\
  \bibinfo {author} {\bibfnamefont {R.~D.}\ \bibnamefont {Pisarski}},\ }\href
  {https://doi.org/10.1016/j.nuclphysa.2010.05.053} {\bibfield  {journal}
  {\bibinfo  {journal} {Nucl. Phys. A}\ }\textbf {\bibinfo {volume} {843}},\
  \bibinfo {pages} {37} (\bibinfo {year} {2010}{\natexlab{a}})},\ \Eprint
  {https://arxiv.org/abs/0912.3800} {arXiv:0912.3800 [hep-ph]} \BibitemShut
  {NoStop}%
\bibitem [{\citenamefont {Andersen}\ and\ \citenamefont
  {Kneschke}(2018)}]{Andersen:2018osr}%
  \BibitemOpen
  \bibfield  {author} {\bibinfo {author} {\bibfnamefont {J.~O.}\ \bibnamefont
  {Andersen}}\ and\ \bibinfo {author} {\bibfnamefont {P.}~\bibnamefont
  {Kneschke}},\ }\href {https://doi.org/10.1103/PhysRevD.97.076005} {\bibfield
  {journal} {\bibinfo  {journal} {Phys. Rev. D}\ }\textbf {\bibinfo {volume}
  {97}},\ \bibinfo {pages} {076005} (\bibinfo {year} {2018})},\ \Eprint
  {https://arxiv.org/abs/1802.01832} {arXiv:1802.01832 [hep-ph]} \BibitemShut
  {NoStop}%
\bibitem [{\citenamefont {Ferrer}\ and\ \citenamefont {de~la
  Incera}(2020)}]{Ferrer:2019zfp}%
  \BibitemOpen
  \bibfield  {author} {\bibinfo {author} {\bibfnamefont {E.~J.}\ \bibnamefont
  {Ferrer}}\ and\ \bibinfo {author} {\bibfnamefont {V.}~\bibnamefont {de~la
  Incera}},\ }\href {https://doi.org/10.1103/PhysRevD.102.014010} {\bibfield
  {journal} {\bibinfo  {journal} {Phys. Rev. D}\ }\textbf {\bibinfo {volume}
  {102}},\ \bibinfo {pages} {014010} (\bibinfo {year} {2020})},\ \Eprint
  {https://arxiv.org/abs/1902.06810} {arXiv:1902.06810 [nucl-th]} \BibitemShut
  {NoStop}%
\bibitem [{\citenamefont {Buballa}\ and\ \citenamefont
  {Carignano}(2015)}]{Buballa:2014tba}%
  \BibitemOpen
  \bibfield  {author} {\bibinfo {author} {\bibfnamefont {M.}~\bibnamefont
  {Buballa}}\ and\ \bibinfo {author} {\bibfnamefont {S.}~\bibnamefont
  {Carignano}},\ }\href {https://doi.org/10.1016/j.ppnp.2014.11.001} {\bibfield
   {journal} {\bibinfo  {journal} {Prog. Part. Nucl. Phys.}\ }\textbf {\bibinfo
  {volume} {81}},\ \bibinfo {pages} {39} (\bibinfo {year} {2015})},\ \Eprint
  {https://arxiv.org/abs/1406.1367} {arXiv:1406.1367 [hep-ph]} \BibitemShut
  {NoStop}%
\bibitem [{\citenamefont {Nakano}\ and\ \citenamefont
  {Tatsumi}(2005)}]{Nakano:2004cd}%
  \BibitemOpen
  \bibfield  {author} {\bibinfo {author} {\bibfnamefont {E.}~\bibnamefont
  {Nakano}}\ and\ \bibinfo {author} {\bibfnamefont {T.}~\bibnamefont
  {Tatsumi}},\ }\href {https://doi.org/10.1103/PhysRevD.71.114006} {\bibfield
  {journal} {\bibinfo  {journal} {Phys. Rev. D}\ }\textbf {\bibinfo {volume}
  {71}},\ \bibinfo {pages} {114006} (\bibinfo {year} {2005})},\ \Eprint
  {https://arxiv.org/abs/hep-ph/0411350} {arXiv:hep-ph/0411350} \BibitemShut
  {NoStop}%
\bibitem [{\citenamefont {Kojo}(2014)}]{Kojo:2014fxa}%
  \BibitemOpen
  \bibfield  {author} {\bibinfo {author} {\bibfnamefont {T.}~\bibnamefont
  {Kojo}},\ }\href {https://doi.org/10.1103/PhysRevD.90.065030} {\bibfield
  {journal} {\bibinfo  {journal} {Phys. Rev. D}\ }\textbf {\bibinfo {volume}
  {90}},\ \bibinfo {pages} {065030} (\bibinfo {year} {2014})},\ \Eprint
  {https://arxiv.org/abs/1406.4630} {arXiv:1406.4630 [hep-ph]} \BibitemShut
  {NoStop}%
\bibitem [{\citenamefont {Kojo}\ \emph
  {et~al.}(2010{\natexlab{b}})\citenamefont {Kojo}, \citenamefont {Pisarski},\
  and\ \citenamefont {Tsvelik}}]{Kojo:2010fe}%
  \BibitemOpen
  \bibfield  {author} {\bibinfo {author} {\bibfnamefont {T.}~\bibnamefont
  {Kojo}}, \bibinfo {author} {\bibfnamefont {R.~D.}\ \bibnamefont {Pisarski}},\
  and\ \bibinfo {author} {\bibfnamefont {A.}~\bibnamefont {Tsvelik}},\ }\href
  {https://doi.org/10.1103/PhysRevD.82.074015} {\bibfield  {journal} {\bibinfo
  {journal} {Phys. Rev. D}\ }\textbf {\bibinfo {volume} {82}},\ \bibinfo
  {pages} {074015} (\bibinfo {year} {2010}{\natexlab{b}})},\ \Eprint
  {https://arxiv.org/abs/1007.0248} {arXiv:1007.0248 [hep-ph]} \BibitemShut
  {NoStop}%
\bibitem [{\citenamefont {Kojo}\ \emph {et~al.}(2012)\citenamefont {Kojo},
  \citenamefont {Hidaka}, \citenamefont {Fukushima}, \citenamefont {McLerran},\
  and\ \citenamefont {Pisarski}}]{Kojo:2011cn}%
  \BibitemOpen
  \bibfield  {author} {\bibinfo {author} {\bibfnamefont {T.}~\bibnamefont
  {Kojo}}, \bibinfo {author} {\bibfnamefont {Y.}~\bibnamefont {Hidaka}},
  \bibinfo {author} {\bibfnamefont {K.}~\bibnamefont {Fukushima}}, \bibinfo
  {author} {\bibfnamefont {L.~D.}\ \bibnamefont {McLerran}},\ and\ \bibinfo
  {author} {\bibfnamefont {R.~D.}\ \bibnamefont {Pisarski}},\ }\href
  {https://doi.org/10.1016/j.nuclphysa.2011.11.007} {\bibfield  {journal}
  {\bibinfo  {journal} {Nucl. Phys. A}\ }\textbf {\bibinfo {volume} {875}},\
  \bibinfo {pages} {94} (\bibinfo {year} {2012})},\ \Eprint
  {https://arxiv.org/abs/1107.2124} {arXiv:1107.2124 [hep-ph]} \BibitemShut
  {NoStop}%
\end{thebibliography}%



\end{document}